\documentclass[doublespacing,onecolumn,extra]{gji}
\usepackage{timet}
\usepackage{amsfonts}
\usepackage{mathrsfs}
\usepackage{graphicx}
\usepackage{bm}
\usepackage{subfigure}
\usepackage{ amssymb }
\usepackage{tikz-cd}
\usepackage{csquotes}

\title[Linear inference problems]
      {Linear inference problems with deterministic constraints} 
\author[David Al-Attar]
  {David Al-Attar   \\
    Bullard Laboratories, University of Cambridge, Madingley Rise, CB3 0EZ, UK. 
  }      
\date{Received ?; in original form ?}
\pagerange{\pageref{firstpage}--\pageref{lastpage}}
\volume{142}
\pubyear{2020}

\newcommand{\dd}{\,\mathrm{d}}
\newcommand{\ddns}{\mathrm{d}}
\newcommand{\ee}{\mathrm{e}}

\newcommand{\tr}{\mathrm{tr}}

\newcommand{\image}{\mathrm{im}}

\newcommand{\eproof}{\begin{flushright}$\blacksquare$\end{flushright}}

\newcommand{\braket}[2]{\left\langle{#1},{#2}\right\rangle}
\newcommand{\cbraket}[2]{\left({#1},{#2}\right)}
\newcommand{\ccbraket}[2]{\left(\left({#1},{#2}\right)\right)}

\newcommand{\mult}[2]{\left\langle{#1}\right\rangle_{#2}}

\newcommand{\Hom}{\mathrm{Hom}}
\newcommand{\vertiii}[1]{{\left\vert\kern-0.25ex\left\vert\kern-0.25ex\left\vert #1 
    \right\vert\kern-0.25ex\right\vert\kern-0.25ex\right\vert}}

\newtheorem{thm}{Theorem}[section]
\newtheorem{lem}{Lemma}[section]
\newtheorem{prop}{Proposition}[section]
\newtheorem{coro}{Corollary}[section]
\newtheorem{asm}{Assumption}[section]

\eqsecnum

\begin{document}

\label{firstpage}

\maketitle

\begin{summary}

  Methods are described for the solution of linear inference problems subject to
  deterministic constraints. The approach  builds on  work
  by \cite{backus1970inferenceI,backus1970inferenceII,backus1970inferenceIII} and
  \cite{parker1977linear}, but a range useful advances are suggested to  address
  both conceptual and practical issues.   The theory   is motivated by, and illustrated with,
  the estimation of a finite number of a function's spherical harmonic coefficients from a
  finite set of its  point values.  Numerical examples are included to demonstrate that
  the methods can be efficiently applied to realistic problems.
\end{summary}

\begin{keywords}
Inverse theory; Statistical methods; Numerical approximations and analysis.
\end{keywords}

\section{Introduction}

\label{sec:intro}

This paper describes   methods for solving linear inference problems.
Any work of this topic  owes a profound debt  to 
\cite{backusgilbert,backus1967numerical,backus1970uniqueness},
\cite{backus1970inferenceI, backus1970inferenceII,
  backus1970inferenceIII,backus1972,backus1988bayesian,backus1989confidence,backus1996trimming},  and
\cite{parker1972inverse,parker1977linear}, along with other notable  contributions including
\cite{SOLA1,SOLA2}, \cite{genovese1996data}, \cite{evans2002inverse},
and \cite{stark2008generalizing}.  At the outset it is worth stating clearly
 what is meant by an   inference problem as opposed to an  inverse problem.
Within each we are given data  that is related in a
specified manner to an unknown model.  In broad terms,
the inverse problem seeks to recover the unknown model  as best as possible. By contrast, 
an inference problem aims merely  to 
estimate  finitely-many  numerical properties of the model that are of interest.

From a solution of an inverse problem one can trivially obtain a corresponding solution
for an associated inference problem. But not \emph{all} solutions of inference problems need be
obtained in this manner, with  such an approach  generally being sub-optimal
both theoretically and computationally. The basic reason is that while
inverse problems with finite data generically have an infinite-fold non-uniqueness, only
a finite-dimensional component of this non-uniqueness is \emph{seen} by the  properties 
estimated within an inference problem. As a result, while uncertainty quantification
within inverse problems remains very difficult (from a
computationally perspective if nothing else),  significantly more can be done within
the   context of an inference problem. This point is especially
relevant to situations where the physics and data are such that no
plausible reconstruction of the model can be obtained \citep[][]{parker1972inverse}.

To give one example, consider lateral density variations
in the lowermost mantle, and particularly with regard to the two
large low shear velocity provinces (LLSVPs) whose nature and geodynamic significance remains
unclear. The geophysical observations
that are  sensitive to any such density variations are principally:
low-frequency free oscillations, body tides, and long-wavelength
geoid anomalies \citep[e.g.][]{ishii1999normal,lau2017tidal}. However, these observations
see only very broad spatial averages of lower-mantle lateral
density variations (along with other parameters). There is no physical basis for believing that
structures do not exist in the lower mantle below the resolvable length-scales. And hence
inversion of this data must always lead to models that are overly smooth and contain
substantial uncertainties which are difficult to quantify. A quite different approach would
be simply to ask what the average density anomalies within the two LLSVPs are. This
is an inference problem in which only two numbers are to be estimated, but if
it could be solved -- meaning  all plausible   average densities are determined --
then the result would be of unquestionable value in understanding the deep Earth.

It might here  be added more generally that  geophysical inference problems are necessarily
associated with  specific quantitative questions about the Earth. In contrast,  
geophysical inverse problems  generally focus on \emph{model building}, this usually being
done in the hope that some new and interesting  qualitative feature will be discovered.
Both approaches have value, of course. And a model building exercise that reveals an
interesting feature might be followed by a more targeted study that asks
specific quantitative questions. Indeed, this is precisely what is suggested above
with regard to lower mantle density. But it is important to remember that inverse and inference problems
are  distinct in their aims; this is reflected in the methods developed,
and the manner in which their efficacy should be judged.

It is the author's experience that the literature on  inference problems is
considerably less well known within  geophysics than that on  inverse problems.
This seems a shame, particularly so because  this approach came first,   from within
our community, and was motivated directly by the grossly under-determined
problems that are  familiar to us, but comparatively rare in other parts of physics.
Perhaps most  have at least \emph{heard} of Backus-Gilbert theory, but often only  as a historical
curiosity that might be cited but need not be understood.  There are
exceptions, of course, including the recent work of \cite{zaroli2016global,zaroli2019seismic} who
applied a variant of Backus-Gilbert theory
to linearised seismic tomography. But even these admirable studies have been done in seeming
ignorance of later work by Backus and others. It is only here that the  qualitative approach
to uncertainties within Backus-Gilbert theory based on resolution length is replaced by a
fully quantitative theory in which the need for suitable prior constraints is made clear.

Why then have inference problems received comparatively little attention in recent times? Here
one can only speculate, but there seem to be two issues. The first is that the numerical cost  of
the methods was thought to be prohibitively high. For example, \cite{parker1977linear}
described a method for solving linear inference problems  with a prior norm bound 
in the error-free case. To do this it is necessary to invert a square matrix with
dimension equal to the sum of the number of data and the number of quantities to be estimated.
At the time these linear systems were simply too large
within any real  application.  Moreover, while Parker discussed in outline how
random data errors could be incorporated into the problem,  the method's numerical implementation
would be both complicated and require a substantial increase in costs. Parker concluded that
his approach, while theoretically superior, was simply not competitive with the
discretised least squares methods that had become popular \citep[e.g][]{wiggins1972general}.
Times change, however, and what was once  impossible numerically is now routine. In particular,
iterative matrix-free
methods mean that there need be no difficulty in solving linear equations in high-dimensional
spaces. Indeed, an important contribution of \cite{zaroli2016global,zaroli2019seismic} was to
show that the oft-stated objection to applying Backus-Gilbert theory to  large-scale problems
can be overcome through the application of modern computational techniques. In a similar manner,
this paper describes a new approach to solving linear inference problems  
that is computationally efficient and well-suited to problems with large data sets.
In the error-free case the theory is equivalent to that in \cite{backus1970inferenceI}
and \cite{parker1977linear}, but significant computational savings are made possible
through both the formulation of the theoretical results and the numerical methods applied.
To incorporate data errors
a new approach is developed that builds on the largely qualitative discussion in \cite{parker1977linear}.
While there is an increase in computational costs over the error-free case, the method
remains practically feasible for realistic applications. 
This new approach also has the advantage of working not just in the case of Gaussian
errors, but also for a wider class of unimodal distributions. 

The second point that seems to  limit work on linear inference problems is  the
perceived difficulty of the  literature.  In almost all geophysical inverse and
inference problems the  model comprises a function  belonging to an infinite-dimensional
vector space. In discussing these problems it is, therefore, necessary to use the  methods and  language
of functional analysis, albeit only at a low level. Within papers on linear inference problems this has always been done, and it
is  true for this paper also. In contrast
much work on geophysical inverse problems assumes from the outset a finite-dimensional model space
\citep[e.g.][]{tarantola2005inverse,wunsch2006discrete, menke2018geophysical}.
This is not, of course, to say that function space methods are not used within
the literature on geophysical inverse theory
\citep[e.g.][]{parker}, but only that such techniques can be avoided by those
who do not know, and will not learn, the necessary ideas.
There is generally no physical reason for working with  a finite-dimensional
model space, however, and this step is done only to make the  mathematics easier. If all
that is desired are models that fit the data,
then there is no harm in doing this so long as the discretised
space is sufficiently large. However, as soon as questions of uncertainty arise,
then by limiting the size of the model space it is inevitable
that errors will be \emph{underestimated}. The extent to which this matters is
necessarily application dependent, but usually cannot be quantified. The issue can
perhaps be mitigated  by allowing the dimension
of the approximating space to vary \citep[e.g.][]{sambridge2006trans}, but in practice
the range of dimensions explored is small.

The passage from finite- to infinite-dimensions is in some respects easy, and yet in
others rather subtle. Properties that can be taken for granted in finite-dimensions
can be lost, while   entirely new phenomena can occur.
For example, in finite-dimensional linear problems  the model space has a unique topology, but in
infinite-dimensions the model space topology must be specified as
an essential part of the problem's formulation. Moreover, there is usually a
\emph{choice} about what topology is selected, and hence one needs  to carefully consider how  this
is done.  In almost all the literature on geophysical inference problems, however, the model space has
been assumed to be a Hilbert space.  But this has largely been done for
convenience, with most geophysical problems being more naturally posed on Banach spaces.
Notable exceptions  are  \cite{evans2002inverse} and \cite{stark2008generalizing} who generalised
Backus-Gilbert estimators from the perspective of statistical inference theory. A specific aim of this paper is the
formulation and characterisation of  linear inference problems in a Banach space. While there
are links to what is done in  \cite{evans2002inverse} and \cite{stark2008generalizing}, the focus
here is primarily on ideas found in Backus' later work done independently of Gilbert. In particular,
three key results given by \cite{backus1970inferenceI} for problems in Hilbert spaces are generalised
and extended. It will be shown that while a complete   theory can be developed in a
Banach space setting, its implementation requires the solution of some very difficult, and perhaps
intractable, optimisation problems \citep[c.f.][]{stark2015constraints}. As a pragmatic step,
the results are then reduced to the Hilbert space case for which the necessary calculations are easy.
In contrast to earlier work, however, this simplifying step is made explicit, and
a  discussion is given as to how an appropriate Hilbert space can be both identified
and  justified.

As noted,  work on linear inference problems necessitates a basic understanding of
functional analysis, and this has likely been an impediment to people who might otherwise
be interested. Moreover, in trying to present new ideas in this field it  cannot be
reasonably assumed that all  readers will be conversant with what has  been done previously.
As a result,  this paper has been written to  be largely self-contained, while appendices are
included to summarise mathematical concepts and notations that may be unfamiliar. In doing this
there is a cost in terms of length, while some potentially  routine material needs to be repeated.  But it
is hoped that this approach makes for an easier  read, though perhaps still not an easy one. Chapter 1 of
\cite{parker} also provides a nice introduction to functional analysis, and covers most
of what it required in this work. Proofs are given either when a result is thought
to be original, or if the argument is short and some useful insight can be gained.
Readers who wish to skip any or all of the proofs should still be able to follow the
 the main arguments.

\subsection{Spectral estimation from point data on a sphere}

\label{ssec:prelim}

\begin{figure}
  \centering
  \includegraphics[width=\textwidth,trim={0cm 3.25cm 0cm 7.25cm},clip]{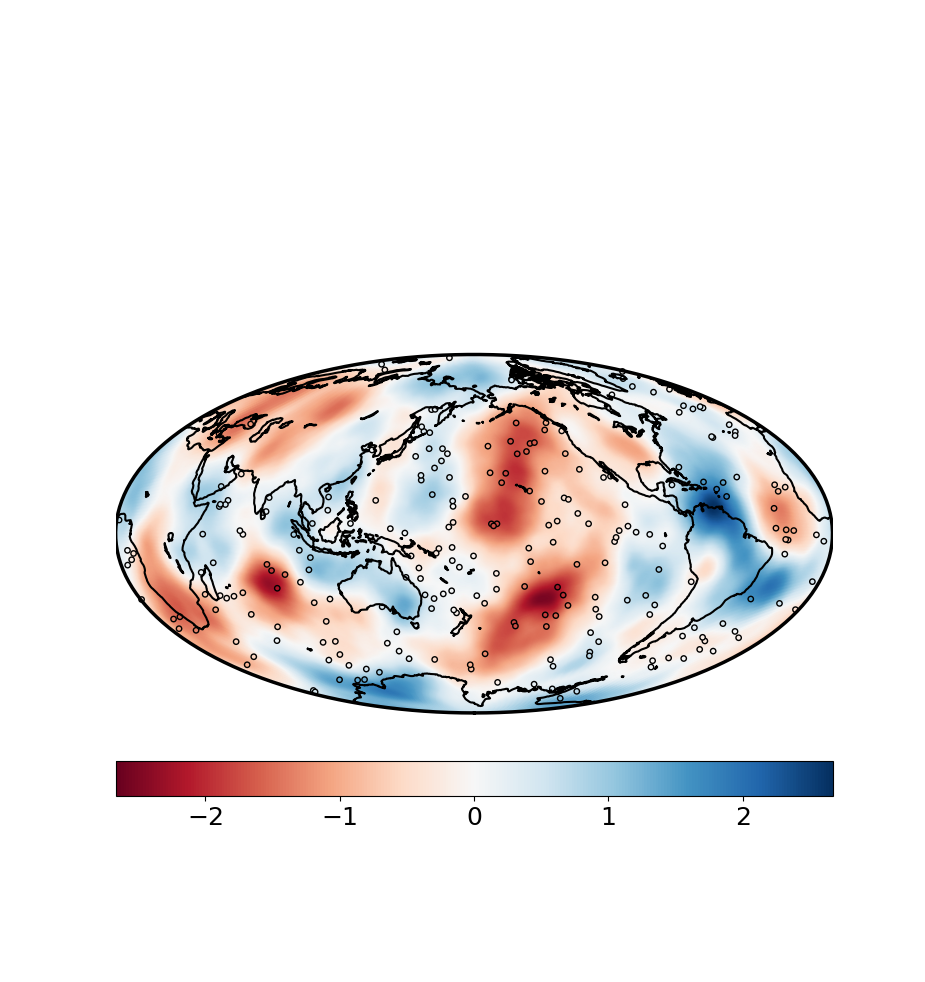}
  \caption{An example of a function on the unit sphere that has been sampled at 250 locations indicated
    by  circular markers. The value of the function at each  location is indicated by
    the colour of the circle using the same scale as the main plot. Coastlines are shown,
    and it can be seen that all sample locations have been chosen to lie below sea level.
    All  map plots later in the paper have been made in an identical manner using the same colour scale.
  }
  \label{fig:data_ne}
\end{figure}

To  explain further  the aims of this paper, it will be useful to turn  to a concrete 
inference problem motivated by the work of \cite{hoggard2016} on dynamic topography.
Let $\{x_{i}\}_{i=1}^{n}$  be a set of distinct points on the unit two-sphere, $\mathbb{S}^{2}$,
and $u:\mathbb{S}^{2}\rightarrow \mathbb{R}$ an unknown continuous function. Suppose
we are given the point values $\{u(x_{i})\}_{i=1}^{n}$,  and from
this data wish to estimate the function's $(l,m)$th spherical harmonic coefficient.
In practice, we might require  simultaneous estimates of some finite number of
spherical harmonic coefficients (e.g. all those at  degree $l$ from which we could
compute the associated power), while all real
data are subject to errors.
Such generalisations will be discussed later, but for the moment we  keep
the problem as simple as possible.  Fig.\ref{fig:data_ne} shows an 
example of the data set along with the underlying function. In accordance with
 \cite{hoggard2016}, all points lie below sea level, and hence there are large regions of the
 domain where the function is completely unconstrained by the data.

 An associated inverse problem can of course also be
 considered in this situation. Namely, the   reconstruction of the unknown function from the given point
 data. Having done this by some means, the associated value of the $(l,m)$th spherical
 harmonic coefficient could then be extracted. This is precisely what 
 \cite{hoggard2016} did via a simple discretised least squares method. A range of other approaches
 are possible for reconstructing the function
 \citep[e.g.][]{valentine2020global}, including    function-space techniques
  designed specifically for interpolation of point data on a sphere
 \citep[e.g.][]{freeden1995survey}. Error estimates
 on the solution of the inverse problem, if obtained, can be propagated through to the
 spherical harmonic coefficient, and hence uncertainties quantified. As noted previously,
 however,  the optimal approach to solving an inference problems is not necessarily  to 
 first solve  an inverse problem.

We denote by $C^{0}(\mathbb{S}^{2})$ the set of all continuous real-valued functions
on $\mathbb{S}^{2}$. Relative to the supremum norm
\begin{equation}
  \label{eq:supnorm}
  \|u\|_{C^{0}(\mathbb{S}^{2})} = \sup_{x \in \mathbb{S}^{2}}|u(x)|, 
\end{equation}
this is a Banach space.  For each $x \in \mathbb{S}^{2}$, the mapping
\begin{equation}
  u \mapsto u(x), 
\end{equation}
is a continuous linear functional on $C^{0}(\mathbb{S}^{2})$, and hence defines an element of the
dual space $C^{0}(\mathbb{S}^{2})'$ that we denote by $\delta_{x}$ and call the Dirac measure at $x$. 
The $i$th datum can then be written in the form
\begin{equation}
  v_{i} = \braket{\delta_{x_{i}}}{u}, 
\end{equation}
where  an angular bracket is used to denote a dual pairing. It is worth commenting that this relation
\emph{might} also be written 
\begin{equation}
  \label{eq:dist}
  v_{i} = \int_{\mathbb{S}^{2}} \delta_{x_{i}}(x) \, u(x) \dd S, 
\end{equation}
using the Dirac delta function
  on $\mathbb{S}^{2}$. Having done this, we could identify
\emph{sensitivity kernel} for the $i$th datum as a delta function based at the observation point.
While this notation is  familiar and suggestive, it does have a downside. While the right hand side of 
eq.(\ref{eq:dist}) \emph{looks} like an inner product in the space $L^{2}(\mathbb{S}^{2})$,
this is not at all the case. There are inference problems on $L^{2}(\mathbb{S}^{2})$ for which the data
takes the form
\begin{equation}
  \label{eq:dist}
  v_{i} = \int_{\mathbb{S}^{2}} k_{i}(x) \, u(x) \dd S, 
\end{equation}
with the $i$th sensitivity kernel, $k_{i}$, is an element of $L^{2}(\mathbb{S}^{2})$. For such
problems, the Hilbert space methods developed by Backus  can be applied directly.
But to confuse the present problem, posed in a more general Banach space, with the $L^{2}$-case can only
lead to confusion \citep[e.g.][Chapter 9]{gubbins2004time}.  Returning to the problem at hand, each spherical
harmonic $Y_{lm}$ can be uniquely associated with
a dual vector (denoted for simplicity with the same symbol) that is defined through
\begin{equation}
  \label{eq:sphfun}
 \braket{Y_{lm}}{u} = \int_{\mathbb{S}^{2}} u\, Y_{lm} \dd S
\end{equation}
 where  $\ddns S$ is the standard surface element on $\mathbb{S}^{2}$. For definiteness,   fully normalised real spherical harmonics
as defined in Appendix B of \cite{DT} are used throughout this paper. 
Given this preamble, our first result shows that  the data alone provide no
information about the  spherical harmonic coefficient of interest:

\begin{prop}
  \label{prop:nonu1}
  For any $v_{1},\dots,v_{n},w \in \mathbb{R}$ there
  exists a (non-unique)  function $u \in C^{0}(\mathbb{S}^{2})$ such that
  \begin{equation}
    v_{i} = \braket{\delta_{x_{i}}}{u},\quad i \in \{ 1,\dots,n\},
  \end{equation}
  while  simultaneously  $w= \braket{Y_{lm}}{u}$. 
\end{prop}

\emph{Proof:} A simple  constructive proof can be given using Urysohn's Lemma
\citep[e.g.][]{kelleygeneral}.
First, we select disjoint open subsets $\{U_{i}\}_{i=1}^{n}$ of $\mathbb{S}^{2}$
such that $x_{1} \in U_{1}, \dots,x_{n} \in U_{n}$. For each $i \in \{1,\dots,n\}$, we then choose $V_{i}$ to be open, properly
contained in $U_{i}$, and have $x_{i} \in V_{i}$.  Urysohn's Lemma yields for each $i \in  \{1,\dots,n\}$
a continuous function $\varphi_{i}:\mathbb{S}^{2}\rightarrow [0,1]$ such that $\varphi_{i} = 1$ within $V_{i}$ and
$\varphi_{i} = 0$ on the complement of $U_{i}$.  We now define  the function
\begin{equation}
  \label{eq:udef1}
  u = \sum_{i=1}^{n}v_{i}\varphi_{i} + \alpha \left(1-\sum_{i=1}^{n}\varphi_{i}\right) Y_{lm}, 
\end{equation}
where $\alpha\in \mathbb{R}$ is to be determined.  By construction  $u(x_{i}) = v_{i}$
for each $i  \in \{1,\dots, n\}$, while on the complement of $\bigcup_{i=1}^{n}U_{i}$
it is proportional to $Y_{lm}$. Evaluating the function's $(l,m)$th spherical harmonic coefficient we find
\begin{equation}
  \label{eq:usph1}
  \braket{Y_{lm}}{u} = \sum_{i=1}^{n}v_{i} \int_{U_{i}} \varphi_{i}\,Y_{lm} \dd S
  + \alpha \left(1 - \sum_{i=1}^{n}\int_{U_{i}} \varphi_{i}\, Y_{lm}^{2} \dd S
  \right), 
\end{equation}
and so long as 
\begin{equation}
  \label{eq:con1}
   \sum_{i=1}^{n}\int_{U_{i}} \varphi_{i}\, Y_{lm}^{2} \dd S \ne 1, 
\end{equation}
we can choose $\alpha$ such that the coefficient is equal to any $w \in \mathbb{R}$.
One way this  condition can  be met  is by taking the open subsets
$U_{1},\cdots,U_{n}$
to be sufficiently small.  Indeed, for any $\delta > 0$ we can clearly select them such that
\begin{equation}
  \label{eq:deldef}
  \int_{U_{i}}\varphi_{i} \dd S < \delta, \quad i \in \{1,\dots,n\},
\end{equation}
from which we obtain
\begin{equation}
  \sum_{i=1}^{n}\int_{U_{i}} \varphi_{i}\, Y_{lm}^{2} \dd S < n\, \delta \, \|Y_{lm}\|^{2}_{C^{0}(\mathbb{S}^{2})}.
\end{equation}
By taking $0 < \delta < (n\,  \|Y_{lm}\|^{2}_{C^{0}(\mathbb{S}^{2})})^{-1}$ we are done.
\eproof

In Section \ref{sec:banach} we  recall and generalise a fundamental result of
\cite{backus1970inferenceI}  showing that
the behaviour in Proposition \ref{prop:nonu1} is typical for linear
inference problems. In light of this result, Backus argued that a prior bound on the
function's norm should be introduced,  and  showed 
that such a prior bound in conjunction with the data
constrains the property of interest to lie within a finite interval.
Following this approach, suppose that we are willing to accept
\begin{equation}
  \label{eq:nbound}
  \|u\|_{C^{0}(\mathbb{S}^{2})} \le r, 
\end{equation}
for a given $r$ which,  to be consistent with the data,  has to satisfy
\begin{equation}
  \label{eq:compat}
  r \ge  \max_{i\in \{1,\dots,n\}}|v_{i}|.
\end{equation}
The set,  $I$, of coefficient values consistent
with the norm-bound is the image of a closed and bounded set under a continuous linear mapping, and  hence
is itself closed and bounded. Similarly, we write $J$ for the set of
coefficient values consistent with \emph{both} the norm-bound and the point
data which, for the same reasons, is closed and bounded, while it
clearly satisfies $J \subseteq I$. The question is whether this inclusion
is proper, in which case  the  data does improve our knowledge of the 
spherical harmonic coefficient. Indeed, what we really hope is that a norm-bound can be chosen
to be so large as to be uncontroversial, but such that the  coefficient is
constrained in a useful manner \citep[c.f.][]{backus1970inferenceI}. Sadly, the following
result shows that $I$ is comprised of limit points of $J$,
and hence  $I=J$ as both sets are closed.

\begin{prop}
  \label{prop:nonu2}
  Suppose we are given   $u' \in C^{0}(\mathbb{S}^{2})$ which satisfies
  eq.(\ref{eq:nbound}) for some $r>0$, and point data
  $v_{1},\dots,v_{n}\in \mathbb{R}$   consistent with eq.(\ref{eq:compat}).
  For any $\epsilon > 0$, there exists a $u \in  C^{0}(\mathbb{S}^{2})$
  that is compatible with the norm-bound,   fits the point data
  \begin{equation}
    v_{i} = \braket{\delta_{x_{i}}}{u},\quad i \in  \{1,\dots,n\}, 
  \end{equation}
  and satisfies the estimate
  \begin{equation}
    \left|\braket{Y_{lm}}{u'-u} \right| < \epsilon.
  \end{equation}

\end{prop}

 \emph{Proof:} We again give a constructive proof making
use of Uyrsohn's lemma, with the open sets $\{U_{i}\}_{i=1}^{n}$, the 
functions $\{\varphi_{i}\}_{i=1}^{n}$, and the parameter $\delta > 0$ 
described in the proof of Proposition \ref{prop:nonu1}.  Consider the function
\begin{equation}
  \label{eq:udef2}    
  u = \sum_{i=1}^{n}\varphi_{i}\, v_{i} + \left(1-\sum_{i}\varphi_{i}\right) u', 
\end{equation}
which  satisfies the given point data. In the complement of the open set
$\bigcup_{i=1}^{n}U_{i}$ we have $u = u'$, and so  $|u|\le r$  in this region.
Turning to the $i$th subset, $U_{i}$, we note that eq.(\ref{eq:udef2}) simplifies to
\begin{equation}
  u = \varphi_{i}\, v_{i} + (1-\varphi_{i})u', 
\end{equation}
and hence for each $x \in U_{i}$ we have
\begin{equation}
  |u(x)| \le  \varphi_{i}(x)\,|v_{i}| + [1-\varphi_{i}(x)]|u'(x)|
   \le  \varphi_{i}(x)\, r + [1-\varphi_{i}(x)] r  =  r, 
\end{equation}
which shows that $u$  is indeed  consistent with the norm-bound. Finally,
we form the estimate
\begin{equation}
  \left|\braket{Y_{lm}}{u'-u} \right| = \left|
  \sum_{i=1}^{n}v_{i}\int_{U_{i}}\varphi_{i} Y_{lm} \dd S - \sum_{i=1}^{n}\int_{U_{i}}\varphi_{i} u' Y_{lm} \dd S
  \right| \le K \delta,
\end{equation}
where  $K > 0$ is some  constant independent of $\delta$.
By choosing $0 < \delta < \frac{\epsilon}{K}$ we are done. \eproof

This  proof  relies crucially  on our ability to continuously deform  functions
in  an arbitrarily small neighbourhood of each observation point.
In doing this we are, of course, making the functions rougher and rougher, but the supremum  norm   
is oblivious to this fact. 
We might  reasonably hope that the situation could  be improved by  posing the problem
in a function space whose norm incorporates 
derivative information. An obvious choice is the  space $C^{k}(\mathbb{S}^{2})$ of
$k$-times continuously differentiable functions for some $k \ge 1$. Indeed,
if bounds are placed on the function's pointwise derivatives up to some finite-order, then the above argument
clearly fails, and it  seems certain that the data must then  provide some meaningful
information about the  spherical harmonic coefficient.

\subsection{Questions arising from the motivating problem}

The preceding example illustrates two  important features of linear inference problems.
First, it shows the essential role of prior constraints on the unknown model. To put matters
starkly,  if we are unwilling provide such information,  then there is usually nothing
that can be learned from the data. How suitable prior constraints can
be found in practice is necessarily application dependent, and so lies
beyond the scope of this paper. In the above problem a bound was placed on the unknown function's
norm. This is an example of a \emph{deterministic constraint}, 
the key point being that  the model is thought to definitely belong to a given
subset of the model space, but  no  further information is provided to distinguish between points
within this subset. This paper  restricts attention to such  deterministic constraints, but
probabilistic constraints can also be introduced and this topic may be discussed in future work.
It is, however worth emphasising that the right question is not whether methods based on deterministic or
probabilistic methods are \emph{better}, but merely ascertaining what can and cannot be done with the
different types of prior information that  one  encounters within  geophysical applications
\cite[c.f.][]{stark2015constraints}.

The second point is that the model space and its topology need to be specified as part of the problem's
formulation. Within the above example,  the data was shown to  provided absolutely no  information if
the  obvious model space was selected along with a prior norm bound, while we
conjectured that switching to a different  model space would obviate this difficulty. 
Such  issues do not seem to  have  received very much attention in the  literature on inference problems,
with most papers either working in a general, but fixed, type of function space, or
with $L^{2}$-spaces even though they do not always
constitute a valid option. Backus was aware of this issue, of course, and tried to
solve it through what he called \emph{quellings} \citep{backus1970inferenceII},
but it can be fairly said that this approach is both difficult and rather limited.
As an initial observation,  the choice of model space is not  completely
arbitrary. Indeed,   for any progress  to be made  with an inference problem  the  data and property mappings
(these terms are defined formally below) must be well-defined and continuous, with these requirements
determining the largest valid model space for the problem. But is  it ever   permissible to work with a
smaller model space? Clearly such a choice cannot  be made simply because it leads to preferable results,
but must somehow   be based on the underlying   physics.   Questions about prior
constraints and the choice of model space are, in fact, closely related. The view this paper aims to
substantiate is that the appropriate model space is  determined by the prior constraints
we are willing to accept.

As a final comment, we note that a solution of the motivating inference problem has been
obtained in $C^{0}(\mathbb{S}^{2})$ subject to a prior norm bound. In this manner
we have extended the ideas of \cite{backus1970inferenceI}, which were developed
in a Hilbert space setting, to a linear inference problem posed in a Banach space.
The methods used, however, were entirely ad hoc, and there is no obvious way that they
can be extended to other linear inference problems. In particular, though
we have conjectured that solution of the present problem posed in $C^{k}(\mathbb{S}^{2})$
for some $k \ge 1$ would yield more interesting results, there seems to be
no method in the existing literature on geophysical inference problems for doing this.
It is to address precisely such questions that
the general theory of this paper is done in a Banach  space setting.

\section{Linear inference problems without data errors}

\label{sec:banach}

Within this section we consider linear inference problems in the absence of random
data errors. While such  errors will be present in all  applications, it is thought that
there is a pedagogical advantage to see first what can be done with perfect data.
Our aim is to develop a  theory that can be applied to a wide range of
geophysical inference problems. In doing this we need to decide, in particular, on the
mathematical structure assumed for the model space. For linear problems, this
means determining the type of topological vector space to consider.
At one extreme, we could allow the model space to be the most general that is conceivable, and hence be
sure the theory  will always be  applicable. However,  it  might then be very difficult
 to establish  useful results. Conversely
if the model space is given a great deal of structure,  then while we may be able to prove many
things,  the results need not be applicable in cases of practical interest.

To chart an appropriate course forward  we must   be guided by neither a desire for 
generality nor convenience, but through concrete geophysical
applications, with the spectral estimation problem of the introduction being a simple but
 representative example. In this instance the obvious model space to work with
was  $C^{0}(\mathbb{S}^{2})$ which is a Banach space, while our discussion
suggested that  $C^{k}(\mathbb{S}^{2})$ for some  $k \ge 1$
might actually be preferable, and again these are Banach spaces.
Indeed, despite  the predominance of Hilbert spaces in the literature,
most geophysical inverse and inference problems are  naturally posed in  Banach spaces.
For example,  very frequently the model space will comprise coefficients
within a partial differential equation. In order for such equations
to be well-posed these  coefficients typically  must  be
continuously differentiable to some finite-order, or perhaps  only  essentially bounded
 if less regularity is required of the solutions;
see \cite{blazek2013mathematical} for a relevant discussion
within a seismological context.
Moreover, considerations below will make clear  that  we  should, in fact,
  think about \emph{Banachable spaces}, which is to
say complete topological vector spaces whose structure can be defined by any one
of a set of equivalent norms \citep[e.g.][]{lang2012fundamentals}. 
It might be reasonably asked if even more general
structures  should be permitted (e.g.  Fr\'{e}chet or inductive limit spaces), 
but this  does not  seem to be the case within  geophysical applications the author is aware of.

\subsection{Formulation of the inference problem}

\label{ssec:blif}

\subsubsection{General theory}

Within a  linear inference problem we  need to introduce three real vector spaces (all vector spaces
within this paper are real, and we will not constantly state this restriction).
First there is the \emph{model space}, denoted  by $E$, and  assumed to be
an infinite-dimensional
 topological vector space. Next  the \emph{data space}  is denoted by $F$, and is a
finite-dimensional  topological vector space. Finally   the \emph{property space} 
will be written
$G$, and is also a finite-dimensional  topological vector space. We assume that
each of these spaces is Banachable, with this concept clarified below. This
implies, in particular, that the spaces are Hausdorff, which
is to say that any two distinct points have disjoint open neighbourhoods.
An $n$-dimensional Hausdorff topological vector space is isomorphic
to $\mathbb{R}^{n}$ with its standard topology \citep[e.g.][Theorem 9.1]{treves}.
There is, therefore, nothing to decide for the data nor
property space from a topological perspective. For the model space, however,
more needs to be said. That $E$ is Banachable means that it is a complete
topological vector space such that for \emph{some} norm $\|\cdot\|_{E}:E\rightarrow \mathbb{R}$ the balls
\begin{equation}
  \label{eq:ball}
  B_{\epsilon} = \{u \in E \,|\, \|u\|_{E} \le  r\}, \quad r > 0,
\end{equation}
form a basis of neighbourhoods of the origin \citep[e.g.][Chapter 11]{treves}.
Such a norm is said to be \emph{compatible} with the topology on $E$.
A second norm $\|\cdot\|'_{E}:E\rightarrow \mathbb{R}$ on $E$ is
\emph{equivalent} to the first if there exist positive constants $c < C$ such that
\begin{equation}
  \label{eq:nequiv}
  c\|u\|_{E} \le \|u\|'_{E} \le C \|u\|_{E}, 
\end{equation}
for all $u \in E$. It follows readily  that $\|\cdot\|_{E}'$ is also
compatible with the  topology on $E$. Given a Banach space, we can
trivially pass to a Banachable space by  forgetting about its
norm  but retaining the underlying topology. Equally, from a Banachable
space we can form a Banach space by selecting any one of its compatible norms.

Within the  inference problem, we are given
a \emph{data mapping} $A \in \Hom(E,F)$ which takes each element $u$ of the model
space to the \emph{data vector} $v = Au$ (see Appendix \ref{ssec:linop} for notations).
We are also given the \emph{property  mapping} $B \in \Hom(E,G)$ that returns the \emph{property vector} $w = Bu$ 
corresponding to the model $u \in E$.  The aim of the inference problem is, in
broad terms, to 
constrain the value of the property vector from knowledge of the data vector.
Throughout this section we  make the following assumption:
\begin{asm}
  \label{asm:sur}
  Both the data mapping and the property mapping 
  are surjective. 
\end{asm}
This assumption is difficult to verify directly. By thinking about  dual spaces, however,
the situation is considerably simplified.
The dual of the model space will be written $E'$, and is Banachable
when equipped with the strong dual topology \citep[e.g.][Chapter 19]{treves}.
To understand this structure, let $\|\cdot\|_{E}$ be a compatible norm for
$E$, and for $u' \in E$ set
\begin{equation}
  \label{eq:dual_norm}
  \|u'\|_{E'} = \sup_{u \in E\setminus\{0\}} \frac{|\braket{u'}{u}|}{\,\,\,\|u\|_{E}}.
\end{equation}
Relative to this \emph{dual norm} it can be shown that $(E',\|\cdot\|_{E'})$ is a
Banach space  \citep[e.g.][Theorem 11.5]{treves}. It is readily checked that
equivalent norms on $E$ lead  to equivalent norms on $E'$, and hence a
unique Banachable structure on the dual space is defined. The same idea applies to
the  data and property spaces.
The dual $A'\in \Hom(F',E')$ of the data mapping is defined  uniquely through
\begin{equation}
  \braket{v'}{Au}= \braket{A'v'}{u}, 
\end{equation}
for all $u \in E$ and $v' \in F'$. For a subspace $V \subseteq F$,  its \emph{polar}
is the subspace  $V^{\circ} \subseteq F'$ defined by
\begin{equation}
  \label{eq:polar}
  V^{\circ} = \{ v' \in F' \,|\, \braket{v'}{v} = 0, \,\forall v \in V\}, 
\end{equation}
which is readily seen to be  closed. The following identity \citep[e.g.][Proposition 23.2]{treves} will be extremely useful
\begin{equation}
  \label{eq:bclim}
  \ker A' = (\image A)^{\circ}.
\end{equation}
See Appendix \ref{ssec:linop} for the definition of the kernel and image of a linear mapping.
As a first application, we have:
\begin{prop}
  \label{prop:asur}
  A mapping $A \in \Hom(E,F)$ with $\dim F < \infty$ is surjective if and only if $\ker A' = \{0\}$.
\end{prop}
\emph{Proof:} Suppose that $A$ is surjective.
Clearly  $F^{\circ}= \{0\}$, and hence eq.(\ref{eq:bclim}) 
implies that the dual mapping $A'$ is injective.
Conversely, if $\ker A' = \{0\}$,  then eq.(\ref{eq:bclim}) 
implies via the Hahn-Banach theorem \citep[e.g.][Chapter 18, Corollary 3]{treves}  that $\image A$ is dense in
$F$. Because, however, $F$ is  finite-dimensional, its only
dense subspace is $F$ itself, and so we  conclude that $A$ is indeed surjective. \eproof
\begin{coro}
  \label{coro:surcon}				
  Assumption \ref{asm:sur} holds if and only if $\ker A' = \{0\}$ and  $\ker B' = \{0\}$.
\end{coro}

It is worth noting that these conditions on the data and property mapping
are stable with respect to small perturbations. This matters
practically because these operators will  be approximated within numerical
calculations, while the data mapping might also be subject to  theoretical errors.
To understand why this holds, suppose that the data mapping undergoes a perturbation
$A \mapsto A + \delta A$ for some
$\delta A\in \Hom(E,F)$. Picking compatible norms on $E$ and $F$,
we know from the Heine-Borel theorem \citep[e.g.][Theorem 9.2]{treves} that the unit
ball in $F'$ is compact, and hence
 $\ker A' = \{0\}$ if and only if
\begin{equation}
  d = \inf_{\|v'\|_{F'}=1}\|A' v'\|_{E'} > 0.
\end{equation}
 Using the triangle inequality we obtain
\begin{equation}
  \|A'v'\|_{E'}  \le \|(A' + \delta A')v'\|_{E'}
  +\|\delta A'v'\|_{E'}, 
\end{equation}
which implies that
\begin{equation}
 \inf_{\|v'\|_{F'}=1} \|(A' + \delta A')v'\|_{E'} \ge d- \sup_{\|v'\|_{F'}=1}\|\delta A'v'\|_{E'}. 
\end{equation}
Making use of the definition of the operator norm in eq.(\ref{eq:op_norm}), it  follows that so long as
\begin{equation}
  \|\delta A'\|_{\Hom(F',E')} < d, 
\end{equation}
 the perturbed data mapping will be surjective.

 We now formalise the idea that an inference problem might be posed with respect to
 several   different choices of model space.  Suppose that
$\tilde{E}$ is a  Banachable space that is embedded properly, continuously,
and densely in $E$. We write $i \in \Hom(\tilde{E},E)$ for the inclusion
mapping, and define the induced linear mappings
\begin{equation}
  \tilde{A} = A i \in \Hom(\tilde{E},F), \quad \tilde{B} = Bi \in \Hom(\tilde{E},G).
\end{equation}
In this manner, we can formulate a new linear inference problem with
$\tilde{E}$ playing the role of the model space, and $\tilde{A}$ and $\tilde{B}$
 the data and property mappings, respectively. This new problem  will be
called a \emph{restriction} of the  original.
\begin{prop}
  \label{prop:rest}
  If Assumption \ref{asm:sur} holds, then it is also  valid for any
  restriction of the linear inference problem.
\end{prop}
\emph{Proof:} We will show that  $\tilde{A}$ is surjective, with
an identical argument applying to $\tilde{B}$. By definition of the dual mapping
we have $\tilde{A}' = i' A'$ where $i' \in \Hom(E',\tilde{E}')$, and so $\tilde{A}' v' = 0$ implies that
$u' = A' v' \in \ker i'$. We know that $A'$ is injective, and so $v'$  vanishes if and only
if $u'$ does. But if $u' \in \ker\,i'$, we have
\begin{equation}
  \braket{i'u'}{\tilde{u}}=\braket{u'}{i\tilde{u}} = 0,
\end{equation}
for all  $\tilde{u} \in \tilde{E}$, and hence  $u' = 0$ as  $\image \,i$ is dense in $E$.
\eproof

The proof of this result shows that density of the embedding $\tilde{E}\hookrightarrow E$
is sufficient for the data and property mappings to remain surjective. Because a finite-dimensional
space can never be dense within an infinite-dimensional one, the chosen definition
precludes the introduction of finite-dimensional model parametrisation. This is not to say that
for such a parametrisation the induced data and property mappings could not be
surjective,  but this would need to be established on a case by case basis.

\subsubsection{Application to the spectral estimation problem}

The  formalism above can  readily be applied to  our motivating problem,
though we take the opportunity to generalise things slightly by  considering simultaneous estimation
of $p \ge 1$ distinct  spherical harmonic coefficients. As the model space we initially take
$C^{0}(\mathbb{S}^{2})$. The data and property spaces are $\mathbb{R}^{n}$
and $\mathbb{R}^{p}$, respectively, with each  given their usual topologies.
Letting $\{f_{i}\}_{i=1}^{n}$ denote the standard basis for $\mathbb{R}^{n}$,
the data mapping for the problem can be written in the form
\begin{equation}
  A u = \sum_{i=1}^{n} \braket{\delta_{x_{i}}}{u} f_{i}, 
\end{equation}
for any $u \in C^{0}(\mathbb{S}^{2})$. For any $v' \in (\mathbb{R}^{n})' \cong \mathbb{R}^{n}$
we then have
\begin{equation}
  \braket{v'}{A u} = \sum_{i=1}^{n}\braket{v'}{f_{i}}\braket{\delta_{x_{i}}}{u}
  = \braket{\sum_{i=1}^{n}\braket{v'}{f_{i}}\delta_{x_{i}}}{u}, 
\end{equation}
which shows that the dual data mapping takes the form
\begin{equation}
  \label{eq:adual}
  A'v' = \sum_{i=1}^{n}\braket{v'}{f_{i}}\delta_{x_{i}}.
\end{equation}
Similarly, we let $\{g_{j}\}_{j=1}^{p}$ be the standard basis for $\mathbb{R}^{p}$ and define the property mapping by
\begin{equation}
  B u = \sum_{j=1}^{p} \braket{Y_{l_{j}m_{j}}}{u} g_{j}, 
\end{equation}
where $\{(l_{1},m_{1}),\dots,(l_{p},m_{p})\}$
are $p$ distinct pairs of spherical harmonic indices. The dual mapping is therefore
\begin{equation}
  \label{eq:bdual}
  B' w' = \sum_{j=1}^{p} \braket{w'}{g_{j}}  Y_{l_{j}m_{j}}, 
\end{equation}
for any $w' \in (\mathbb{R}^{p})' \cong \mathbb{R}^{p}$.
From eq.(\ref{eq:adual}) and (\ref{eq:bdual}) it follows that
\begin{equation}
  \image A' = \mathrm{span}\{\delta_{x_{1}},\dots,\delta_{x_{n}}\}, \quad
  \image B' = \mathrm{span}\{Y_{l_{1}m_{1}},\dots,Y_{l_{p}m_{p}}\}.
\end{equation}
As seen in Proposition \ref{prop:asur}, the  surjectivity for the data
and property mappings is equivalent to the linear independence of these two finite-dimensional
sets of dual vectors \citep[c.f.][]{backus1970inferenceI}. This   is shown in the following result, and hence
Assumption \ref{asm:sur} is valid for the problem at hand.  In fact, 
a stronger linear independence condition is established that will be needed later. 

\begin{prop}
  \label{prop:linind}
  Let $\{x_{i}\}_{i=1}^{n}$ be distinct points on $\mathbb{S}^{2}$, and $\{Y_{l_{j}m_{j}}\}_{j=1}^{p}$
  a collection of distinct spherical harmonics. The  dual vectors
  \begin{equation}
    \{\delta_{x_{1}},\dots,\delta_{x_{n}},Y_{l_{1}m_{1}},\dots,Y_{l_{p}m_{p}}\}
  \end{equation}
  are linearly independent in $C^{0}(\mathbb{S}^{2})'$. 
\end{prop}

\emph{Proof:} We need to show that the  equality
\begin{equation}
  \label{eq:linind}
  \sum_{i=1}^{n} \alpha_{i}\delta_{x_{i}} + \sum_{j=1}^{p}\beta_{j}Y_{l_{j}m_{j}} = 0, 
\end{equation}
holds in $C^{0}(\mathbb{S}^{2})'$ only if the coefficients
$\alpha_{1},\dots,\alpha_{n},\beta_{1},\dots,\beta_{p}$ vanish. Once again we apply
Uyrsohn's lemma using the notations
described in the proof of Proposition \ref{prop:nonu1}.
Acting the left hand side of eq.(\ref{eq:linind}) on $\varphi_{i} \in C^{0}(\mathbb{S}^{2})$ we find
\begin{equation}
   \alpha_{i} + \sum_{j=1}^{p}\beta_{j}\int_{U_{i}} \varphi_{i}\, Y_{l_{j}m_{j}} \dd S = 0, 
\end{equation}
and  note that  second term on the left hand
side can be made as small as we like by letting $\delta$ tend to zero.
It follows that  $\alpha_{i}=0$  for each $i \in \{1,\dots,n\}$.  It remains to show
that the dual vectors $\{Y_{l_{j}m_{j}}\}_{j=1}^{p}$ are linearly independent
in $C^{0}(\mathbb{S}^{2})'$,  but this is immediate due to the spherical harmonic orthogonality relation
\begin{equation}
  \int_{\mathbb{S}^{2}} Y_{lm} Y_{l'm'} \dd S = \delta_{ll'}\delta_{mm'}. 
\end{equation}
\eproof

So far we have made no explicit use of the norm in eq.(\ref{eq:supnorm}). Indeed,
all that has been required is that the data and property mappings
are continuous relative to the underlying topology on $C^{0}(\mathbb{S}^{2})$.
Let $\rho \in C^{0}(\mathbb{S}^{2})$ be a function satisfying
\begin{equation}
  c \le \rho(x) \le C, 
\end{equation}
for all $x \in \mathbb{S}^{2}$ and some positive constants $c\le C$. We can
then define a new norm on $C^{0}(\mathbb{S}^{2})$ by
\begin{equation}
  \|u\|_{C^{0}(\mathbb{S}^{2})}' = \sup_{x \in \mathbb{S}^{2}} | \rho(x)u(x)|, 
\end{equation}
which is clearly  equivalent to that in eq.(\ref{eq:supnorm}). Both these
norms define the same  topology, and we have no clear  reason
to use one over the other. Here it might be argued that the original
norm is preferable because, being rotationally invariant, it is  simpler. While this
is a reasonable point, it is not unanswerable. For example, within a geophysical application
we might wish for the norm 
to   somehow reflect  differences between oceanic and continental regions.
Ultimately, a  specific choice of norm is not \emph{required} to formulate the inference problem, and hence
the model space  is more naturally regarded as   Banachable. 

Building on this point, suppose that we wished
to formulate the inference problem in $C^{k}(\mathbb{S}^{2})$ for some $k \ge 1$.
Because $\mathbb{S}^{2}$ is a non-trivial manifold, the definition of
a suitable norm for this space is somewhat involved. One approach is
to introduce a metric on $\mathbb{S}^{2}$, and to make use of the associated
covariant derivative. It can then be shown that $C^{k}(\mathbb{S}^{2})$ is complete
relative to the norm
\begin{equation}
  \label{eq:cknorm}
  \|u\|_{C^{k}(\mathbb{S}^{2})} = \sup_{l \le k}\sup_{x \in \mathbb{S}^{2}} \|\nabla ^{l}u\|_{T^{0}_{l}\mathbb{S}^{2}}, 
\end{equation}
where $\nabla^{l}u$ denotes the $l$-fold covariant derivative of a function $u$, and
$\|\cdot\|_{T^{0}_{l}\mathbb{S}^{2}}$ is any pointwise-norm on the bundle of $l$-times
covariant tensors (the precise meaning of these geometric terms is not required in what follows).
Within this construction a number of choices have obviously been made, but from a topological perspective
all that matters is that pointwise-derivatives up to order $k$ are included. Thus,  again, these
function spaces   are most naturally regarded as Banachable.
From the expressions  in eq.(\ref{eq:supnorm}) and (\ref{eq:cknorm}),
it is immediate that
\begin{equation}
  \|u\|_{C^{0}(\mathbb{S}^{2})} \le C \|u\|_{C^{k}(\mathbb{S}^{2})}, 
\end{equation}
for all $u \in C^{k}(\mathbb{S}^{2})$ and some constant $C >0$, and hence the
embedding $C^{k}(\mathbb{S}^{2}) \hookrightarrow C^{0}(\mathbb{S}^{2})$ is continuous.
Clearly there exist continuous functions that are not $k$-times continuously differentiable,
and so the embedding is  proper,
while its density follows from the  fact that the space, $C^{\infty}(\mathbb{S}^{2})$, of smooth functions
is dense  in $C^{k}(\mathbb{S}^{2})$ for any $k\ge 0$.
We conclude that the  spectral estimation problem does indeed have a well-defined 
restriction to $C^{k}(\mathbb{S}^{2})$ for  each $k \ge 1$ , while
Assumption \ref{asm:sur} remains valid thanks to Proposition \ref{prop:rest}.

\subsection{Backus' theorems}

\label{ssec:backus}

\subsubsection{General theory}

In this subsection we reformulate three fundamental results of  \cite{backus1970inferenceI}
on linear inference  problems
within the setting of Banachable spaces. As a starting point, we recall a useful
 factorisation of the data mapping. Considering the equation
\begin{equation}
  v = Au, 
\end{equation}
for given $v \in F$, the  surjectivity of $A$ implies that there always
exists a solution. In fact, if $u \in E$ solves this problem, so
does $u + u_{0}$ for any $u_{0} \in \ker A$. We will say that
two elements $u_{1},u_{2} \in E$ are equivalent modulo $\ker A$ if
and only if $u_{1}-u_{2} \in \ker A$. This defines an equivalence
relation which we denote by $u_{1} \sim_{\ker A} u_{2}$. The set of
all such equivalence classes  forms the \emph{quotient
  space}, $E/\ker A$, which has the structure of a vector space in an obvious way. 
We write $\pi_{A}:E\rightarrow E/\ker A$ for the
\emph{quotient mapping} which takes an element of $E$ to its corresponding equivalence class.
Clearly $\pi_{A}$ is linear and has kernel equal to $\ker A$.
Because $A$ is continuous,  its kernel is closed. 
It follows that $E/\ker A$ can be made into a Hausdorff topological
vector space by declaring its open subsets to be images of  open subsets
in $E$ under the quotient mapping \citep[e.g.][Proposition 4.5]{treves}.
In fact, because $E$ is Banachable, the same is true of $E/\ker A$.
To describe this structure, let $\|\cdot\|_{E}$ be a compatible norm  for $E$.
We then define the corresponding
\emph{quotient norm} on $E/\ker A$ through
\begin{equation}
  \label{eq:quotnorm}
  \|\pi_{A} u\|_{E/\ker A} = \inf_{u_{0} \in \ker A} \|u+u_{0}\|_{E}, 
\end{equation}
which can be shown to be compatible with the topology on the quotient space  \citep[e.g.][Chapter 11]{treves}.
It is readily seen  that equivalent norms on $E$  lead to  equivalent
norms on the quotient space, and hence $E/\ker A$ is Banachable. Given this
structure, the quotient mapping is  trivially  continuous, while  the
data mapping can be uniquely written
\begin{equation}
  \label{eq:afac}
  A = \hat{A} \,\pi_{A}, 
\end{equation}
where $\hat{A} \in \Hom(E/\ker A,F)$ has a continuous
inverse \citep[e.g.][Proposition 4.6]{treves}. In summary, these results show that from
the equation $v = Au$ we can uniquely and  continuously recover the equivalence
class of $u$ modulo $\ker A$.

We recall that the linear inference problem aims to estimate the property vector $w = Bu$ given
the data vector $v = Au$. It is natural to ask whether this problem can ever be solved exactly,
with a necessary and sufficient condition given by:

\begin{thm}
  \label{thm:bbackus1} The property vector can be recovered
  continuously from the data vector if and only if
  $\image B'  \subseteq \image A'$.
\end{thm}

\emph{Proof:}
Within Lemma \ref{lem:kereq} below it is shown that the stated
inclusion is equivalent to
\begin{equation}
  \label{eq:kercon}
  \ker A \subseteq \ker B.
\end{equation}
Suppose that the property vector can be computed from the data in
a unique and continuous manner.  It follows that there is a continuous
function $f:F\rightarrow G$ such that $f(Au) = Bu$ for all $u \in E$. Clearly
this mapping is linear,   and hence for some  $C \in \Hom(F,G)$
we have $B = CA$. This identity implies that $\ker A \subseteq \ker B$, and so
the necessity of eq.(\ref{eq:kercon}) is established.
To show that the condition is  sufficient
we first obtain the following factorisation of the property mapping
\begin{equation}
  \label{eq:bfac}
  B = \bar{B}\,\pi_{A},
\end{equation}
for some $\bar{B} \in \Hom(E/\ker A,G)$. Eq.(\ref{eq:kercon}) implies that
if $u_{1} \sim_{\ker A} u_{2}$ then $u_{1} \sim_{\ker B} u_{2}$.
A linear mapping $\bar{B}: E/\ker A \rightarrow G $
can, therefore, be uniquely defined by $\bar{B} (\pi_{A} u) = B u$. To show that this
mapping is continuous, we need to demonstrate that the inverse image under $\bar{B}$ of an
open subset in $G$ is open in $E/\ker A$. But this inverse image is,
due to the continuity of $B$, the image under $\pi_{A}$ of an open subset in
$E$, and hence  open by definition in the quotient topology. Applying eq.(\ref{eq:afac})
to the equation $v = Au$  we can write $\pi_{A} u = \hat{A}^{-1} v$,  and  using eq.(\ref{eq:bfac})
the property vector is given by  $w =\bar{B} \hat{A}^{-1} v$. \eproof

\begin{lem}
  \label{lem:kereq} The inclusion $\image B'  \subseteq \image A'$ holds
  if and only if $\ker A \subseteq \ker B$.
\end{lem}

\emph{Proof:}
To simplify the argument we assume that $E$ is reflexive,
this meaning that the canonical inclusion of
$E$ into its  bidual $E'' = (E')'$ is surjective \citep[e.g.][Definition 36.2]{treves}.
Non-reflexive  spaces
do  occur  within applications with  $C^{k}(\mathbb{S}^{2})$ for any $k \in \mathbb{N}$ being  pertinent
examples. The necessary modifications to the argument  are only a little fiddly, but
not  difficult. Applying eq.(\ref{eq:bclim}) to the dual data and property mappings we obtain
\begin{equation}
  \label{eq:bclim_dual}
  \ker A = (\image A')^{\circ}, \quad  \ker B = (\image B')^{\circ}.
\end{equation}
Suppose that $U_{1}$ and $U_{2}$ are two subspaces of $E$ with $U_{1} \subseteq U_{2}$.
From the definition of the polar of a subspace, it is clear that
  $U_{2}^{\circ} \subseteq U_{1}^{\circ}$ within $E'$.
Applying this identity to the inclusion $\ker A \subseteq \ker B$  and using
eq.(\ref{eq:bclim_dual}) we then obtain
\begin{equation}
 [(\image B')^{\circ}]^{\circ} \subseteq  [(\image A')^{\circ}]^{\circ}. 
\end{equation}
The bipolar $[(\image B')^{\circ}]^{\circ}$ is
equal to the closure of  $\image B'$\citep[e.g.][Proposition 35.3]{treves}, but
$\image B'$ is already closed because it is  finite-dimensional.
The same argument applies to the dual of the data mapping,
and hence we have shown that $\ker A \subseteq \ker B$  
implies $\image B' \subseteq \image A'$. Conversely, we can start with
$\image B' \subseteq \image A'$, take polars, and apply eq.(\ref{eq:bclim_dual})
to obtain $\ker A \subseteq \ker B$.
\eproof

Unfortunately, a moment's reflection shows that  the condition in
Theorem \ref{thm:bbackus1} will almost surely fail within
applications. This is because the inclusion of one finite-dimensional
linear subspace inside another is  unstable with respect to 
perturbations; consider, for example, the probability that a randomly chosen
line through the origin in $\mathbb{R}^{3}$ lies within a given plane.
Instead, we should generically expect that:
\begin{asm}
  \label{asm:tran} 
  The data and property spaces are \emph{transversal}, by which we mean
  $\image A^{'} \cap \image B^{'} = \{0\}$.
\end{asm}
Given this transversality condition is met, we now generalise to
Banachable spaces a second  result of
\cite{backus1970inferenceI} showing that there is absolutely nothing that can be learned about
the property vector from the data alone.

\begin{thm}
  \label{thm:bbackus2} For any $v \in F$ and $w \in G$ there exist infinitely many
  model vectors $u \in E$ such that $v = Au$ and $w = Bu$.
\end{thm}

\emph{Proof:} Following \cite{parker1977linear}, we introduce the \emph{joint
  data-property mapping} $C \in \Hom(E,F\oplus G)$ such that
\begin{equation}
  \label{eq:jdpdef}
  Cu = (A u)\oplus (B u), 
\end{equation}
where $\oplus$ denotes the direct sum. The claimed result
is equivalent to $C$ being surjective with a non-trivial
kernel. The latter statement is clear from the dimension
of the spaces involved, and so we need only establish the former.
Recalling  Proposition \ref{prop:asur}, we know that $C$ is surjective if
and only if its dual $C'$ has a trivial kernel. By definition, we have
\begin{equation}
  \braket{u}{C'(v'\oplus w')} = \braket{Cu}{v'\oplus w'} = \braket{Au}{v'} + \braket{B u}{w'}
   = \braket{u}{A'v'+B' w'}, 
\end{equation}
for all $u \in E$, $v' \in F'$, and $w' \in G'$. It follows that the dual mapping $C'$
takes the form
\begin{equation}
  C'(v'\oplus w') = A'v' + B'w'.
\end{equation}
If $v'\oplus w' \in \ker C'$ is non-trivial, we  must have
\begin{equation}
  A'v' + B'w' = 0. 
\end{equation}
Due to Corollary \ref{coro:surcon},  neither term on
the left hand side can vanish, and so there has to be a non-zero
element of $\image A'\cap \image B'$. But this contradicts
Assumption \ref{asm:tran}, and so we  conclude that $\ker C'$ is indeed trivial. \eproof

\begin{coro}
  \label{coro:tran}
  If Assumption \ref{asm:tran} holds for a linear inference problem,
  then it holds for any restriction of the problem.
\end{coro}

\emph{Proof:} Theorem \ref{thm:bbackus2} shows
that, given Assumption \ref{asm:sur}, the transversality condition is equivalent to the joint
data-property mapping $C\in \Hom(E,F\oplus G)$ being surjective. The argument
within the proof of Proposition \ref{prop:rest} implies, however,
that if $C$ is surjective, then the same is true of the corresponding
mapping $\tilde{C} \in \Hom(\tilde{E},F\oplus G)$ for the restricted
 problem.  \eproof

\begin{coro}
  \label{coro:tran1}
  Assumption \ref{asm:tran} is stable with respect to small perturbations. This is to
  say that if it holds for $(A,B) \in \Hom(E,F) \times \Hom(E,G)$,
  then  there is a open neighbourhood of $(A,B)$  for which it remains true. Here
  $\Hom(E,F)$ and $\Hom(E,G)$ carry their usual operator norm topologies, and
  $\Hom(E,F) \times \Hom(E,G)$ the product topology.
\end{coro}

\emph{Proof:} This follows from the preceding proof along with the discussion
after Corollary \ref{coro:surcon}. \eproof

 It is now  clear that to learn
 anything about the property vector from the data a suitable  prior constraint on the model must be provided.
 Within \cite{backus1970inferenceI}  a bound on the model norm was introduced, while subsequent work by
\cite{backus1970inferenceII,backus1972} and \cite{parker1977linear} discussed
the use of  distinct but related constraints.
These  ideas are subsumed within
the following  result which, importantly, depends only on the model space topology.
Before stating it we recall 
\citep[e.g.][Chapter 14]{treves} that a subset $U\subseteq E$ of a Banachable space
is bounded if and only if for some compatible norm $\|\cdot\|_{E}:E\rightarrow \mathbb{R}$
there is a positive constant $K$ such that
\begin{equation}
  \|u\|_{E} \le K,
\end{equation}
for all $u \in U$. It is readily checked that if this condition holds
for one norm, then it holds for all equivalent norms. Moreover,  the image
of a bounded set under a  continuous linear mapping is clearly bounded. 

\begin{thm}
  \label{thm:bbackus3}
Given a  \emph{constraint set}  $U \subseteq E$, the condition  $u \in U$
  is compatible with data $v \in F$ if and only if  $U \cap \pi_{A}^{-1}\{\hat{A}^{-1}v\} $ is
  non-empty. When this  requirement is met, a necessary and sufficient
  condition for the property vector to be restricted
  to a bounded subset of the property space is that $\pi_{B}(U\cap \pi_{A}^{-1}\{\hat{A}^{-1}v\} )$
  is bounded in $E/\ker B$.
\end{thm}

\emph{Proof:} Using  eq.(\ref{eq:afac}) we obtain
$\pi_{A} u = \hat{A}^{-1}v$, and hence the model lies in the closed affine subspace
$\pi_{A}^{-1}\{\hat{A}^{-1}v\}$. It follows trivially that the prior constraint
is  compatible with the data if and only if  $U \cap \pi_{A}^{-1}\{\hat{A}^{-1}v\} $
is  non-empty. When this holds, the collection of property vectors compatible with both the
data and the constraint can be written
\begin{equation}
  \label{eq:wconst}    
   \hat{B}\, \pi_{B}(U \cap \pi_{A}^{-1}\{\hat{A}^{-1}v\} ), 
\end{equation}
where, by analogy with eq.(\ref{eq:afac}), we have  used a factorisation $B = \hat{B}\,\pi_{B}$ 
such that $\hat{B}\in \Hom(E/\ker B,G)$ has a continuous inverse. The  subset above is, therefore,
bounded in $G$ if and only if $\pi_{B}(U \cap \pi_{A}^{-1}\{\hat{A}^{-1}v\} )$
is bounded in $E/\ker B$. \eproof

Given only a prior constraint $u \in U$, the set of possible property vectors
is given by $B U = \hat{B} \,\pi_{B} U$ which is bounded
in $G$ if and only if $\pi_{B}U$ is bounded in $E/\ker B$.
Assuming this constraint is compatible with the data $v \in F$,
the trivial inclusion
\begin{equation}
  \label{eq:tbound}
  \pi_{B}(U \cap \pi_{A}^{-1}\{\hat{A}^{-1}v\} ) \subseteq \pi_{B} U, 
\end{equation}
implies firstly that the boundedness of $\pi_{B}U$  is sufficient  for the
combination of the constraint and the data to restrict the property
vector to a bounded subset. Moreover, we see that use of
the data improves our knowledge of the property vector
relative to the prior constraint  if and only if the inclusion in
eq.(\ref{eq:tbound}) is proper.  Importantly,  while
$U \cap \pi_{A}^{-1}\{\hat{A}^{-1}v\} $ will generically be a proper subset of
$U$, Proposition \ref{prop:nonu2} shows that it can happen that
both $U\cap \pi_{A}^{-1}\{\hat{A}^{-1}v\}$ and $U$   project onto the \emph{same}
subset of $E/\ker B$.
Such behaviour occurs when each point in $U$ can be obtained
from one in $U \cap \pi_{A}^{-1}\{\hat{A}^{-1}v\} $ by adding a suitable element of $\ker B$,
 this being  essentially what was done within the proof of  Proposition \ref{prop:nonu2}.

\subsubsection{Application to the spectral estimation problem}

To apply these ideas to our motivating problem we need to
determine whether or not Assumption \ref{asm:tran} holds.
Taking the model space to be $C^{0}(\mathbb{S}^{2})$, and recalling
eq.(\ref{eq:adual}) and (\ref{eq:bdual}), the
transversality condition takes the form
\begin{equation}
  \mathrm{span}\{\delta_{x_{1}},\dots,\delta_{x_{n}}\}\cap
   \mathrm{span}\{Y_{l_{1}m_{1}},\dots,Y_{l_{p}m_{p}}\} = \{0\}, 
\end{equation}
but this has already been demonstrated in Proposition \ref{prop:linind}.
Applying Theorem \ref{thm:bbackus2}  we can, therefore, extend the conclusions of
Proposition \ref{prop:nonu1} to the case of any finite-number
of spherical harmonic coefficients. From Corollary \ref{coro:tran},
it follows that the same result holds for any restriction of the inference problem
including, in particular, the use of $C^{k}(\mathbb{S}^{2})$ for some $k \ge 1$.

Within eq.(\ref{eq:nbound}) we defined a  constraint set in $C^{0}(\mathbb{S}^{2})$ by
\begin{equation}
  U = \{u \in C^{0}(\mathbb{S}^{2}) \,|\, \|u\|_{C^{0}(\mathbb{S}^{2})} \le r\}, 
\end{equation}
which for $r\ge \sup\{v_{1},\dots,v_{n}\}$  is  compatible with the given data. Because
this choice of constraint set is bounded,  the same is trivially  true of
$U \cap \pi_{A}^{-1}\{\hat{A}^{-1}v\} $
and hence also its projection onto $E/\ker B$. Theorem \ref{thm:bbackus3} therefore
tells us that this prior constraint along with the data restricts
the spherical harmonic coefficients to a bounded set. It is important to note,
however,  that though this constraint set has been defined in terms of a particular
norm, it is bounded relative to any equivalent norm, and hence
the conclusion obtained from Theorem \ref{thm:bbackus3} is independent of the specific
norm selected.
Generalising  slightly the proof of Proposition \ref{prop:nonu2}, it can be shown that
the equality
\begin{equation}
   \pi_{B}(U \cap \pi_{A}^{-1}\{\hat{A}^{-1}v\} )  =  \pi_{B} U, 
\end{equation}
holds, and hence  the data provide no new information
over the prior constraint when the problem is posed in $C^{0}(\mathbb{S}^{2})$.

\subsection{Constraint sets defined by compatible norms}

\label{ssec:norm_bound_comp}

\subsubsection{General theory}

Using  Theorem \ref{thm:bbackus3} we can  recover Backus' original result
by considering constraint
sets defined in terms of a compatible norm on the model space.
Given such a norm $\|\cdot\|_{E}:\rightarrow \mathbb{R}$ on $E$, we
define the closed ball
\begin{equation}
  B_{r}(u_{0}) = \{u \in E \,|\, \|u-u_{0}\|_{E} \le r\},
\end{equation}
of radius $r>0$  and centre   $u_{0}\in E$. Using this notation, we then have:

\begin{prop}
  \label{prop:nbound}
  For  any $u_{0} \in E$ there exists an $r > 0$ for which  the constraint
  $u \in B_{r}(u_{0})$ is   compatible with the data $v \in F$. For such an $r > 0$,
  the constraint and  data  restrict the property vector to a bounded subset of $G$.
\end{prop}

\emph{Proof:} By  translation the general case can be reduced to $u_{0} = 0$.
Considering the equation $v = Au$ for given $v \in F$, we know from eq.(\ref{eq:afac})
that  $\pi_{A } u = \hat{A}^{-1}v$. 
Letting $\|\cdot\|_{E/\ker A}$ denote the quotient norm in eq.(\ref{eq:quotnorm}) induced by $\|\cdot\|_{E}$ we note that $\|\hat{A}^{-1}v\|_{E/\ker A}$ is equal to
the infimum of $\|u\|_{E}$ as $u$ ranges over all solutions of $v = Au$. It follows that if
 $r > \|\hat{A}^{-1}v\|_{E/\ker A}$ the
subset $B_{r}(0)$ has non-empty intersection with the closed affine subspace
$\pi_{A}^{-1}\{\hat{A}^{-1}v\}$. Moreover,
as $B_{r}(0)$ is  bounded the same holds  for 
$ B_{r}(0) \cap \pi_{A}^{-1} \{\hat{A}^{-1}v\}$ and hence  its projection onto $E/\ker B$.\eproof

To apply this result practically we need a  method for determining which values of
$w \in G$ are compatible with both the constraint $u \in B_{r}(0)$ and the data $v\in F$. This
can be done using Parker's
joint data-property mapping $C \in \Hom(E,F\oplus G)$ introduced within in the proof of Theorem
\ref{thm:bbackus2}. Making use of a factorisation analogous to that in eq.(\ref{eq:afac}) we have
\begin{equation}
  C  = \hat{C}\, \pi_{C}, 
\end{equation}
where $\pi_{ C}\in \Hom(E,E/\ker C)$ is the quotient mapping and $\hat{C} \in \Hom(E/\ker C,F\oplus G)$
has a continuous inverse. Given $w \in G$  we know that $\|\hat{C}^{-1}(v\oplus w)\|_{E/\ker C}$
is equal to the infimum of $\|u\|_{E}$ as $u$ ranges over all solutions of $Cu = v\oplus w$. Holding
the data fixed, it follows that the property vector is acceptable  
if and only if
\begin{equation}
  \label{eq:wset}
  \|\hat{C}^{-1}(v\oplus w)\|_{E/\ker C} \le r.
\end{equation}
If, therefore,
we can calculate  $  \|\hat{C}^{-1}(v\oplus w)\|_{E/\ker C}$ for given $v \in F$
and $w \in G$, we could check whether the condition in eq.(\ref{eq:wset})
is satisfied, and hence delimit the  subset of the property space in which
$w$ must lie. Moreover,  the mapping $w \mapsto \|\hat{C}^{-1}(v\oplus w)\|_{E/\ker C}$ is
 continuous and convex, and so
eq.(\ref{eq:wset}) defines a bounded and convex subset.
The value of  $  \|\hat{C}^{-1}(v\oplus w)\|_{E/\ker C}$ can, in principle, be found
by solving the  following convex optimisation problem with linear constraints:
\begin{equation}
  \label{eq:banopt}
  \inf_{u \in E} \|u\|_{E} \quad \mbox{s.t.}\quad Cu = v \oplus w.
\end{equation}
Geometrically this problem is equivalent to finding the infimum  distance from the
origin to a  closed affine subspace. In a general Banach space, 
this infimum need not be attained for any point within the affine subspace,
and even if it is, the point need not be unique \citep[][]{james1964weakly}. 
In the present context, however, this
does not really matter as it is only the infimum \emph{value}  that is of interest.
Nonetheless, numerical optimisation within a general Banach space
can be very   challenging, and  it may not be
possible to solve the problem practically.

The key property of such  a constraint set $B_{r}(0)$ 
is its boundedness, but it has others that are worth noting. First,
due to the norm being positively
homogeneous of order one, it follows that
if $u \in B_{r}(0)$ then so is $\lambda u$ for all $|\lambda| \le 1$.
Such a set is said to be \emph{balanced} \citep[e.g.][Definition 3.2]{treves}.
Next,  the triangle inequality shows that $ B_{r}(0)$ is convex. The
intersection of two convex sets is convex, and the same is true for the image of a convex set under
a  linear mapping. Thus, the convexity of the constraint set alone 
is sufficient to guarantee that the resulting subset of property space is  convex.
Finally, because the norm $\|\cdot\|_{E}$ is compatible with the topology on $E$,
the constraint set $B_{r}(0)$ is a closed neighbourhood of zero. Suppose, conversely, that
 $U\subseteq E$ is a balanced, convex, and  closed neighbourhood
of zero. It might then be asked if there is a compatible norm on $E$ such that $U$ is a closed ball of some
radius $r > 0$. The answer is yes, with  $U$ being the closed unit-ball relative to   its
\emph{Minkowski functional}  \citep[e.g.][Proposition 7.5]{treves} which is defined by
\begin{equation}
  p_{U}(u) = \inf\{\mu > 0\,|\, \mu u \in U\}.
\end{equation}
We conclude that a prior constraint $u \in B_{r}(0)$ defined in terms of a compatible norm
is logically equivalent to selecting as constraint set a bounded, balanced, convex, and
closed neighbourhood of zero.

\subsubsection{Application to the spectral estimation problem}

Working with $C^{0}(\mathbb{S}^{2})$ as the model space, we have already shown that
the data provide no information over that given by a prior norm bound.
If instead we took the model space to be $C^{k}(\mathbb{S}^{2})$ for some $k \ge 1$
it seems certain that  the inclusion in eq.(\ref{eq:tbound}) would be proper,
with the data then providing useful information on the desired
spherical harmonic coefficients. To quantify this idea, we could select a norm for
$C^{k}(\mathbb{S}^{2})$ along with a compatible prior bound, and  try to solve the
constrained optimisation problem in eq.(\ref{eq:banopt}) for different property
vectors.  Within this non-reflexive Banach space, however, there seems to be no practical
 method for doing this.

\subsection{Constraints defined in terms of a restriction of the inference problem}

\label{ssec:norm_bound_res}

\subsubsection{General theory}

Let $\tilde{E}$ be a Banachable space that is embedded continuously, properly, and densely
within $E$. As before we write $i\in \Hom(\tilde{E},E)$ for the inclusion
mapping, and define $\tilde{A} = A i$, $\tilde{B} = Bi$ as the data and
property mappings for the restriction of the inference problem.
For given data $v \in F$, let $\tilde{U} \subseteq \tilde{E}$ be a constraint set such
that the conditions in Theorem \ref{thm:bbackus3} are met for the restricted problem, and define
$U = i \tilde{U}\subseteq E$.  To avoid unwieldy notations, we  write
\begin{equation}
 \tilde{U}_{v} = \{\tilde{u} \in \tilde{E} \,|\, v = \tilde{A} \tilde{u} \}, 
\end{equation}
for the affine subspace in $\tilde{E}$ consistent with the data $v \in F$, and let
  $U_{v}$ be the corresponding affine subspace in $E$. By construction, it is clear that
$U \cap U_{v}   = i (\tilde{U} \cap\tilde{U}_{v})$, 
and as the subset on the right hand side is non-empty, the
constraint $u \in U$ is compatible with the data. Acting $B$ on
this equality, we see that
\begin{equation}
  B(U \cap U_{v} ) = (B\,i) (\tilde{U} \cap \tilde{U}_{v})
  = \tilde{B}(\tilde{U} \cap \tilde{U}_{v}),
\end{equation}
which shows that identical bounds on the property vector are obtained if we either apply  the
constraint $\tilde{u} \in \tilde{U}$ to the restriction of the inference problem,
or   use the induced constraint $u \in i \tilde{U}$  within the problem's original formulation. 
Restricting an inference problem to a smaller model space is,
therefore,  equivalent leaving the model space unchanged but
appropriately limiting the choice of constraint set. This
is a central result in this paper,  showing concretely  what is meant in the
introduction by saying that it is the prior constraints that determine
the choice of model space.

As an application of these ideas, we now
 consider how a Hilbert space structure might be
introduced into an inference problem. Indeed, the presence of such a structure will
be crucial for the remainder of the paper. We let $E$ denote the original Banachable model
space, and suppose that $E_{0}$ is a dense subspace.
On this subspace, we assume that  a symmetric and non-negative bilinear form is defined
\begin{equation}
  b:E_{0}\times E_{0} \rightarrow \mathbb{R}. 
\end{equation}
It can be shown that $u\mapsto b(u,u)$ vanishes on a linear subspace in $E_{0}$
that we denote by $\ker b$ \citep[e.g.][Chapter 7]{treves}. On the quotient space $E_{0}/\ker b$, this
bilinear form induces a well-defined inner product. In general, 
$E_{0}/\ker b$ will not be complete, but there is a standard procedure
by which it can be completed \citep[e.g.][Theorem 5.2]{treves}. In
this manner we obtain a Hilbert space that will be denoted by $\tilde{E}$.
Two  bilinear forms $b_{1}$ and $b_{2}$  on $E_{0}$ give
rise to isomorphic Hilbert spaces if 
\begin{equation}
c\,  b_{1}(u,u) \le b_{2}(u,u) \le C\, b_{1}(u,u), 
\end{equation}
 for all $u \in E_{0}$ and  some constants $0<c<C$. In practical applications there is usually
no definitive reason for choosing between such bilinear forms,
and so we instead focus on the common \emph{Hilbertable} structure they define.
 It might be thought that $\tilde{E}$  would be
 a dense subspace of  $E$, but    this will generally not hold. Indeed, due to
 both the passage to a quotient space and the completion process, there need be
 no simple relation between elements of $\tilde{E}$ and $E_{0}$.
 In some cases, however, there does exist
a proper, continuous, and dense embedding $\tilde{E}\hookrightarrow E$,
and we then  say that the inference problem has a \emph{Hilbertable restriction}. In such a situation we might also believe that the image of some $\tilde{U}\subseteq \tilde{E}$
under the inclusion mapping is an appropriate constraint set for the inference problem.
Due to   the above discussion, 
it follows that there is  nothing lost by simply working with $\tilde{E}$ as the
model space and taking $\tilde{U}$ to be the constraint set. The key point here is
that even though the natural model space in most geophysical inference problems is 
not  Hilbertable, the introduction of such a structure can be
justified so long as we believe that a suitable  prior constraint holds.

\subsubsection{Application to the spectral estimation problem}

We first show that the introduction of a Hilbertable structure to
an inference problem can fail. We take $C^{0}(\mathbb{S}^{2})$ to be the
model space, while the dense subspace is
simply $C^{0}(\mathbb{S}^{2})$ itself. The 
following symmetric bilinear form is then well-defined
\begin{equation}
  b(u,u')  = \int_{\mathbb{S}^{2}} u\,u' \dd S,
\end{equation}
and is clearly non-negative. Passing to the appropriate quotient  and
forming its completion we arrive at the familiar Hilbert space $L^{2}(\mathbb{S}^{2})$.
Further background on this space can be found in  Appendix \ref{ssec:sqint},  but
for the moment we need only recall that point-values of elements of $L^{2}(\mathbb{S}^{2})$
are not defined. Our motivating problem cannot, therefore,
be meaningfully formulated in $L^{2}(\mathbb{S}^{2})$.

To show that this problem does have a Hilbertable restriction we make use
of the Sobolev spaces $H^{s}(\mathbb{S}^{2})$ for appropriate values
of the exponent $s \in \mathbb{R}$. The definition and key properties of these
spaces can be found in Appendix \ref{sec:fspace}, and here we
only summarise the necessary facts.  We take $C^{0}(\mathbb{S}^{2})$ to be the 
model space, while the relevant dense subspace is now
$C^{\infty}(\mathbb{S}^{2})$. For $u \in C^{\infty}(\mathbb{S}^{2})$
we can define 
\begin{equation}
  u_{lm} = \int_{\mathbb{S}^{2}} u \, Y_{lm} \dd S, 
\end{equation}
for $l \in \mathbb{N}$ and $-l\le m \le l$. It can be shown that
the function 
\begin{equation}
  l \mapsto \sup_{-l\le m \le l}|u_{lm}|,
\end{equation}
decreases faster than any polynomial, and hence the inner product
\begin{equation}
  b(u,u') = \sum_{lm}  \left[1+ \lambda^{2}\,l(l+1)\right]^{s} u_{lm} u'_{lm}, 
\end{equation}
is well-defined for $u,u' \in C^{\infty}(\mathbb{S}^{2})$ any $\lambda > 0$ and $s \in \mathbb{R}$.
Completing $C^{\infty}(\mathbb{S}^{2})$ relative to this inner product we arrive at the Sobolev space
$H^{s}(\mathbb{S}^{2})$.  From Proposition \ref{prop:stopo}  we see that the   topology
of this  space depends only on the value of the exponent $s$, with different
choices of  $\lambda$ leading only to equivalent  inner product.
Of crucial importance for applications is the  Sobolev embedding theorem
which shows that if $s > 1 + k$, for $k \in \mathbb{N}$, there is a continuous, proper, and
dense embedding of $H^{s}(\mathbb{S}^{2})$ into $C^{k}(\mathbb{S}^{2})$.  It follows
that $H^{s}(\mathbb{S}^{2})$ with $s > 1$   provides a well-defined Hilbertable restriction of
our motivating problem, while the application of
Proposition \ref{prop:rest} and Corollary \ref{coro:tran} implies that
Assumptions \ref{asm:sur} and \ref{asm:tran} continue to hold.

\subsection{Solution of linear inference problems in Hilbert spaces}

\label{ssec:hilbert_sol}

\subsubsection{General theory}

We now specialise the approach in Section \ref{ssec:norm_bound_comp}
to problems in Hilbert spaces. Specifically, we consider a Hilbertable restriction of
a linear inference problem  with the constraint set being a closed ball
relative to a compatible inner product. This inner product singles out
a specific Hilbert space structure on the model space that will be
used throughout this section. Inner products must  also be selected
for the data and property spaces, but we will see that these latter choices
have no effect on the final results.  We start by summarising some properties
of Hilbert spaces that will be needed both here and elsewhere in  the paper.
A central result for Hilbert spaces is the
\emph{Riesz representation theorem} \cite[e.g.][Theorem 12.2]{treves}.
Applied to the model space $E$, it states that there is a unique
continuous linear mapping $\mathscr{J}_{E} \in \Hom(E',E)$ with continuous inverse such that
for each $u' \in E'$ and all $u \in E$ we have
\begin{equation}
  \braket{u'}{u} = \cbraket{\mathscr{J}_{E}u'}{u}_{E},  \quad \|\mathscr{J}_{E}u'\|_{E} = \|u'\|_{E'}.
\end{equation}
It follows, in particular, that  Hilbert spaces are reflexive. 
For each dual vector $u' \in E'$, we call $\mathscr{J}u'$ its $E$-\emph{representation}.
In terms of the data mapping $A \in \Hom(E,F)$, we can now define its \emph{adjoint}
$A^{*} \in \Hom(F,E)$ through the requirement that
\begin{equation}
  \cbraket{Au}{v}_{F} = \cbraket{u}{A^{*}v}_{E}, 
\end{equation}
for all $u \in E$ and $v \in F$. From this definition we see that the
adjoint and dual of the data mapping are related by
\begin{equation}
  A^{*} = \mathscr{J}_{E} A' \mathscr{J}_{F}^{-1}, 
\end{equation}
with $\mathscr{J}_{F}\in \Hom(F',F)$ defined through
the Riesz representation theorem applied to the data space.  Using
this identity, it follows that the surjectivity
and transversality assumptions on $A$ and $B$ can be  equivalently expressed in
terms of their adjoint mappings.
For a linear subspace  $U\subseteq E$ its orthogonal complement $U^{\perp}$ is
the subspace of $E$ defined by
\begin{equation}
  U^{\perp} = \{u' \in E \,|\, \cbraket{u}{u'}_{E} = 0,\, \forall u \in U\}, 
\end{equation}
and  is  related to the polar of $U$ through
\begin{equation}
  \label{eq:perppol}
U^{\perp} =  \mathscr{J}_{E}  U^{\circ}.
\end{equation}
A special case of the Hilbert space \emph{projection theorem} \citep[e.g.][Theorem 12.1]{treves} shows that for
each closed linear subspace $U\subseteq E$ there is an associated orthogonal  decomposition
\begin{equation}
  E = U \oplus U^{\perp}, 
\end{equation}
where $\oplus$ denotes the direct sum of two linear subspaces as defined in Appendix \ref{sec:funal}.
Making use of this result in conjunction with eq.(\ref{eq:bclim}) and eq.(\ref{eq:perppol})
we arrive at the useful decompositions
\begin{equation}
  \label{eq:dporth}
 F = \image A \oplus \ker A^{*}, \quad G = \image B \oplus \ker B^{*}, 
\end{equation}
of the data and property spaces. Because Hilbert spaces are reflexive,
we can apply the same idea to the adjoint operators to obtain
\begin{equation}
  \label{eq:modorth}
  E = \image A^{*} \oplus \ker A = \image B^{*} \oplus \ker B. 
\end{equation}

 From Section \ref{ssec:backus}
we know that the equation $v = Au$ for $u \in E$ given $v \in F$ is
under-determined. Within a Hilbert space there is a simple method for
obtaining the unique solution of this problem having the smallest norm:
\begin{prop}
  \label{prop:minnorm}
  For any $v \in F$, the equation $v = Au$ admits the general
  solution
  \begin{equation}
    u = A^{*}(AA^{*})^{-1} v + u_{0}, 
  \end{equation}
  where $u_{0}$ is an arbitrary element of $\ker A$. Of all these solutions
  $u = A^{*}(AA^{*})^{-1} v$ has the smallest norm.
\end{prop}
\emph{Proof:} The assumption that $A$ is surjective means 
the  equation $v = Au$ has solutions for any $v \in F$.
Using the first orthogonal decomposition of $E$ in eq.(\ref{eq:modorth}),
we can write $u = A^{*}\tilde{v} + u_{0}$ for some $\tilde{v} \in F$ and $u_{0} \in \ker A$, and hence obtain
\begin{equation}
  \label{eq:vtmp}
  v = AA^{*} \tilde{v}.
\end{equation}
If we can  show that $AA^{*} \in \Hom(F)$ is continuously invertible, we
arrive at the desired general solution.
Because $AA^{*} \in \Hom(F)$ with $F$ finite-dimensional,
we need only prove that $\ker AA^{*}$ is trivial. Supposing $v_{0} \in \ker AA^{*}$ we have
\begin{equation}
 0 =   \cbraket{AA^{*}v_{0}}{v_{0}}_{F} = \cbraket{A^{*}v_{0}}{A^{*}v_{0}}_{E} = \|A^{*}v_{0}\|_{E}^{2}.
\end{equation}
and hence $A^{*}v_{0} = 0$.
The surjectivity of $A$ along with the orthogonal
decomposition of $F$ in eq.(\ref{eq:dporth}) implies that $\ker A^{*}$
is trivial, and so $v_{0} = 0$ as desired. Finally, we
again use the orthogonal decomposition of $E$ to write
\begin{equation}
  \|u\|_{E}^{2} = \|A^{*}(AA^{*})^{-1}v\|_{E}^{2} + \|u_{0}\|_{E}^{2}, 
\end{equation}
which is   minimised by taking $u_{0} = 0$. \eproof
\begin{coro}
  \label{cor:proj}
  The orthogonal projection operator $\mathbb{P}_{\image A^{*}} $
  onto $\image A^{*}$ is given by $A^{*}(AA^{*})^{-1}A$, while the orthogonal
  projection operator   onto $\ker A$ can be written $\mathbb{P}_{\ker A} = 1- A^{*}(AA^{*})^{-1}A$.
\end{coro}
Minimum norm solutions can be obtained by solving the finite-dimensional
system of linear equations in eq.(\ref{eq:vtmp})
either directly or with iterative methods. In practice, however, $AA^{*}$
can be poorly conditioned meaning that standard  iterative methods such as linear conjugate gradients
can be slow to converge. A useful alternative     is based on  gradient-based optimisation
of the least-squares functional
\begin{equation}
  \label{eq:lsf}
  J(u) = \frac{1}{2}\|Au-v\|^{2}_{F}, 
\end{equation}
defined on the model space.
Given the first orthogonal decomposition of $E$ in eq.(\ref{eq:modorth}), we can again set
$u = A^{*}\tilde{v} + u_{0}$  with $\tilde{v} \in F$ and $u_{0} \in \ker A$, and hence
write the functional in the form
\begin{equation}
  J(A^{*}\tilde{v} + u_{0}) = \frac{1}{2}\|AA^{*}\tilde{v} - v\|_{F}^{2}. 
\end{equation}
The minimum value of $J$ is, therefore, equal to zero, this being   obtained
whenever $u = A^{*}(AA^{*})^{-1} v + u_{0}$ with $u_{0} \in \ker A$.
Moreover, the derivative of $J$ is readily seen to be
\begin{equation}
  DJ(u) = A^{*}(Au-v), 
\end{equation}
where the Riesz representation theorem has been implicitly used to identify
$DJ(u)$ with an element of the model space. Noting that the derivative  takes values
in $\image A^{*}$, it follows that by applying
 gradient-based optimisation  to
$J$ starting from the zero vector we converge
precisely to the minimum norm solution \citep{kammerer1972convergence}.
Here it is, of course, critical that  descent directions are
formed from linear combinations of the gradient at the current or past stages of
the iterative process, but this holds for standard algorithms like non-linear conjugate
gradients or L-BFGS \citep[e.g.][]{nocedal2006numerical}.

Returning to the  approach in Section \ref{ssec:norm_bound_comp}, we can
test the compatibility of a  property vector $w \in G$ with the
data $v \in F$ and prior constraint $u \in B_{r}(0)$ by  computing the
infimum norm of  solutions to $Cu = v\oplus w$, where
$C \in \Hom(E,F\oplus G)$ is the joint data-property mapping
defined in eq.(\ref{eq:jdpdef}).
In a Hilbert space this problem is readily solved by applying
Proposition \ref{prop:minnorm} to arrive at the 
 unique model for which this infimum norm is attained
\begin{equation}
  \tilde{u} = C^{*}(CC^{*})^{-1} (v\oplus w),
\end{equation}
and substituting into  eq.(\ref{eq:wset})  we obtain the simple condition
\begin{equation}
  \label{eq:mincon}
  \cbraket{(CC^{*})^{-1} (v\oplus w)}{v\oplus w}_{F\oplus G} \le r^{2}.
\end{equation}
which defines  a closed subset in $G$ whose boundary is a
hyperellipsoid.

\begin{prop}
  The subset defined by eq.(\ref{eq:mincon}) is independent of the
  choice of inner products on $F$ and $G$.
\end{prop}

\emph{Proof:}  Let $\ccbraket{\cdot}{\cdot}_{F\oplus G}$ denote an equivalent
choice of inner product on $F\oplus G$. From the Riesz representation theorem
it is easily shown that there exists a unique positive-definite and self-adjoint
operator $S \in \Hom(F\oplus G)$ such that for all 
$v\oplus w, v'\oplus w' \in F\oplus G$ we have
\begin{equation}
  \ccbraket{v\oplus w}{v'\oplus w'}_{F\oplus G}
  = \cbraket{S(v\oplus w)}{v'\oplus w'}_{F\oplus G}.
\end{equation}
Writing $C^{\dagger}$ for the adjoint of $C$ relative to the new inner product it follows
that $C^{\dagger} = C^{*} S$, and hence
\begin{equation}
  \ccbraket{(CC^{\dagger})^{-1} (v\oplus w)}{v\oplus w}_{F\oplus G}
  = \cbraket{S(CC^{*} S)^{-1} (v\oplus w)}{v\oplus w}_{F\oplus G}
  =   \cbraket{(CC^{*})^{-1} (v\oplus w)}{v\oplus w}_{F\oplus G}.
\end{equation}
\eproof

An alternate form of eq.(\ref{eq:mincon}) can
be obtained that is preferable  computationally.
Starting from the requirement $v = Au $ and applying
Proposition \ref{prop:minnorm}, we know that the model
takes the form
\begin{equation}
  u = \tilde{u} + u_{0}, 
\end{equation}
where $\tilde{u} = A^{*}(AA^{*})^{-1}v$ is the minimum norm solution,
and $u_{0} \in \ker A$. Acting the property mapping on
this solution we obtain
\begin{equation}
  w = \tilde{w} + B|_{\ker A} u_{0}, 
\end{equation}
where $\tilde{w} = B \tilde{u}$ is the property mapping corresponding
to the minimum norm solution, and $B|_{\ker A} \in \Hom(\ker A,G)$
is the restriction of the property mapping to
$\ker A$. Letting $i_{\ker A}$ denote the inclusion mapping from $\ker A$ into $E$,
we  note  that its adjoint is given by $\mathbb{P}_{\ker A}$ regarded as
a mapping from $E$ onto $\ker A$. The restriction of the property mapping to
$\ker A$ can then be written
\begin{equation}
  B|_{\ker A} = B \,i_{\ker A}, 
\end{equation}
and taking adjoints we obtain
\begin{equation}
  \label{eq:bra}
  B|_{\ker A}^{*} = \mathbb{P}_{\ker A} B^{*}.
\end{equation}
Recalling eq.(\ref{eq:modorth}), it follows that a non-zero element
of $\ker B|_{\ker A}^{*} $ can exist if and only if $\image A^{*}$ and $\image B^{*}$
have a non-trivial intersection which  contradicts Assumption \ref{asm:tran}.
We  conclude that $B|_{\ker A}$ is surjective and hence
$B|_{\ker A}B|_{\ker A}^{*} \in \Hom(G)$  is positive-definite.
Associated with $B|_{\ker A} \in \Hom(\ker A,G)$ we can form the  orthogonal decomposition  
\begin{equation}
  \ker A = \ker B|_{\ker A} \oplus \image B|_{\ker A}^{*}, 
\end{equation}
which implies that the  part of the model in $\ker A$ can be written
\begin{equation}
  u_{0} = B|_{\ker A}^{*}w' + u_{00}, 
\end{equation}
for some $w' \in G$ and $u_{00} \in \ker B|_{\ker A}$, and hence
\begin{equation}
  \label{eq:wptmp}
  w = \tilde{w} + B|_{\ker A}B|_{\ker A}^{*}w'.
\end{equation}
Solving the latter equation for $w'$ we arrive at
\begin{equation}
  \label{eq:udec}
  u = \tilde{u} + B|_{\ker A}^{*}(B|_{\ker A}B|_{\ker A}^{*})^{-1}(w-\tilde{w}) + u_{00}, 
\end{equation}
and as each term lies in an orthogonal subspace,
eq.(\ref{eq:mincon}) is reduced to
\begin{equation}
  \label{eq:pbound1}
  \cbraket{(B|_{\ker A}B|_{\ker A}^{*})^{-1}(w-\tilde{w})}
  {w-\tilde{w}}_{G}  \le r^{2}-\|\tilde{u}\|^{2}_{E}.
\end{equation}
This result is  equivalent to  eq.(4) of \cite{backus1970inferenceI}
and eq.(B2) in \cite{parker1977linear}, though a direct algebraic verification will
not be given. In the absence of any data, eq.(\ref{eq:pbound1}) reduces to
\begin{equation}
  \label{eq:pbound2}
  \cbraket{(BB^{*})^{-1}w}
  {w}_{G} \le r^{2},
\end{equation}
which defines a closed subset of $G$ whose boundary is a hyperellipsoid
centred about the origin.  Necessarily we have 
\begin{equation}
\cbraket{(BB^{*})^{-1}w}
  {w}_{G} \le   \|\tilde{u}\|^{2}_{E} +   \cbraket{(B|_{\ker A}B|_{\ker A}^{*})^{-1}(w-\tilde{w})}
  {w-\tilde{w}}_{G}, 
\end{equation}
for all $w \in G$. If this inequality were strict, it would mean that the
data provides additional information over the assumed constraint.
In fact, it is not difficult to see that this will hold in all but one situation.
First, we consider the coefficients of the  quadratic terms in a fixed
direction. If the coefficient belonging to the left hand side were larger than that
on the right, then for large enough $w$ the inequality would fail to hold.
The quadratic part of the function on the right
hand side must, therefore, grow at least as quickly as that on the left. The worse
case scenario is when the quadratic parts of the two functions agree,
this meaning that we have $BB^{*} = B|_{\ker A}B|_{\ker A}^{*}$. Using
eq.(\ref{eq:bra}),  this can happen if and only
if $\image B^{*} \subseteq \ker A$ which, while possible, would be
an unlikely and unfortunate coincidence. But even if equality does hold between
the quadratic terms, the constant $\|\tilde{u}\|_{E}^{2}$
on the right hand side should, in general, be sufficient
to ensure that the inequality is strict. Indeed, this
can only fail if $\tilde{u} = 0$, which requires the
data itself to vanish.


\begin{figure}
  \centering
  \begin{minipage}{1.0\linewidth}
    \includegraphics[width=0.49\textwidth,trim={1cm 7.25cm 1cm 7.25cm},clip]{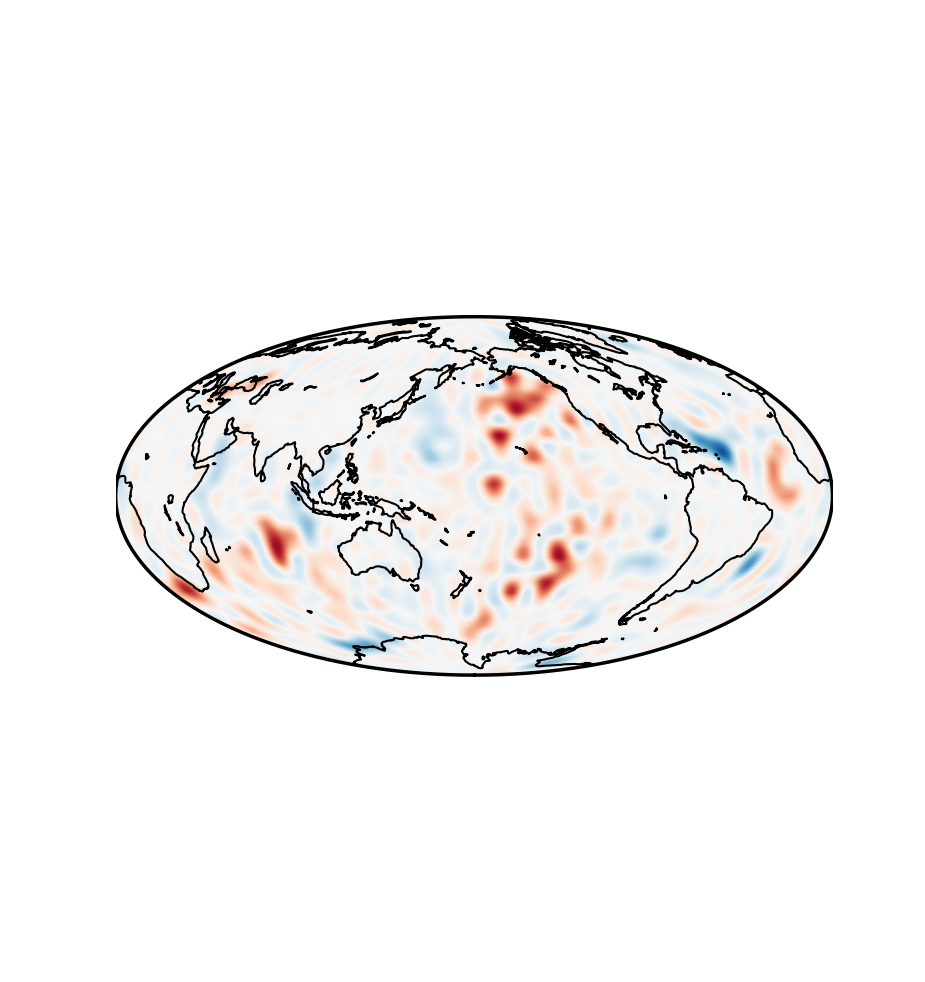}
    \includegraphics[width=0.49\textwidth,trim={1cm 7.25cm 1cm 7.25cm},clip]{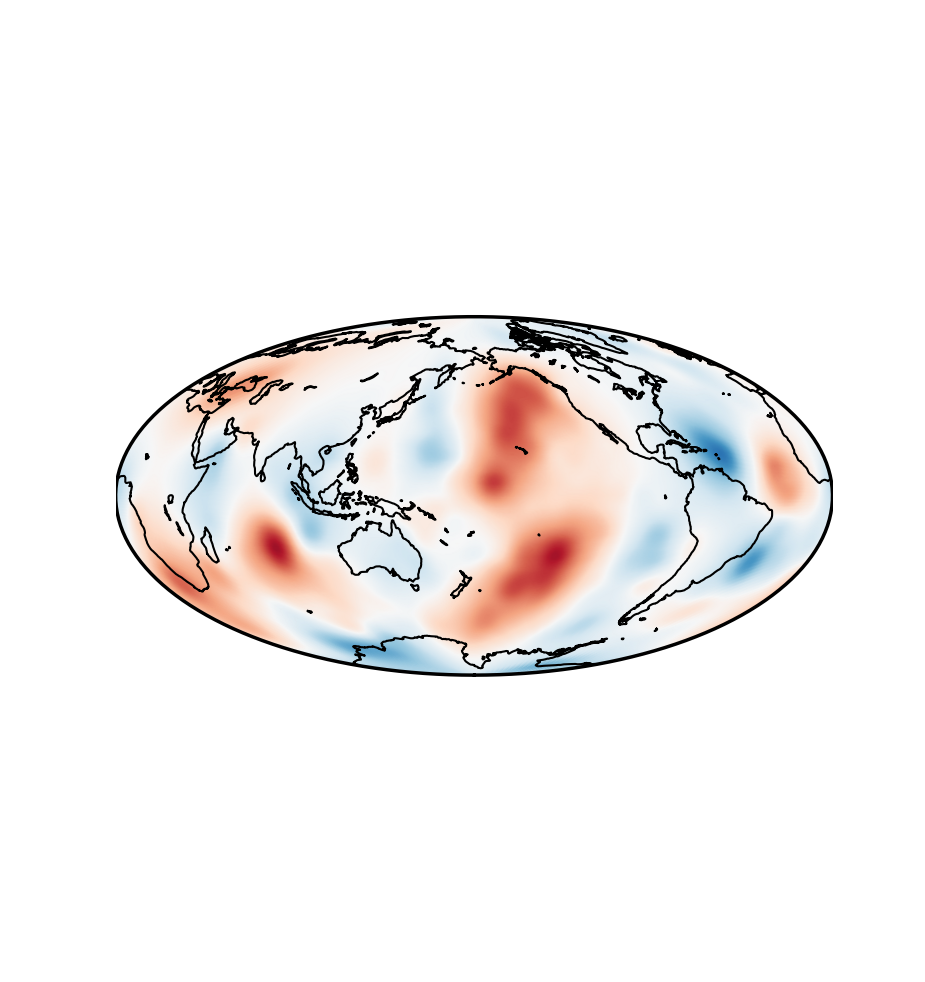}
  \end{minipage}
  \begin{minipage}{1.0\linewidth}
    \includegraphics[width=0.49\textwidth,trim={1cm 7.25cm 1cm 7.25cm},clip]{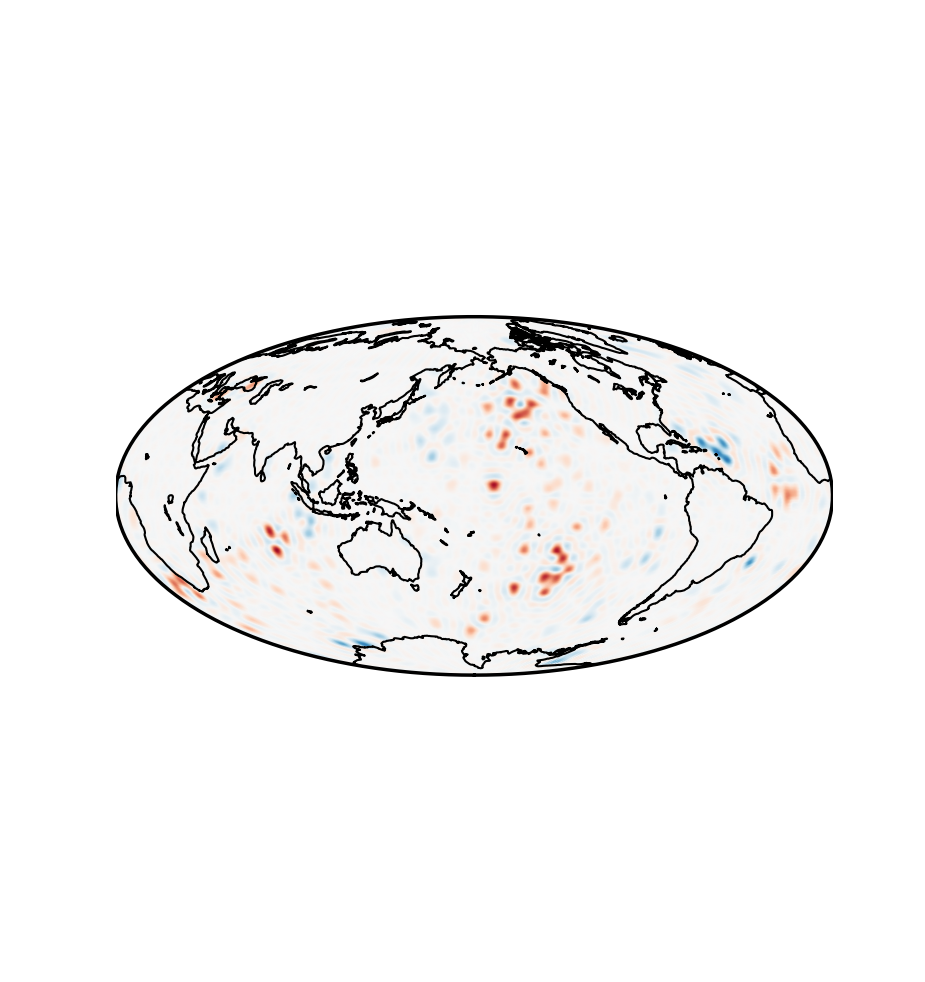}
    \includegraphics[width=0.49\textwidth,trim={1cm 7.25cm 1cm 7.25cm},clip]{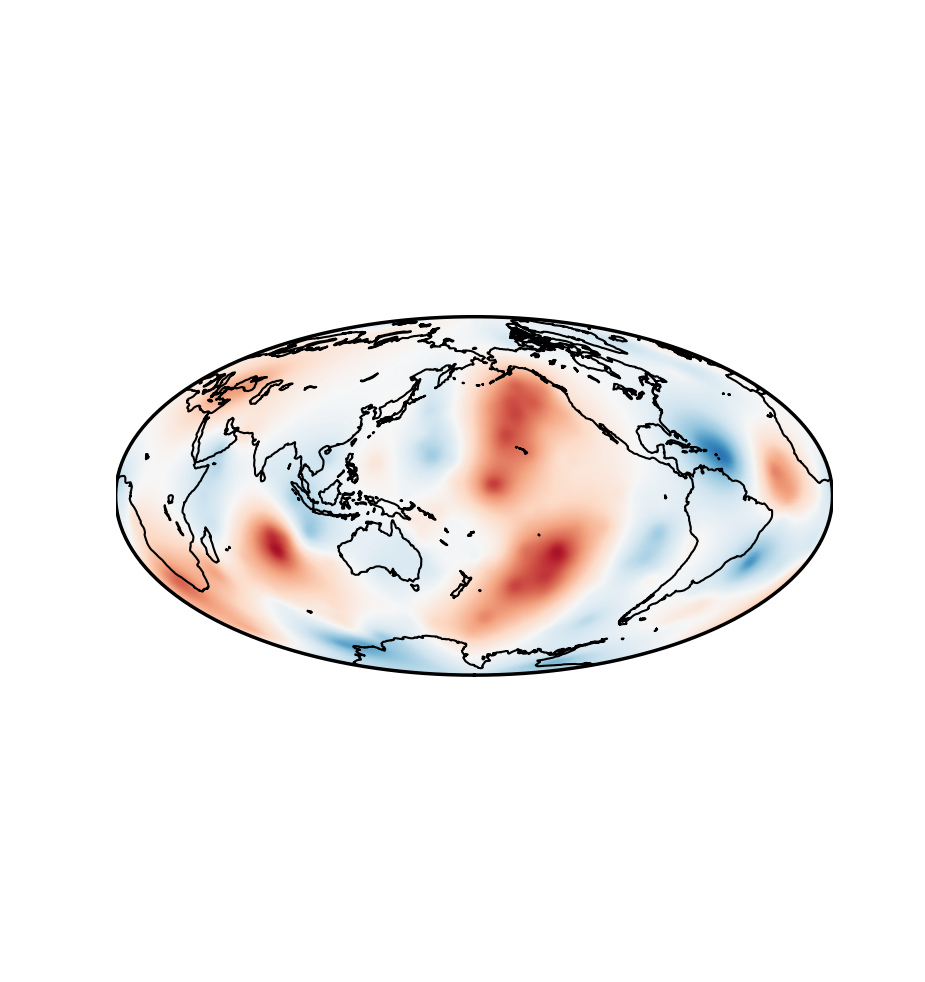}
  \end{minipage}
  \begin{minipage}{1.0\linewidth}
    \includegraphics[width=0.49\textwidth,trim={1cm 7.25cm 1cm 7.25cm},clip]{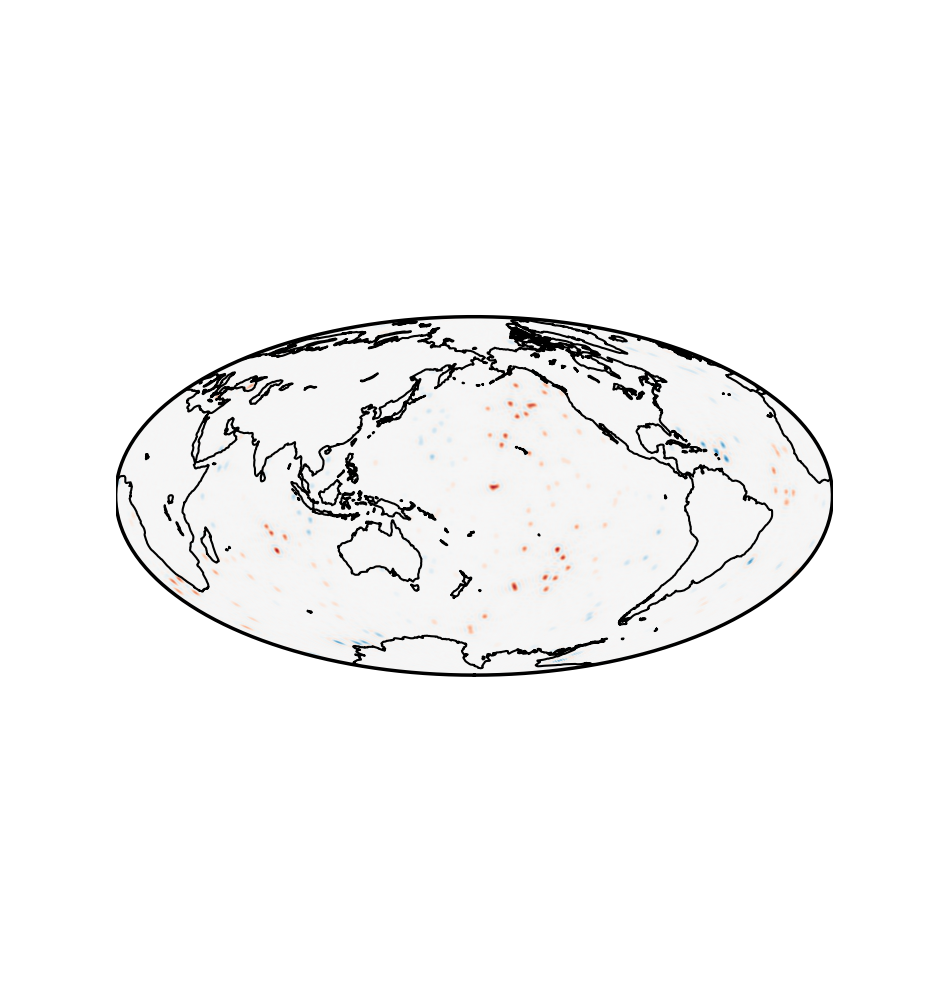}
    \includegraphics[width=0.49\textwidth,trim={1cm 7.25cm 1cm 7.25cm},clip]{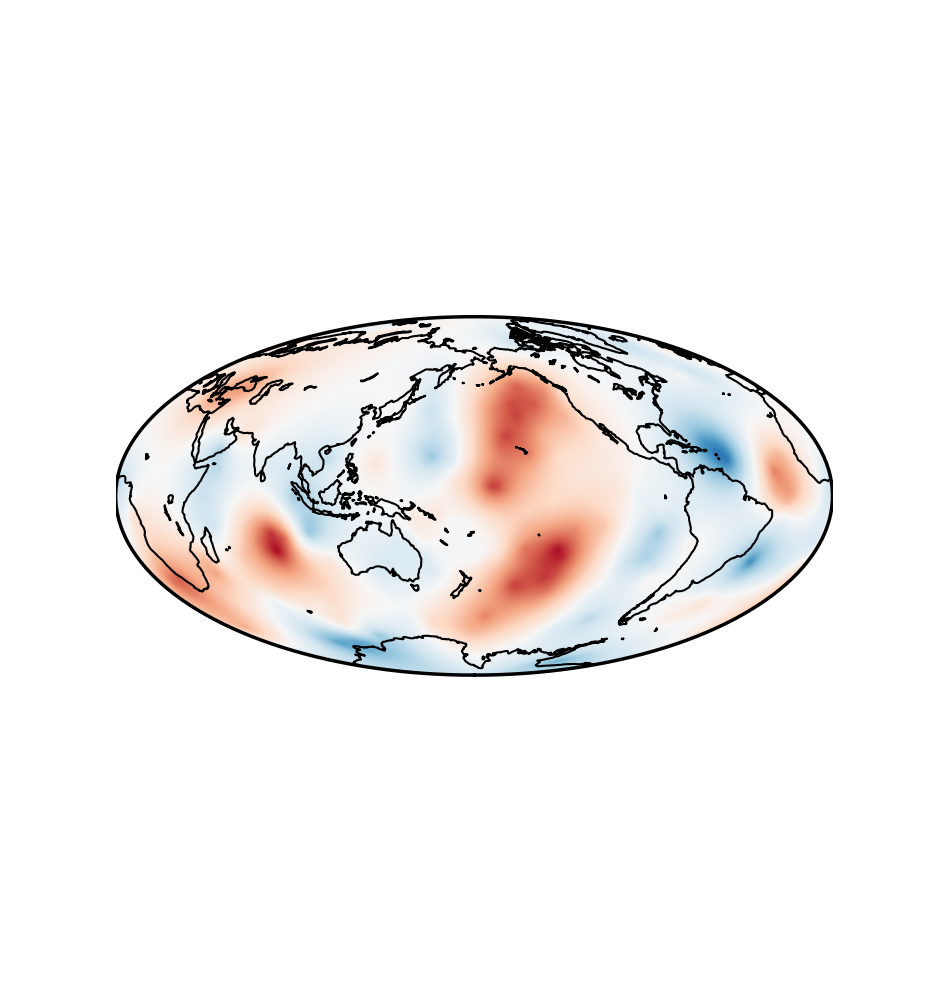}
  \end{minipage}
  \begin{minipage}{1.0\linewidth}
    \includegraphics[width=0.49\textwidth,trim={1cm 7.25cm 1cm 7.25cm},clip]{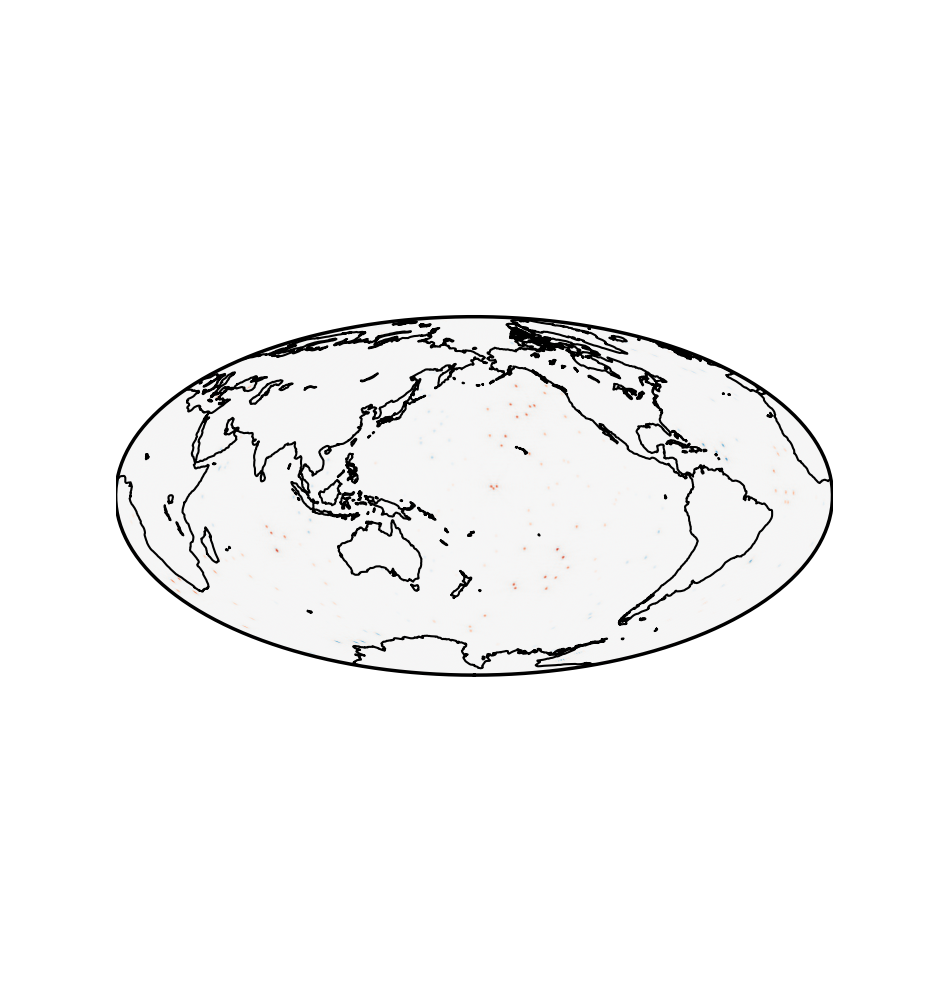}
    \includegraphics[width=0.49\textwidth,trim={1cm 7.25cm 1cm 7.25cm},clip]{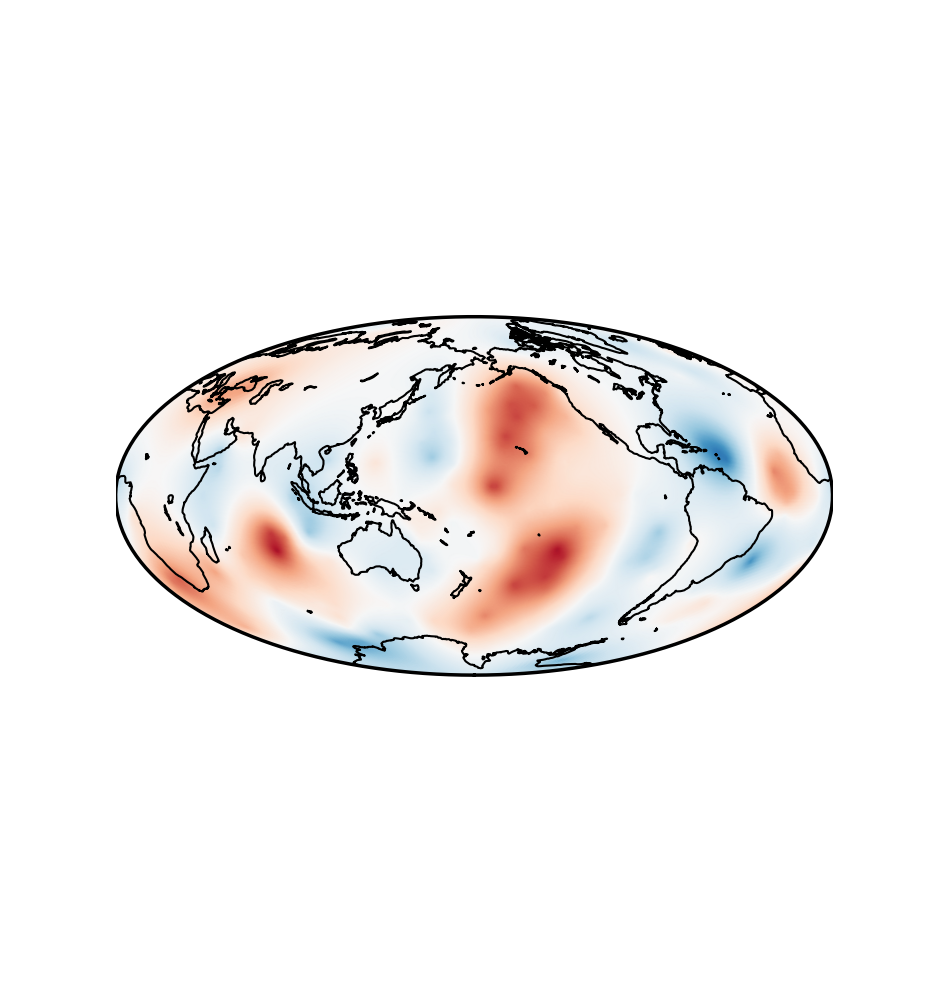}
  \end{minipage}
  \vspace{0.00mm}
  \caption{Minimum norm solutions at different truncation degrees obtained from the
    point data shown in Fig.\ref{fig:data_ne}.
    For  clarity  data locations and values are not displayed, but in
    each  case the point-values are fit to to within a relative error
    of $10^{-8}$.    On the left
    hand side we show, in order from top to bottom, solutions obtained
    using an $L^{2}(\mathbb{S}^{2})$ formulation at truncation degrees
    $32$, $64$, $128$, and $256$.  The right hand side then shows the analogous
    solutions calculated  in $H^{s}(\mathbb{S}^{2})$ with $s = 1.5$ and $\lambda = 0.2$,
    and it is only in this case that pointwise convergence occurs. 
  }
  \label{fig:minnorm1}
\end{figure}


From the inequality in eq.(\ref{eq:pbound1}) we can see that the bounding hyperellipsoid has  centre
at $\tilde{w}$, and while its  shape is determined by the positive-definite operator
$B|_{\ker A}B|_{\ker A}^{*}$. To apply this result practically we first determine the property
vector $\tilde{w}$ corresponding to the minimum norm solution  of
$v = Au$. We then  calculate the components of the operator
$B|_{\ker A}B|_{\ker A}^{*}$ which, using eq.(\ref{eq:bra}), can be written
\begin{equation}
  B|_{\ker A}B|_{\ker A}^{*} = B \,\mathbb{P}_{\ker A} B^{*}.
\end{equation}
Computing the action of the property mapping or its adjoint is trivial, and
so the main cost is associated with the orthogonal projection onto $\ker A$.
Letting $\{g_{j}\}_{j=1}^{p}$ be a basis for $G$, we  set $u_{j} = B^{*}g_{j}$
and $v_{j} = A u_{j}$. Recalling Corollary \ref{cor:proj}, it follows readily that
\begin{equation}
  B \,\mathbb{P}_{\ker A} B^{*} g_{j} = B(u_{j} - \tilde{u}_{j}), 
\end{equation}
where $\tilde{u}_{j}$ is the minimum norm solution of $v_{j} = A u$. In this manner,
we see that the bounding hyperellipsoid defined through eq.(\ref{eq:pbound1})
can be determined at the cost of $\dim G+1$
minimum norm solutions.

Once the vector $\tilde{w} \in G$ and the operator $B|_{\ker A}B|_{\ker A}^{*} \in \Hom(G)$
have been calculated, it can be readily checked whether a given property vector
$w \in G$ is consistent with eq.(\ref{eq:pbound1}). For $\dim G \le 2$ a graphical
representation of this subset is trivial, but this cannot be done in
higher-dimensions. It is, therefore, useful
to determine the  values obtained by a certain functional on $G$ as $w$ varies over the
subset. As a simple example, we consider the linear functional
$w \mapsto \cbraket{w'}{w}_{G}$ for given $w' \in G$. It is clear geometrically
that this functional can be extremised subject to eq.(\ref{eq:pbound2})
only when $w$ lies on the bounding hyperellipsoid. Applying the method of Lagrange
multipliers \citep[e.g.][]{luenberger1997optimization} it is readily seen
that the two stationary points occur at
\begin{equation}
  w = \tilde{w} \pm \sqrt{\frac{r^{2}-\|\tilde{u}\|_{E}^{2}}{\cbraket{B|_{\ker A}B|_{\ker A}^{*}w'}
  {w'}_{G}}}B|_{\ker A}B|_{\ker A}^{*}w', 
\end{equation}
and hence the functional's minimum and maximum values are
\begin{equation}
  \label{eq:1dminmax}
  \cbraket{w'}{w}_{G} = \cbraket{w'}{\tilde{w}}_{G}
  \pm \sqrt{(r^{2}-\|\tilde{u}\|_{E}^{2})\cbraket{B|_{\ker A}B|_{\ker A}^{*}w'}
  {w'}_{G}}.
\end{equation}
A similar approach can be applied to non-linear functionals,
though  the  optimisation problems become more complicated.

\subsubsection{Application to the spectral estimation problem}

Using eq.(\ref{eq:delrep1}), point evaluation of a function in $H^{s}(\mathbb{S}^{2})$ can be written
\begin{equation}
  u(x) = \cbraket{\hat{\delta}_{x}}{u}_{H^{s}(\mathbb{S}^{2})}, 
\end{equation}
where $\hat{\delta}_{x}$ is the $H^{s}(\mathbb{S}^{2})$-representation of
the Dirac measure at $x \in \mathbb{S}^{2}$. It follows that the
data mapping can be written
\begin{equation}
  \label{eq:dmap}    
  A u = \sum_{i=1}^{n}   \cbraket{\hat{\delta}_{x_{i}}}{u}_{H^{s}(\mathbb{S}^{2})}\,f_{i}, 
\end{equation}
where $f_{1},\dots,f_{n}$ denote the standard basis vectors
for the data space $\mathbb{R}^{n}$.  The adjoint data mapping
is  given by
\begin{equation}
  A^{*}v = \sum_{i=1}^{n} \cbraket{f_{i}}{v}_{\mathbb{R}^{n}}\, \hat{\delta}_{x_{i}}. 
\end{equation}
We write $\hat{Y}_{lm}$ for the $H^{s}(\mathbb{S}^{2})$-representation
of the $(l,m)$th spherical harmonic functional
\begin{equation}
  u \mapsto \int_{\mathbb{S}^{2}} u\, Y_{lm} \dd S.
\end{equation}
The property mapping is then given by
\begin{equation}
  B u= \sum_{j=1}^{p}  \cbraket{\hat{Y}_{l_{j}m_{j}}}{u}_{H^{s}(\mathbb{S}^{2})}\,g_{j}, 
\end{equation}
where $g_{1},\dots,g_{p}$ denote the standard basis vectors
for the property space $\mathbb{R}^{p}$, while its adjoint is
\begin{equation}
  B^{*}w = \sum_{j=1}^{p} \cbraket{g_{j}}{w}_{\mathbb{R}^{p}}\,\hat{Y}_{l_{j}m_{j}}.
\end{equation}

\begin{figure}
  \centering
  \begin{minipage}{1.0\linewidth}
        \includegraphics[width=0.49\textwidth,trim={1cm 7.25cm 1cm 7.25cm},clip]{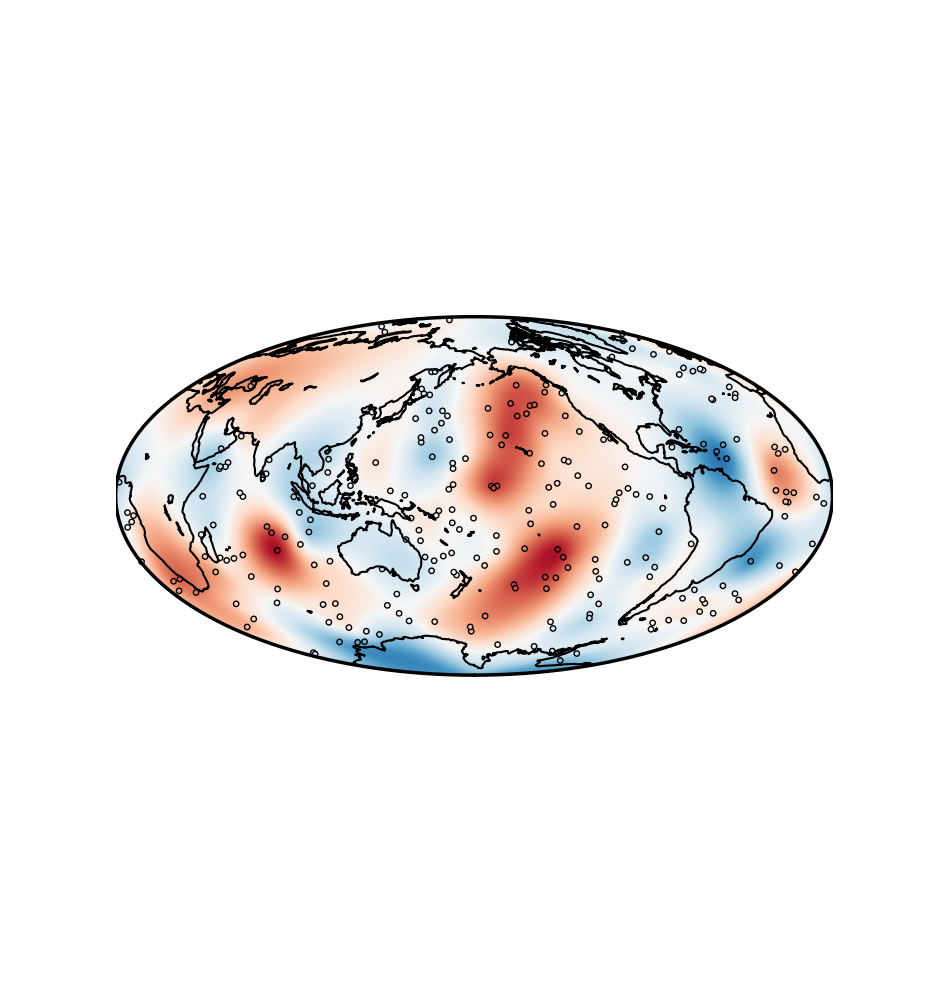}
        \includegraphics[width=0.49\textwidth,trim={1cm 7.25cm 1cm 7.25cm},clip]{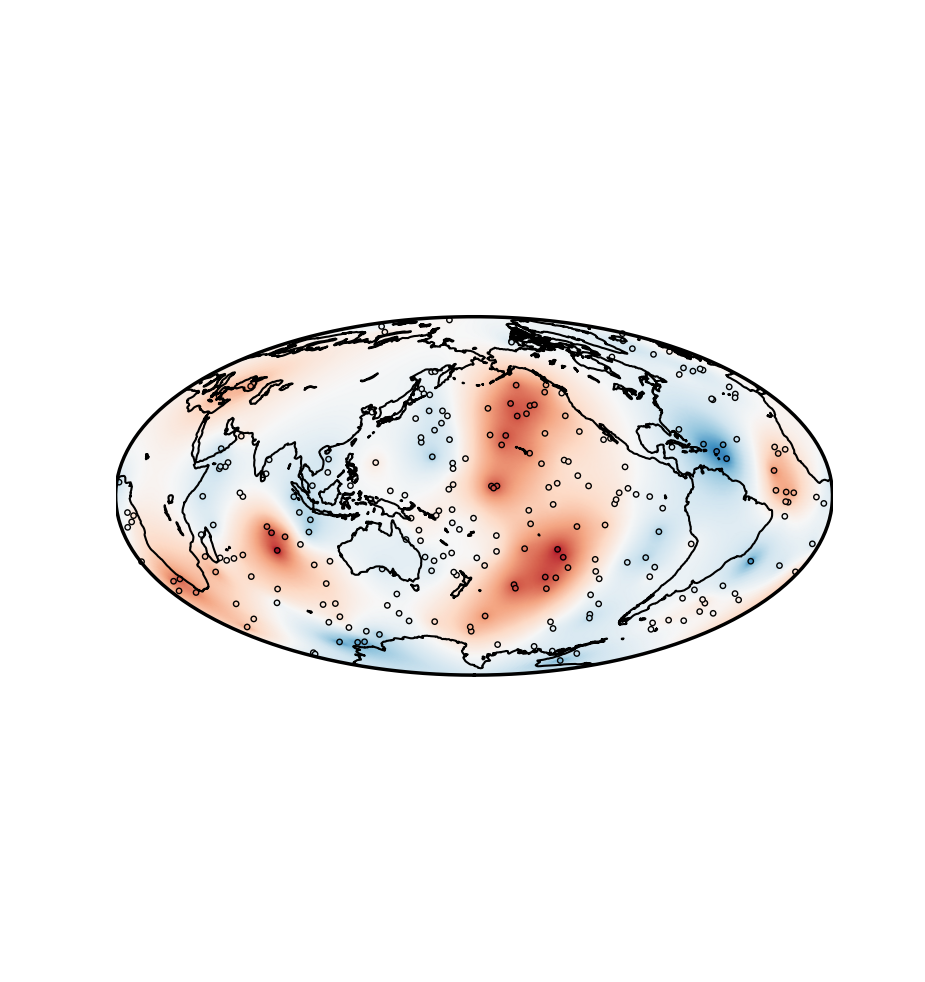}
  \end{minipage}
    \begin{minipage}{1.0\linewidth}
        \includegraphics[width=0.49\textwidth,trim={1cm 7.25cm 1cm 7.25cm},clip]{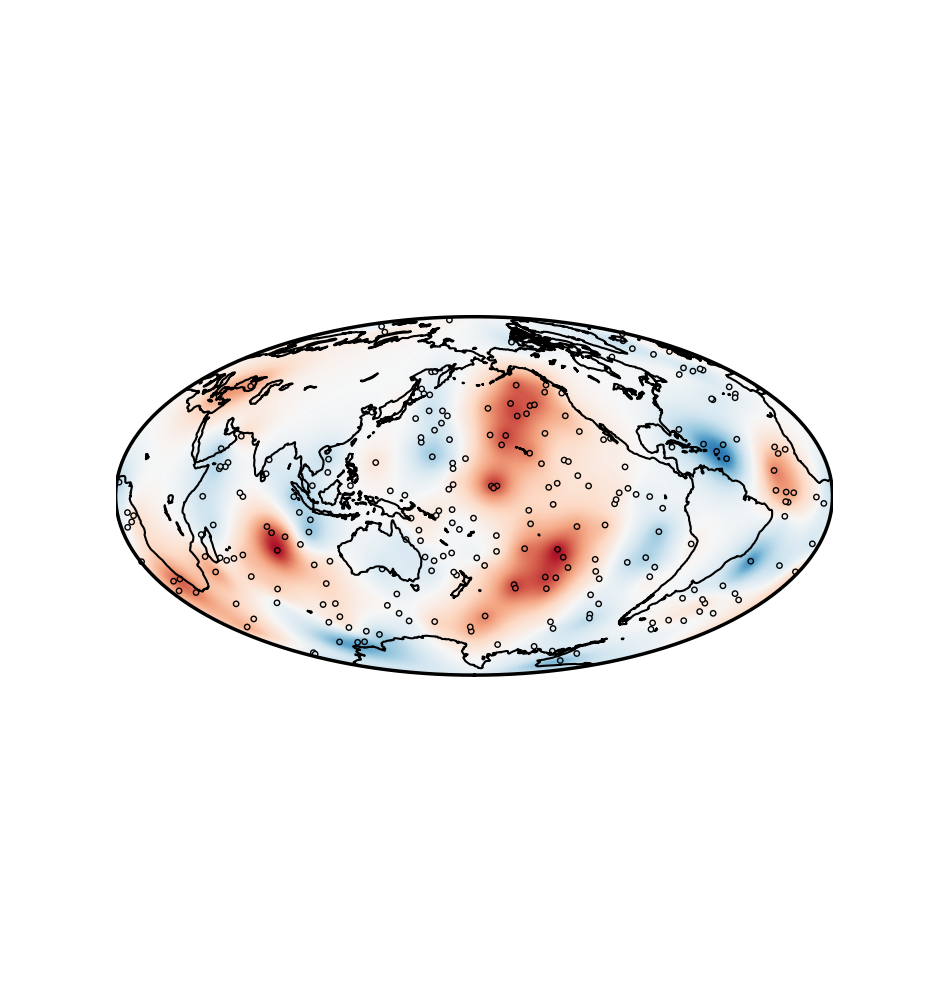}
        \includegraphics[width=0.49\textwidth,trim={1cm 7.25cm 1cm 7.25cm},clip]{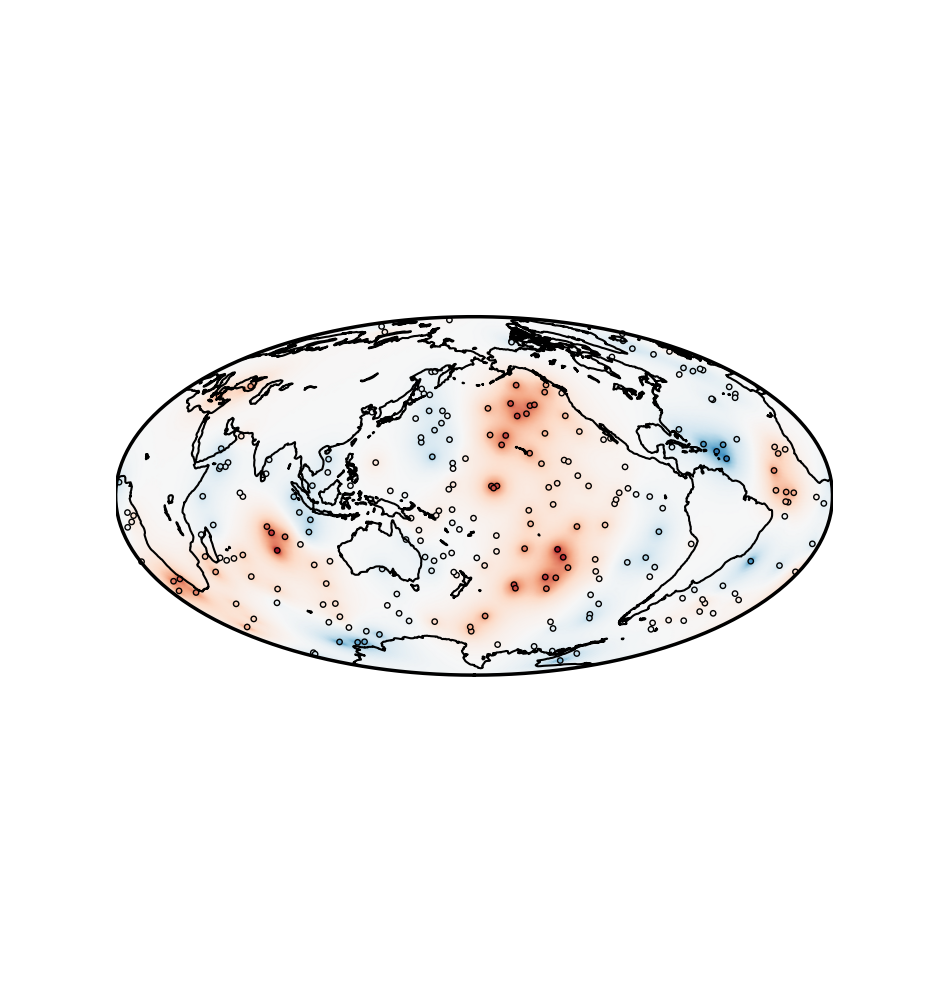}
  \end{minipage}
  \caption{
    Minimum norm solutions obtained from the data in Fig.\ref{fig:data_ne} using different
    values for the Sobolev parameters. For the top left model the values used
    were $s = 2.0$ and $\lambda = 0.5$. For the top right model the values used
    were $s = 1.1$ and $\lambda = 0.5$. For the bottom left model the values used
    were $s = 2.0$ and $\lambda = 0.1$. For the bottom right model the values used
    were $s = 1.1$ and $\lambda = 0.1$.  As a general trend, as either $s$ or $\lambda$
    decrease the models become more localised about the observation points. 
 }  
  \label{fig:minnorm2}
\end{figure}

As discussed in Appendix \ref{ssec:sobnum1}, within numerical calculations we make
use of truncated spherical harmonic expansions, and
hence replace the  data mapping $A$ in eq.(\ref{eq:dmap}) with the following approximation
\begin{equation}
  A_{L}u = \sum_{i=1}^{n}  \cbraket{\hat{\delta}_{x_{i},L}}{u}_{H^{s}(\mathbb{S}^{2})}\,f_{i},
\end{equation}
where $ \hat{\delta}_{x,L}$ is the truncated $H^{s}(\mathbb{S}^{2})$-representation
of a Dirac measure at degree $L$. Using  eq.(\ref{eq:delanorm}) it is readily shown that
\begin{equation}
  \label{eq:aer}
  \|A-A_{L}\|_{\Hom(H^{s}(\mathbb{S}^{2}),\mathbb{R}^{n})} \le n
  \left(\sum_{l=L+1}^{\infty}\frac{2l+1}{4\pi} \mult{l}{\lambda}^{-2s}\right)^{\frac{1}{2}}, 
\end{equation}
where $\|\cdot\|_{\Hom(H^{s}(\mathbb{S}^{2}),\mathbb{R}^{n})}$ denotes the operator-norm
defined in eq.(\ref{eq:op_norm}).
By taking the truncation degree sufficiently high, the term on the right hand side can be
made as small as we wish, and hence minimum norm solutions obtained for the discretised problem
 converge to the correct model in $H^{s}(\mathbb{S}^{2})$ as the truncation degree is increased.

As a first  example, we  show  why it
is necessary to formulate  inference problems using an appropriate function space.  We noted above
that $L^{2}(\mathbb{S}^{2})$ is not suitable for the problem at hand  because elements
of this space do not have well-defined point values.  Once, however,  things have been
approximated  using truncated spherical harmonic expansions, there is nothing to stop us  calculating  minimum norm solutions relative to the structure induced from $L^{2}(\mathbb{S}^{2})$.
Moreover, so long as the truncation degree is sufficiently high, this approach
yields models that fit the data to numerical precision. 
But as is seen in the left column of  Fig.\ref{fig:minnorm1},
as the truncation degree is increased a sequence of models is obtained that
does not converge in a pointwise sense.   In stark contrast, the right hand column in
Fig.\ref{fig:minnorm1}
shows the rapid convergence   obtained from the same data when minimum norm
solutions are sought within an appropriate choice of Sobolev space.
For reference, these solutions were obtained using the gradient-based minimisation approach
discussed above using the  L-BFGS algorithm \citep[e.g.][]{nocedal2006numerical}
coupled to the robust line search method of \cite{more1994line}. Moreover,
the action of data and property mappings and their adjoints have been implemented in
a matrix-free manner using fast spherical harmonic transformations. As a result,
the method can be readily applied to situations involving high truncation degrees and/or
large data sets.

Next we show in Fig.\ref{fig:minnorm2} four   minimum norm
solutions obtained from the data in Fig.\ref{fig:data_ne} using  different values of
the  Sobolev parameters $s$ and $\lambda$. In each case the truncation degree was taken sufficiently
high that convergence has been achieved to a relative error less than $10^{-8}$. The  differences
between these results emphasises  that  minimum norm solutions depend on the
inner product chosen for the model space, and hence none of these models has
any particular interest. In fact, it is not difficult to show  that \emph{any} model
that fits the data is the minimum norm solution relative to some compatible choice
of inner product for the model space. The calculation of
minimum norm solutions is simply a necessary step within the
implementation of our broader theory once a suitable  prior norm bound
has been selected.

\begin{figure}
  \centering  
  \includegraphics[width=\textwidth,trim={0 0.55cm 0 2.25cm},clip]{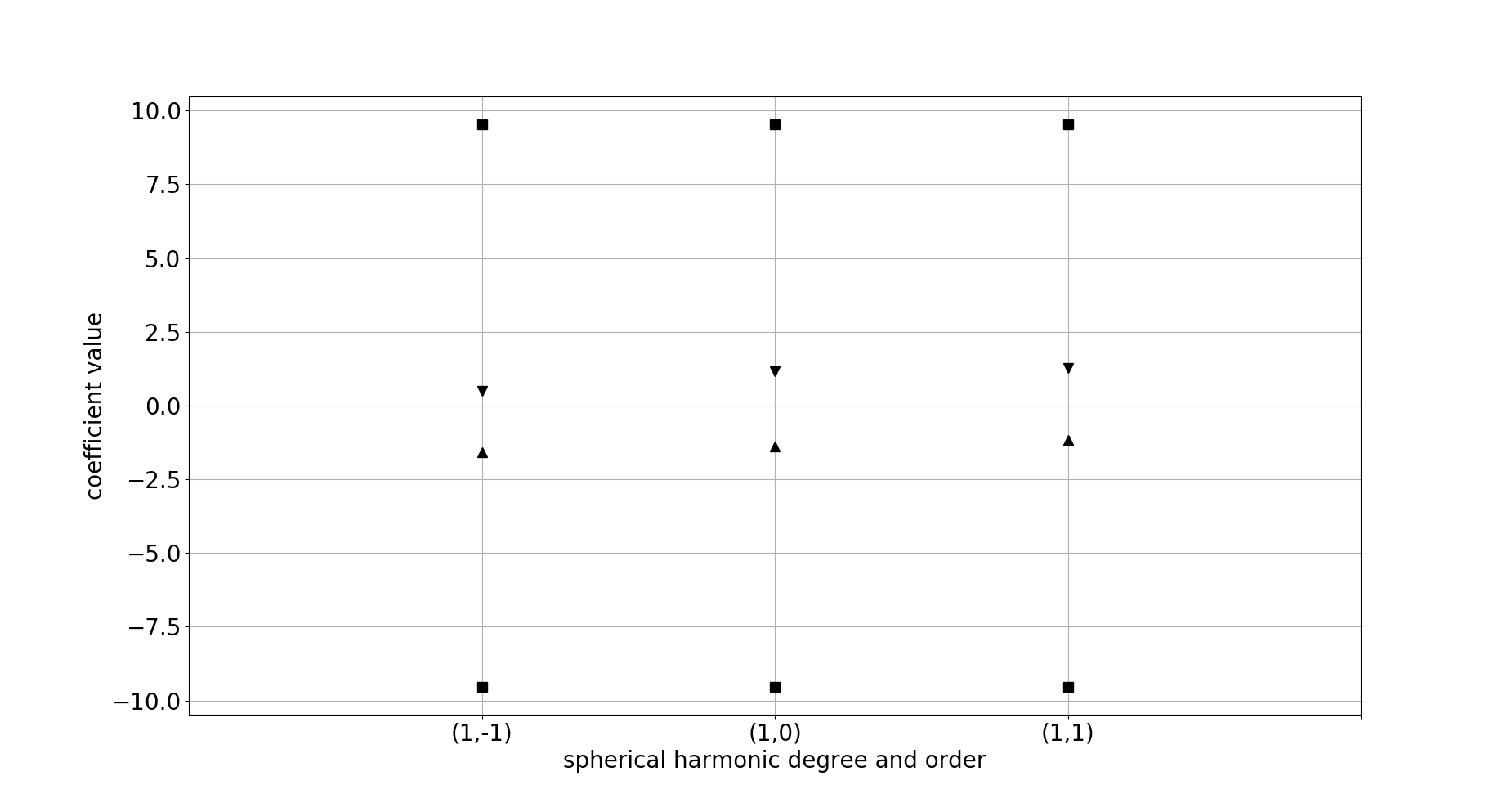}
  \caption{
    Limits on the degree one spherical harmonic coefficients obtained from the
    data shown in Fig.{\ref{fig:data_ne}} in conjunction with the prior
    norm bound $\|u\|_{H^{s}(\mathbb{S}^{2})} \le \frac{5}{4}\|\tilde{u}\|_{H^{s}(\mathbb{S}^{2})} $
    with $\tilde{u}$ the minimum norm solution, and Sobolev parameters $s = 2$ and $\lambda = 0.25$.
    For each coefficient the black squares
    indicate the limits placed by the prior norm bound, while the
    triangles show the smaller interval once the data is incorporated. 
 }  
  \label{fig:bene1d_1}
\end{figure}

\begin{figure}
  \centering  
  \includegraphics[width=\textwidth,trim={0 1.25cm 0 2.25cm},clip]{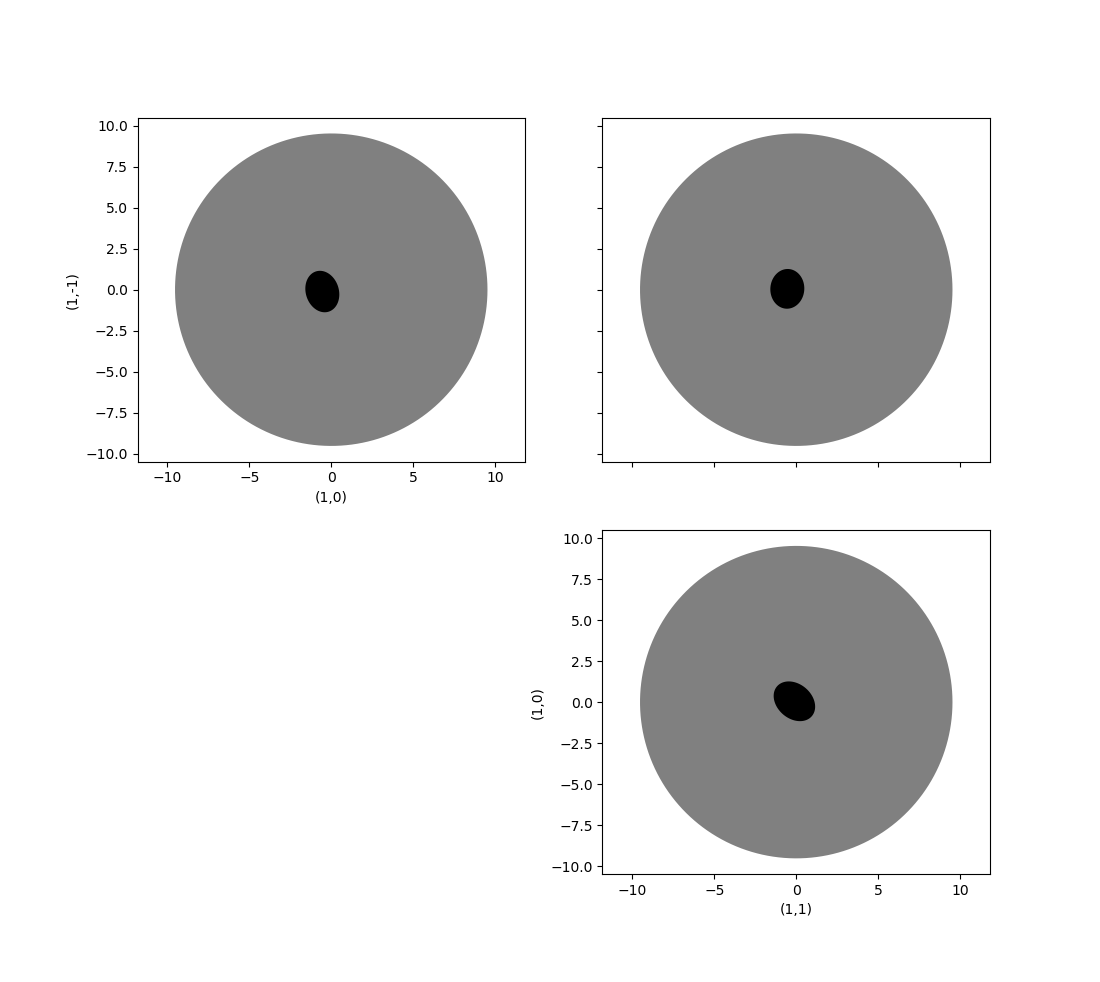}
  \caption{
    As for Fig.\ref{fig:bene1d_1}, but here showing two-dimensional slices
    through the property space that indicate the pair-wise trade offs
    between the different coefficients. In each plot the axes are labelled
    with the degree and order of the coefficients considered. The
    gray shaded region then shows the limits placed by the
    prior norm bound, while the smaller black region is that resulting
    from the incorporation of the data.
 }  
  \label{fig:bene2d}
\end{figure}

\begin{figure}
  \centering  
  \includegraphics[width=\textwidth,trim={0 0.55cm 0 2.25cm},clip]{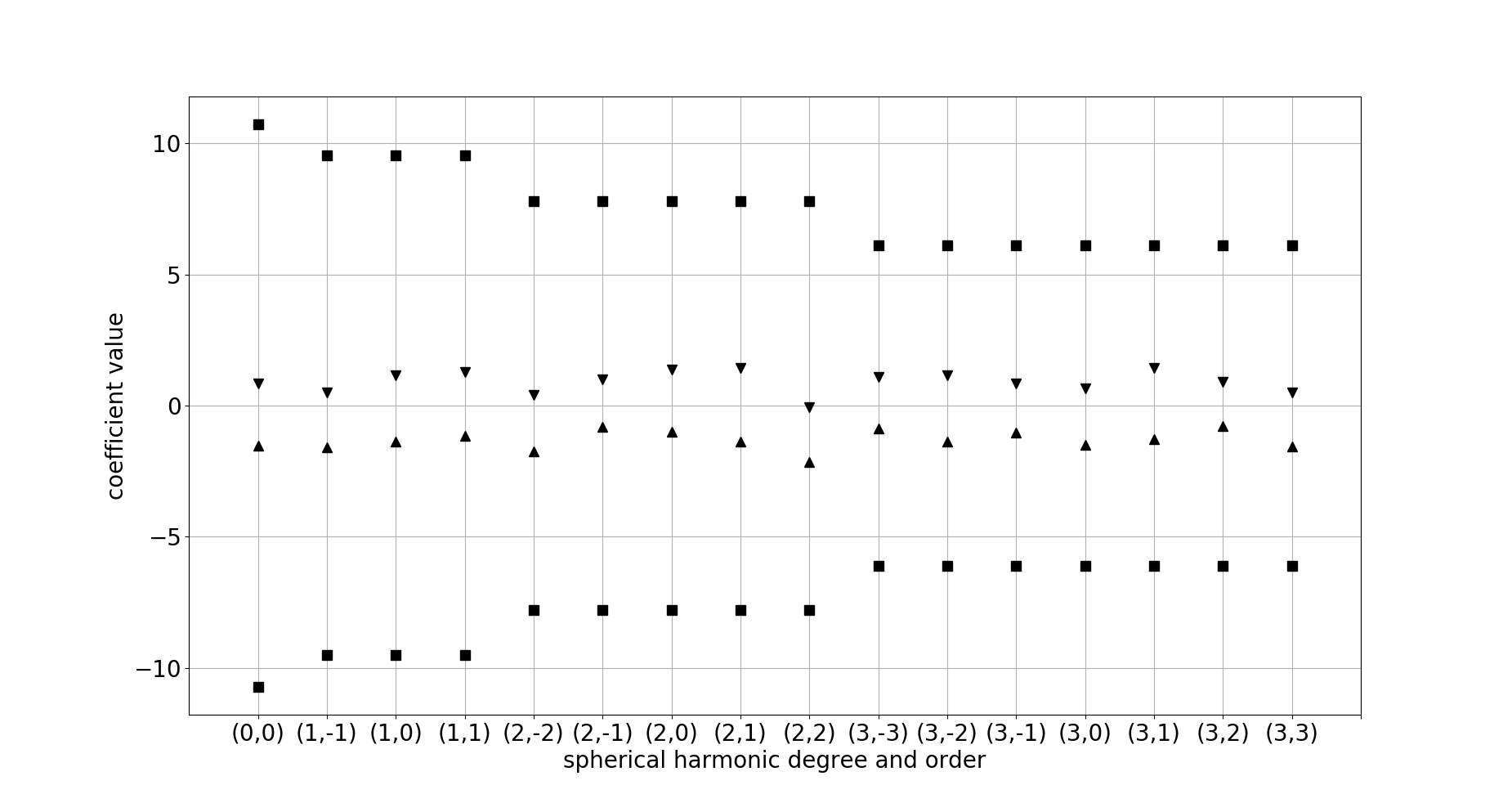}
  \caption{
    As for Fig.\ref{fig:bene1d_1}, but now showing all spherical harmonic coefficients having
    degree $l \le 3$.
 }  
  \label{fig:bene1d_2}
\end{figure}

As our final example for this section we consider the implementation of
eq.(\ref{eq:pbound1}). Again, we
use the data shown in Fig.\ref{fig:data_ne}, and  choose to
work in the Sobolev space with $s = 2$ and $\lambda = 0.25$. A prior
norm bound  is required, and  to insure compatibility with the data we take
the radius of the constraint set to be
\begin{equation}
  r= \frac{5}{4}\|\tilde{u}\|_{H^{s}(\mathbb{S}^{2})}, 
\end{equation}
where $\tilde{u}$ is the minimum norm solution. In a real
application such a norm bound would, of course, need to be carefully justified. As the
property space we consider all degree $l = 1$ spherical harmonic coefficients, and
hence $\dim G = 3$. The main cost of implementing eq.(\ref{eq:pbound1}) is  the calculation of
$3 + 1$ minimum norm solutions. We
show in Fig.\ref{fig:bene1d_1} the limits placed the individual coefficient
obtained using  eq.(\ref{eq:1dminmax}). The plot also shows
permissible values using the prior constraint alone, and
it can be seen that the data has substantially improved our knowledge of
these coefficients. This situation contrasts markedly with the
the behaviour of the problem when posed in $C^{0}(\mathbb{S}^{2})$, and substantiates our hope that the
incorporation of derivative information into the model space topology would  be useful.
 A limitation of Fig.\ref{fig:bene1d_1}
is that trade-offs between the uncertainties in the coefficients
cannot be seen. As a step towards doing this, we show in Fig.\ref{fig:bene2d} the pair-wise trade-offs
between the different coefficients obtained using eq.(\ref{eq:1dminmax}).
We finally show in Fig.\ref{fig:bene1d_2} the result
obtained when the property space is expanded to comprise all coefficients of degree
less than or equal to three. In this manner we see that relatively
large property spaces can be handled, though the visualisation of the
trade-offs between the coefficients becomes more challenging as $\dim G$ increases.

\subsection{Backus-Gilbert estimators}

\label{ssec:bgest}

Prior to Backus' independent work on inference problems a related but different approach was
described by \cite{backus1967numerical,backusgilbert,backus1970uniqueness},
with these ideas later being usefully adapted within the SOLA method of \cite{SOLA1,SOLA2}.
The extension of Backus-Gilbert estimators to Banach spaces has been discussed by
 \cite{stark2008generalizing}, and here a broadly similar approach is used. 
For  arbitrary $C \in \Hom(F,G)$ we can map the data $v = Au$
into the property vector $Cv = CA u$; in words, an estimate  of the property vector
is  sought by forming appropriate linear combinations of the data. From Theorem \ref{thm:bbackus1}
 there is no  $C$ such that $CA = B$, and hence the property
vector cannot be recovered exactly.   Nevertheless,  if we can find $C$ such that
$CA$ approximates $B$  suitably, then $Cv$ might  still provide a useful
estimate of the property vector.
To proceed, for given $C \in \Hom(F,G)$ we set $\tilde{w} = Cv$, and write
\begin{equation}
  \label{eq:bgexp}
  w = \tilde{w} + H u, 
\end{equation}
where $w$ is the true value of the property vector and  $H = B - CA$.
Noting that $H' = B'- A'C'$, the transversality assumption implies
that $\ker H'$ is trivial, and hence
$H$ is surjective for any choice of $C$. It follows
that the property vector can take any value in $G$
if no prior constraints are placed on the model.  Indeed,
this is just another way of proving Theorem \ref{thm:bbackus2}.
Assuming, therefore, that $u \in U$ for a given constraint
set, the property vector satisfies
\begin{equation}
  w \in \tilde{w} + H U = \{\tilde{w} + H u \,|\, u \in U\}.
\end{equation}
The validity of this result depends, of course, on  $v = Au$
having solutions in $U$. Supposing that this holds, 
$ \tilde{w} + H U$ and $BU$   have  non-empty intersection, and hence the
property vector is contained in
\begin{equation}
  \label{eq:bebound1}
   (\tilde{w} + H U)\cap BU, 
\end{equation}
which is bounded if  the same
is true of  $\pi_{B}U$. Different
choices of  $C \in \Hom(F,G)$  lead to different  estimates
and, having quantified what constitutes
a ``good subset'', we could seek an optimal value for this linear mapping.

To proceed we  assume for simplicity that
$E$ is Hilbertable and that the prior constraint takes the form
$u \in B_{r}(0)$ for some compatible choice of inner product. 
The image of the closed ball $ B_{r}(0)$ under the  affine mapping
$u \mapsto \tilde{w} + Hu$ is a closed set whose boundary is a hyperellipsoid. Indeed, for a point
$w$ in this set we have $w-\tilde{w} = Hu$ for some $u \in E$, 
and hence, using Proposition \ref{prop:minnorm}, we obtain
\begin{equation}
  u = H^{*}(HH^{*})^{-1}(w-\tilde{w}) + u_{0},
\end{equation}
with $u_{0} \in \ker H$. But as  $u \in B_{r}(0)$ it follows that
\begin{equation}
  \cbraket{(HH^{*})^{-1}(w-\tilde{w})}{w-\tilde{w}}_{G} \le r^{2}.
\end{equation}
We would like this  subset to  be as small
as possible, but what is meant by 
small in this context  must be decided. When $\dim G = 1$, the
subset degenerates to an interval, and so we  should clearly
 minimise its length. 
A simple extension of this idea  is to consider the arithmetic 
average of the squared-lengths of the  hyperellipsoid's principle axes. We therefore seek 
$C \in \Hom(F,G)$ such that
\begin{equation}
  \label{eq:bgopt}
  J(C) = \tr[(B-CA)(B-CA)^{*}], 
\end{equation}
is minimised. Differentiating and setting the result equal to zero,
the optimal linear estimator is readily found to be
\begin{equation} 
 \label{eq:bgoptr}
  C = B A^{*}(AA^{*})^{-1}.
\end{equation}
Recalling Proposition \ref{prop:minnorm}, $\tilde{w} = C v$
is exactly what would be obtained from the minimum
norm solution of  $v = Au$. Moreover,
we have
\begin{equation}
  H = B-CA = B[1- A^{*}(AA^{*})^{-1}A] =
  B\, \mathbb{P}_{\ker A}, 
\end{equation}
which, using  eq.(\ref{eq:bra}), implies that
\begin{equation}
  HH^{*} = B\,\mathbb{P}_{\ker A} B^{*} = B|_{\ker A}B|_{\ker A}^{*}.
\end{equation}
The optimal Backus-Gilbert estimator, therefore, leads to the
following restriction on the property vector
\begin{equation}
  \label{eq:bgbound}
  \cbraket{(B|_{\ker A}B|_{\ker A}^{*})^{-1}(w-\tilde{w})}
  {w-\tilde{w}}_{G}  \le r^{2}, 
\end{equation}
which is  identical to eq.(\ref{eq:pbound1}) but for
the lack of a term on the right hand side associated with the minimum
norm value. This absence makes perfect sense  because the optimality condition
in eq.(\ref{eq:bgopt}) is defined without reference to the  data.  As a consequence Backus-Gilbert
estimators will generically overestimate uncertainty  relative to
applications of Backus' later theory.

\section{Linear inference problems with  data errors}

\label{sec:error}

In this section we show how our previous results  can be extended
to account for random data errors. While most   ideas apply
to problems with Banachable model spaces, the practical implementation
of the method  again only seems feasible if an
appropriate Hilbert space structure can be introduced. It is worth
emphasising that our discussion  is not limited to the case of
Gaussian errors, with a wide range of unimodal
distributions being accommodated at little to no additional cost.
Broadly similar methods can be applied in the context of Backus-Gilbert
estimators, and these are discussed briefly at the end of the section.

\subsection{Formulation of the problem}

The vector spaces and operators introduced in Section \ref{ssec:blif}
carry over directly. The surjectivity condition on the data mapping will,
however, be dropped to allow for data-redundancy. Instead we require the following:
\begin{asm}
  \label{asm:err}
  The property mapping is surjective, while the transversality condition
  $\image A'\cap \image B' = \{0\}$  holds.
\end{asm}
To account for data errors, we generalise the
relationship between the unknown model $u \in E$ and the observed data
$v \in F$  to 
\begin{equation}
  \label{eq:dmer}
  v = Au + z, 
\end{equation}
where $z$  is a realisation of an $F$-valued random variable.
The probability distribution from which the error term  is drawn will be denoted
by $\nu$ and is assumed to be known exactly. 
The data is, therefore, a realisation of a random variable  
whose probability distribution is  determined by $\nu$ along with the
unknown value $Au$. As ever, the inference problem aims to use the
data to constrain the value of the property vector $w = Bu$ subject to
the model satisfying  $u \in U$ for a given constraint set $U \subseteq E$.

The data $v \in F$ does not now  tell us  $Au$ directly,
and we must  use  our knowledge of $\nu$ to  determine
which values for $Au$ are plausible. We seek an approach that
would be  infrequently wrong under hypothetical repetitions, taking this
behaviour as characteristic of a sound statistical procedure  -- see \cite{mayo2018statistical}
for an interesting and  nuanced discussion of such issues. To this end, we define  a
\emph{confidence set} for $\nu$
having a  \emph{confidence level},  $1-\alpha$, as a  subset $V\subseteq F$  such that
\begin{equation}
  \nu(V) = 1-\alpha, 
\end{equation}
with $\alpha \in [0,1]$.  The interpretation 
is simply that if realisations are repeatedly drawn from $\nu$,  the results will lie
in $V$ with a relative frequency that tends to the confidence level.
For a given distribution there will generally be
many different confidence sets having the same confidence level, and
additional criteria must be invoked to single out one of practical interest.
Nonetheless, having fixed an appropriate choice of $V\subseteq F$,  for each
realisation $v$ of the data
we know that $z = v - Au$
is a realisation of the random error, while the condition $ v - Au \in V$ is equivalent
to $Au \in  v - V$. It follows that whatever the true value of $u$, if, hypothetically,
the data generated from this model could be observed many times,
then the relative frequency at which $Au \in v - V$ holds  would tend to $1-\alpha$.  From
the  data $v \in F$ we, therefore, choose to  infer that  $Au \in v - V$.
This   may, of course, be incorrect in any given instance, but such is the
nature of statistics.

An appealing feature of this approach is that deterministic errors in the
data mapping can, in principle, be incorporated with relative ease. Suppose
that $a:E\rightarrow V$ denotes the exact (and possibly non-linear) data mapping,
and hence  eq.(\ref{eq:dmer}) should be replaced by
\begin{equation}
  v = a(u) + z. 
\end{equation}
If $A \in \Hom(E,F)$ is the linear and approximate data mapping to be used, then we can write
\begin{equation}
  v = Au + [a(u)-Au] + z.
\end{equation}
Let  us assume there is a bounded subset $V_{d} \subseteq F$ such that
\begin{equation}
  a(u)-Au \in V_{d}, 
\end{equation}
for all models $u$ within the constraint set $U\subseteq E$.
Adapting the above argument, it follows that from the observed data $v \in F$ we can
conclude that $Au \in v -V'$ where $V' =  V+V_{d}$  may be viewed as a modified
confidence set. Here, of course, we see  the familiar idea that theoretical
uncertainties can be pragmatically   addressed  by  increasing
the magnitude of the    random data errors.

\subsection{Generalising Backus' theorems}

\label{ssec:gback}

Our aim is to determine the property vector $w = Bu$ as best possible from the information
$u \in U$ and  $Au \in v - V$, 
where $U \subseteq E$ a constraint set for the model, and $V\subseteq F$ a confidence
set for the data error. Given that the data mapping need not  be
surjective, the observed data $v \in F$  may not belong to $\image A$.
For example, suppose that the distribution $\nu$ has
a  probability density function  $p:V\rightarrow \mathbb{R}$
relative to a volume measure $\ddns v$ on $F$ such that
\begin{equation}
  \label{eq:pdf}
  \nu(V) = \int_{V}p(v) \dd v
\end{equation}
for any  subset $V\subseteq F$. If $\image A$
is a proper subspace of $F$ it has zero-volume, and so
\begin{equation}
  \nu(\image A) = 0, 
\end{equation}
which is to say that the observed data will almost surely 
lie outside of the image of the data mapping. It is worth noting that
such a probability density function need not always exist. For example,
the error free-case can be recovered by taking $\nu$ to be
a Dirac measure at the origin. In any case, what really matters is  that
\begin{equation}
  (v-V)\cap \image A \neq \emptyset, 
\end{equation}
this meaning that there exist models that fit the observed data in a statistically
acceptable manner.

To allow for $A$ to be non-surjective, the factorisation of the data mapping in eq.(\ref{eq:afac})
is readily generalised to
\begin{equation}
  \label{eq:afac2}
  A = i_{\image A}\, \hat{A} \,\pi_{A}, 
\end{equation}
where $\pi_{A}$ is the quotient mapping onto $E/\ker A$,
$\hat{A} \in \Hom(E/\ker A,\image A)$ is continuously invertible,
and $i_{\image A} \in \Hom(\image A,F)$ is the inclusion mapping
\citep[e.g.][Proposition 4.6]{treves}.
 For each point $\tilde{v} \in (v-V)\cap \image A $
we can associate a closed affine subspace $\pi_{A}^{-1}\{\hat{A}^{-1}\tilde{v}\}$
within the model space.  Making use of Assumption \ref{asm:err}, the
proof of Theorem \ref{thm:bbackus2} is then readily generalised
to show that for any $\tilde{v} \in (v-V)\cap \image A $ and  $w \in G$ there are
infinitely many models such that $Au = \tilde{v}$ and $Bu = w$.  Thus,
as must be  expected, the incorporation of data errors does nothing to lessen the need
for prior constraints within the inference problem.  Using eq.(\ref{eq:afac2}) we see that
the subset
\begin{equation}
U_{v-V} =   \pi_{A}^{-1} \hat{A}^{-1}[(v-V)\cap \image A], 
\end{equation}
contains all models that fit the data in a statistical sense.
The data is, therefore, compatible with the prior constraint
$u \in U$ if and only if $U \cap U_{v-V}$ is non-empty, this
result generalising the first part of Theorem \ref{thm:bbackus3}.
To extend the second part of that theorem, we  use eq.(\ref{eq:bfac})
to conclude that  $B (U\cap U_{v-V})$ is bounded if and only if the same is
true of  $\pi_{B} (U\cap U_{v-V})$. For later reference, we note that
if the constraint set $U$ and the confidence set $V$ are convex,
then the same is true of  $B (U\cap U_{v-V})$,
this following because the intersection of two convex sets is convex,
while convexity is preserved under images and inverse images of linear mappings.
Finally, the discussion in Section \ref{ssec:norm_bound_res}  on the relationship
between prior constraints
and the  choice of model space  carries over unchanged.

\begin{figure}
  \centering
  \includegraphics[width=0.9\textwidth,trim={1cm 0.0cm 1cm 1.0cm},clip]{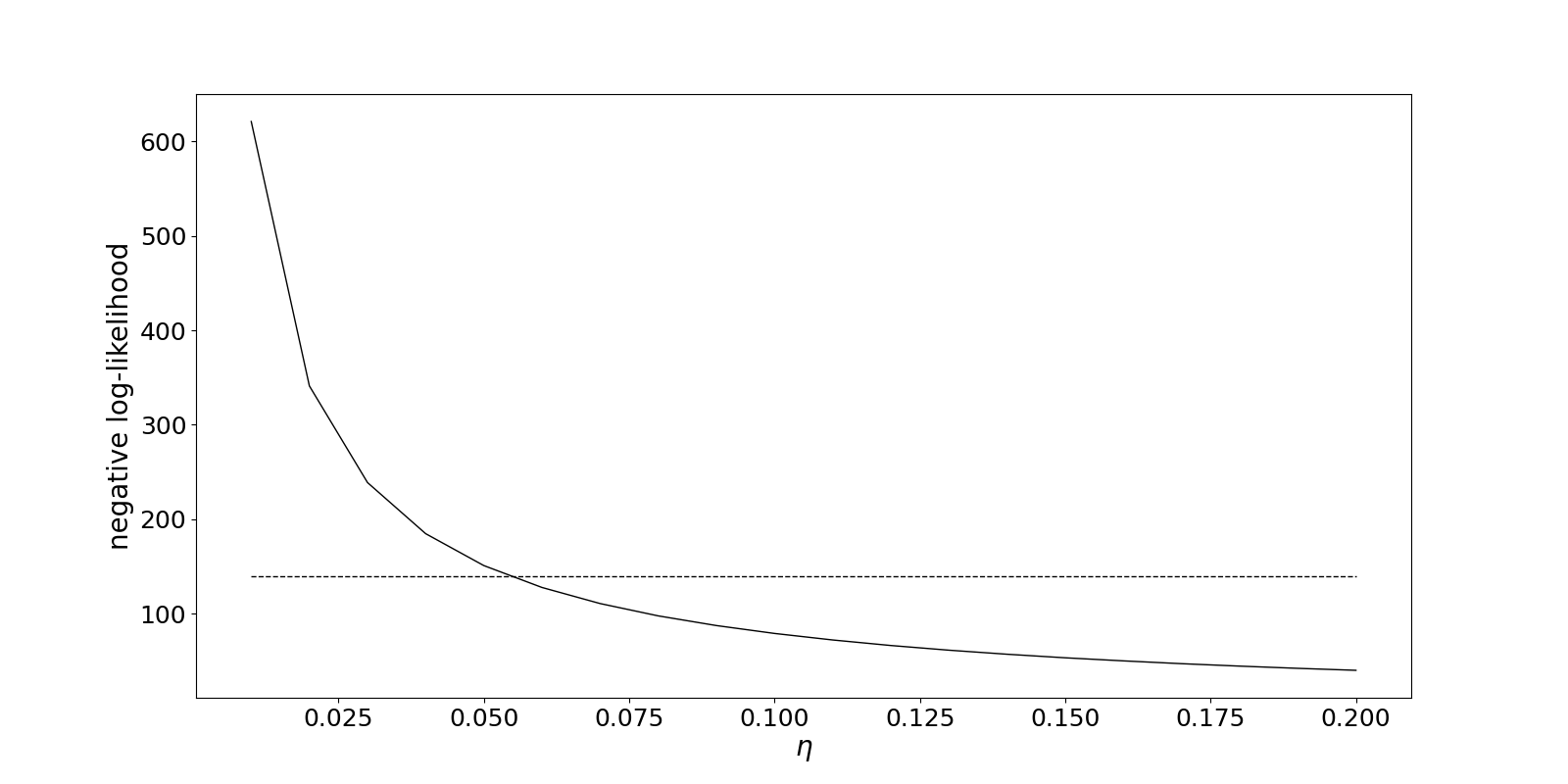}
  \caption{A graphical illustration of how eq.(\ref{eq:lag2}) can be  solved. For each value of
    $\eta > 0$ we  solve a convex optimisation problem associated with the function  $u \mapsto J(u,\eta)$
    defined in eq.(\ref{eq:lag1}), 
    and the plot shows the negative log-likelihood for the models so obtained. The dashed line indicates
    the squared-radius of the desired confidence set, and  the intersection
    of these curves   gives the value of $\eta$ corresponding to the model with smallest norm that
    fits the data in a statistically acceptable fashion. These calculations 
    were done for the data set shown in Fig.\ref{fig:data_ne} but with uncorrelated
    Gaussian errors added to each datum as described in Section \ref{ssec:hilap}.
    The confidence set is that defined in eq.(\ref{eq:lscon}) with confidence level $0.9$,
    while Sobolev parameters $s = 2$ and $\lambda = 0.25$ were selected.
  }
  \label{fig:NLL}
\end{figure}

\subsection{Implementation in  Hilbert spaces}

\label{ssec:hiler}

\subsubsection{General theory}

We assume that $E$ is Hilbertable and that the constraint set takes the form
$u \in B_{r}(0)$ relative to a compatible choice of inner product. Balls not
centred at the origin are readily handled by translation. Using
an orthonormal basis, we can establish an isomorphism between $F$
and $\mathbb{R}^{n}$. On the latter space we have the usual
Lebesgue measure, and this can be trivially pulled-back
to a translation-invariant volume measure $\ddns v $ on $F$ which
is independent of the choice of orthonormal basis.
To define a suitable
confidence set, let us suppose that $\nu$ can be expressed in terms
of a probability density function $p:F\rightarrow \mathbb{R}$
relative to $\ddns v$ which  takes the form
\begin{equation}
  p(v) = a \,\ee^{-l(v)}, 
\end{equation}
where $a$ is a normalisation constant  and $l:F\rightarrow \mathbb{R}$ is known as the
\emph{negative log-likelihood}.
We assume that $l$ is non-negative,  continuously differentiable and 
 strictly convex. One example is the ubiquitous Gaussian distribution which has
\begin{equation}
  l(v) = \frac{1}{2}\cbraket{R^{-1}(v-\bar{v})}{v-\bar{v}}_{F},
\end{equation}
where the covariance $R \in \Hom(V)$ is  self-adjoint and positive-definite, and $\bar{v}$ is the
expectation. For such distributions, a sensible and convenient choice of  confidence set is
defined in terms of the negative log-likelihood by
\begin{equation}
  \label{eq:lscon}
  V = \{ v \in F \,|\, l(v) \le s^{2}\},
\end{equation}
where $s>0$ is fixed uniquely by the requirement that the
confidence level $1-\alpha$ be met.
Because $l$ is continuous and strictly
convex, it follows  that $V$ is a closed and convex set.  The boundary
of the confidence set is, by definition, equal to the inverse image
\begin{equation}
  \label{eq:csbnd}
  \partial V = l^{-1}(\{s^{2}\}). 
\end{equation}
As a final assumption, we require that the derivative $Dl:
V\rightarrow V'$
is nowhere vanishing on $\partial V$ and hence, by the regular value theorem
\cite[e.g][]{spivak1970comprehensive}, this boundary is a closed
submanifold in $F$ with a continuous outward unit normal vector field.
It is worth emphasising that these  conditions on $p$  hold for a  wide range of  unimodal distributions,
and not only for the Gaussian distribution.

\begin{figure}
  \centering  
  \includegraphics[width=\textwidth,trim={0 0.55cm 0 2.25cm},clip]{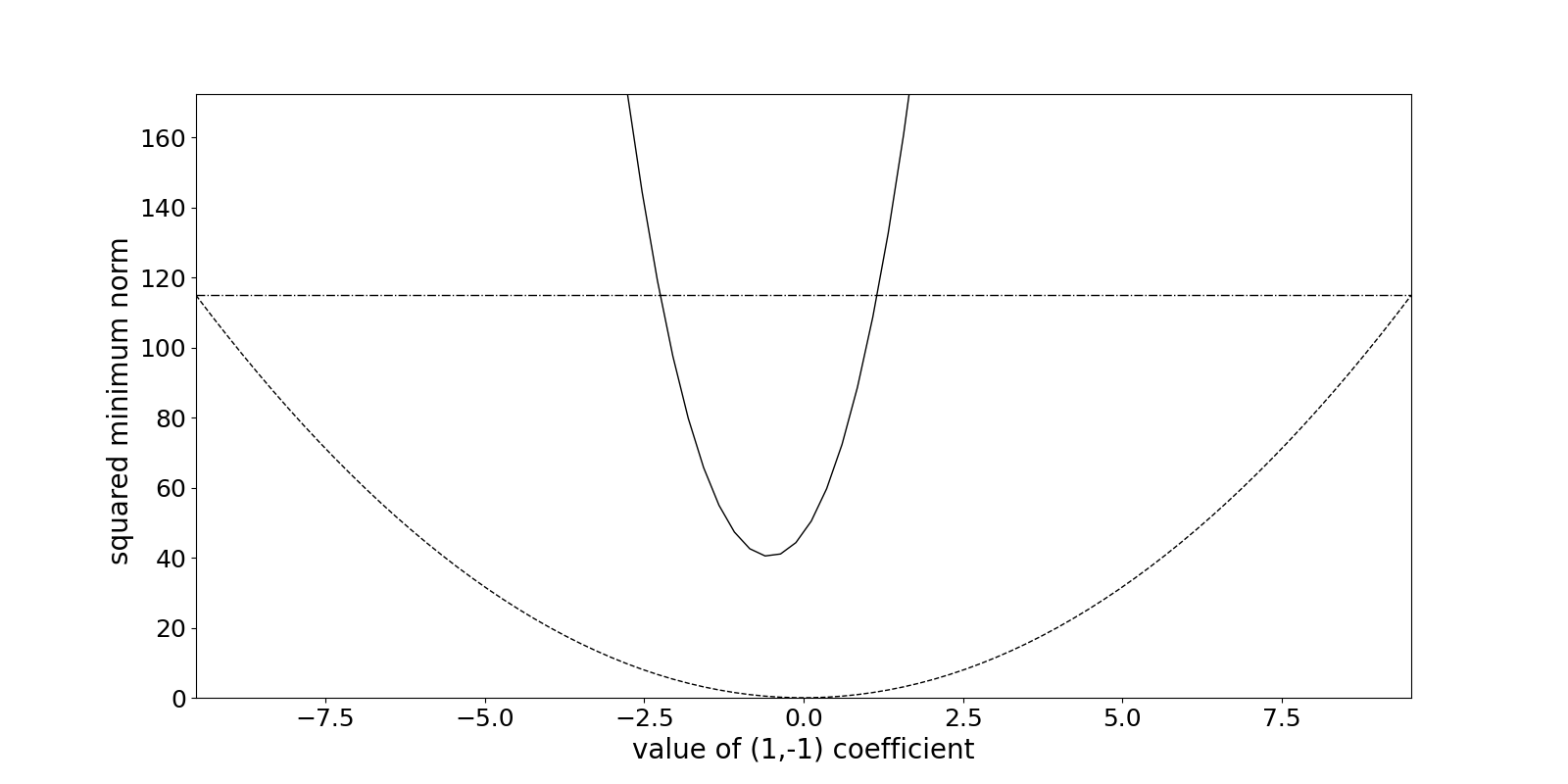}
  \caption{
    An illustration of how bounds are placed on a component of the property vector,
    with the data, confidence set, and model space parameters as described in Fig. \ref{fig:NLL}.
    The horizontal dot-dashed line shows the squared-value of the prior norm bound,
    this  being equal to that used in the error-free case shown
    in Figs. \ref{fig:bene1d_1},  \ref{fig:bene2d}, and \ref{fig:bene1d_2}.
    The dashed curve then shows the squared minimum norm value for a model
    whose $(1,-1)$ spherical harmonic coefficient is
    given by the ordinate, while the solid curve shows the corresponding
    value when the model is also required to fit the data in a statistically
    acceptable sense. The intersections between dashed and dot-dashed curves
    gives the interval in which the prior constraint requires the $(1,-1)$
    coefficient to lie, while the smaller interval defined by the intersections
    of the solid and dot-dashed curves shows what happens once the data is
    used.
 }  
  \label{fig:becmn}
\end{figure}

Given these preliminaries, we first consider how the compatibility of the
prior constraint with the data can be established. To this end we
define $U_{v-V} = A^{-1}(v-V)$ and wish to determine whether its intersection
with the constraint set $B_{r}(0)$ is non-empty.  A necessary and sufficient condition for this
to hold is  given by  
\begin{equation}
  \label{eq:dccompe}
  \inf_{u \in U_{v-V}}\|u\|_{E} \le r. 
\end{equation}
Note that, by definition, the infimum of an empty set of real numbers is equal to positive infinity,
and hence this condition is meaningful even if $U_{v-V}$  is empty. Supposing for the
moment that $U_{v-V}$ is non-empty, we know that this set is closed and  convex.
The general form of the projection theorem in Hilbert spaces \citep[e.g.][Theorem 12.1]{treves}
shows that there is a unique point in $U_{v-V}$ such that the infimum value
of the norm is attained. If $0 \in U_{v-V}$ then the infimum vanishes, and
eq.(\ref{eq:dccompe}) is trivially satisfied. Otherwise it is clear
geometrically that the unique minimum lies somewhere on the boundary $\partial U_{v-V}$.
In terms of the negative log-likelihood we have
\begin{equation}
  U_{v-V} = \{ u \in E \,|\, l(v-Au) \le s^{2}\}, 
\end{equation}
and noting that $u \mapsto l(v-Au) $ is convex, it follows that
\begin{equation}
  \partial U_{v-V} = \{ u \in E \,|\, l(v-Au) = s^{2}\}.
\end{equation}
Applying the Lagrange multiplier theorem \citep[e.g.][]{luenberger1997optimization},
we are led to introduce the functional
\begin{equation}
  \label{eq:lag1}
  J(u,\eta) = \frac{1}{2}\|u\|_{E}^{2} + \eta \left[l(v-Au) -s^{2}\right],
\end{equation}
such that desired minimum can be obtained by solving the simultaneous equations
\begin{eqnarray}
  D_{u}J(u,\eta) = 0, \quad  D_{\eta}J(u,\eta)  = 0,
\end{eqnarray}
which take the concrete form
\begin{eqnarray}
  \label{eq:lag2}
 u - \eta\, A^{*}\mathscr{J}_{F} Dl(v-Au) = 0, \quad l(v-Au) - s^{2} = 0,
\end{eqnarray}
where we recall that $\mathscr{J}_{F} \in\Hom(F',F)$ is defined through the
Riesz representation theorem. If $u$ is the unique minimum point, then the first
equation can be interpreted geometrically as saying that from
this point on $\partial U_{v-V}$ we can reach the origin by
moving a signed distance $\eta$ along the boundary's outward
unit normal. By assumption $U_{v-V}$ is disjoint from the origin,
and hence $\eta$ must be positive. Moreover,
for each fixed $\eta > 0$ the first equation defines the
unique minimum point of the convex functional $u \mapsto J(u,\eta)$.
In this manner, the above set of non-linear 
equations can be reduced to root-finding in a single positive real variable. Indeed,
for fixed $\eta > 0$, we can apply gradient-based methods to uniquely solve the convex
optimisation problem for $u$. The result may then be substituted into the second part
of eq.(\ref{eq:lag2})
to check if the constraint is met. In considering this root finding problem the following
 is useful:

\begin{lem}
  Let $u_{\eta}$ denote the unique minimum point of $u\mapsto J(u,\eta)$ for
  fixed $\eta > 0$.  The function
  \begin{equation}
     \eta \mapsto l(v-A u_{\eta}),
  \end{equation}
  is non-increasing for all $\eta > 0$.
\end{lem}    

\emph{Proof:} Differentiating  $\eta \mapsto l(v-A u_{\eta})$ we obtain
\begin{equation}
  \label{eq:deriv_tmp}
  \frac{\ddns}{\ddns \eta}l(v-Au_{\eta}) = -\cbraket{A^{*}\mathscr{J}_{F}\,Dl(v-A u_{\eta})}{u_{\eta}'}_{E}, 
\end{equation}
where $u'_{\eta} = \frac{\ddns u_{\eta}}{\ddns \eta}$. Next by implicitly differentiating
the first identity in eq.(\ref{eq:lag2}) we find 
\begin{equation}
  \left[1 + \eta A^{*}\mathscr{J}_{F} D^{2}l(v-A u _{\eta}) A\right] u'_{\eta} =
  A^{*}\mathscr{J}_{F} Dl(v-A u_{\eta}).
\end{equation}
Because the negative log-likelihood is strictly convex, its Hessian $D^{2}l(v) \in \Hom(F,F')$
is symmetric and positive-definite at each point. It follows that the linear operator
on the left-hand side is also positive-definite, and hence we obtain
\begin{equation}
   u'_{\eta} =
   \left[1 + \eta A^{*}\mathscr{J}_{F} D^{2}l(v-A u _{\eta}) A\right]^{-1} A^{*}
   \mathscr{J}_{F} Dl(v-A u_{\eta}) = \frac{1}{\eta}\left[1 + \eta A^{*}\mathscr{J}_{F} D^{2}l(v-A u _{\eta}) A\right]^{-1} u_{\eta}.
\end{equation}
Substituting into eq.(\ref{eq:deriv_tmp}) we then have
\begin{equation}
  \frac{\ddns}{\ddns \eta}l(v-Au_{\eta}) = -\frac{1}{\eta^{2}}\cbraket{u_{\eta}}
       {\left[1 + \eta A^{*}\mathscr{J}_{F} D^{2}l(v-A u _{\eta}) A\right]^{-1} u_{\eta}}_{E}, 
\end{equation}
which shows that the derivative is non-positive.
\eproof

\begin{figure}
  \centering  
  \includegraphics[width=\textwidth,trim={0 0.55cm 0 2.25cm},clip]{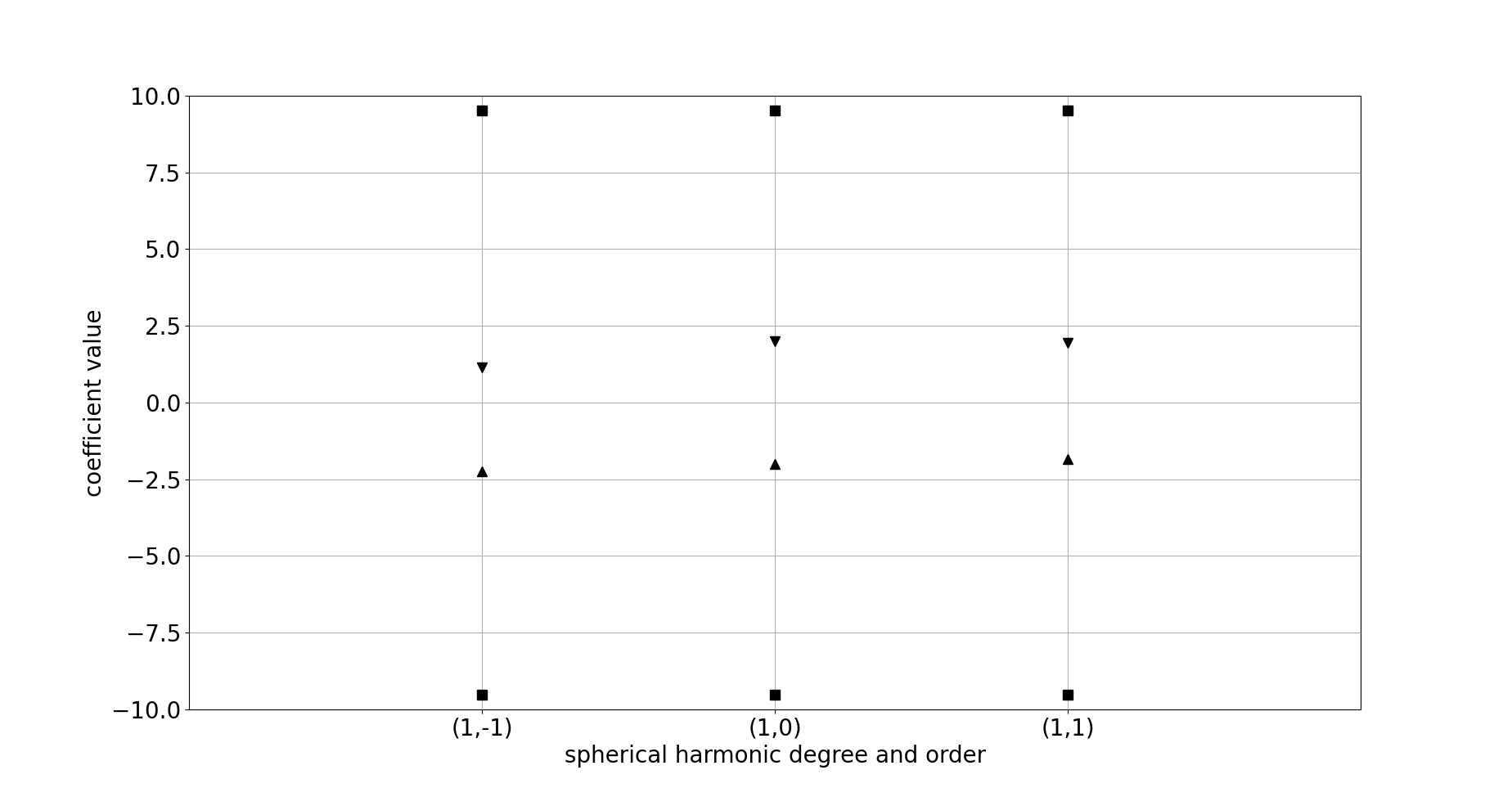}
  \caption{
    As for Fig.\ref{fig:bene1d_1}, but showing constraints on the degree one
    coefficients once data errors are considered. For these calculations
    the Sobolev space parameters and  prior norm bound were taken equal to
    those used in the error-free case, while a confidence level of $0.9$
    was selected.
 }  
  \label{fig:bewe1d}
\end{figure}

It follows that as $\eta$ tends to infinity the value of
$l(v-A u_{\eta})$ approaches a positive lower bound which might
be larger than $s^{2}$. In such a case there is  no value of $\eta > 0$
for which the second part of eq.(\ref{eq:lag2}) holds, this occurring precisely
when $\image A \cap (v-V)$ is empty. The above method also, therefore,
provides a constructive means for checking the validity of this assumption.
It is interesting to  note that the constrained optimisation problem
associated with eq.(\ref{eq:lag1}) is  identical in form to that within
the \emph{Occam's inversion}  of \cite{constable1987occam}. In the present context, however, the model 
 obtained  has  no intrinsic  value nor interest.

Suppose that, by the above method, we have verified that the
constraint and data are compatible in the stated statistical sense. We then
wish to know whether a given property vector $w \in G$
is consistent with both the constraint and data. For a fixed property vector $w \in G$,
Proposition \ref{prop:minnorm} shows that we must restrict attention to model vectors of the form
\begin{equation}
  u = \tilde{u} + u_{0}, 
\end{equation}
where $\tilde{u} = B^{*}(BB^{*})^{-1}w$ and $u_{0} \in \ker B$. The condition $u \in B_{r}(0)$ then implies
$u_{0} \in B_{r_{0}}(0)\subseteq \ker B$ where we have set
\begin{equation}
  r_{0}^{2} = r^{2}-\|\tilde{u}\|^{2}_{E},
\end{equation}
this radius being well-defined if and only if $w$ is compatible with the
prior constraint. Similarly,  the inclusion $Au \in v-V$ is equivalent
to $A|_{\ker B}u_{0} \in v_{0} - V$ where $v_{0} = v-A \tilde{u}$. This new problem
is of exactly the same form as the one  just solved. To work practically within the
subspace $\ker B$  all that must be done
is to replace each  occurrence of the adjoint operator $A^{*}$ by
\begin{equation}
  \label{eq:ara}
  A|_{\ker B}^{*} = \mathbb{P}_{\ker B} A^{*}, 
\end{equation}
with this identity being obtained by analogy with eq.(\ref{eq:bra}).

To conclude this discussion, we note that as both the constraint set and
confidence set are  convex, the same is true of the
subset of acceptable property vectors. While the above method shows how the
inclusion of a vector $w \in G$ within this subset can be established, 
it has not led to a simple method for delimiting the subset's boundary.
This contrasts with the error-free case in Section \ref{ssec:hilbert_sol}
where the boundary was shown to be a hyperellipsoid that could be readily calculated.
In fact, for each point $\tilde{v} \in \image A\cap(v-V)$, the method of Section
\ref{ssec:hilbert_sol} could be applied using the surjective
mapping $\hat{A}\,\pi_{A} \in \Hom(E,\image A)$ defined through
eq.(\ref{eq:afac2}), and would lead to  a subset in $G$ with hyperellipsoidal boundary
comprising \emph{some} of the  property vectors that are compatible with both the data
and prior constraint.
Taking the union of these subsets as $\tilde{v}$ ranges over $\image A\cap(v-V)$
we would arrive at the  subset containing \emph{all} acceptable property vectors. Looking back at
eq.(\ref{eq:pbound1}), we note that as $\tilde{v}$ varies
  the size and centre of the  hyperellipsoids  change, but
 not their shape nor orientation. Nevertheless, a union  of such subsets cannot, in general,
 be expected to   take any simple geometric form.

\subsubsection{Application to the spectral estimation problem}

\label{ssec:hilap}
To apply these ideas practically, we must first specify a
probability distribution $\nu$ for the random data errors. For simplicity
 $\nu$ was taken to be a Gaussian distribution with diagonal covariance matrix and
vanishing expectation.  The standard deviation at each measurement location was
drawn randomly from $[0.05,0.2]$ using a uniform distribution, with  these
errors lying between two and ten percent of the underlying functions maximum absolute value.
Having specified the distribution,  a confidence level of $0.9$ was selected,
and the radius $s$ of the confidence set determined according to eq.(\ref{eq:lscon}).
In the case of Gaussian errors, this can be done using the  chi-squared statistic.
More generally, random samples can be drawn from the distribution $\nu$ and used to
determine an empirical cumulative distribution function for the random variable $v \mapsto l(v)$ from which the appropriate value of $s$ is trivially found.

\begin{figure}
  \centering  
  \includegraphics[width=\textwidth,trim={0 1.25cm 0 2.25cm},clip]{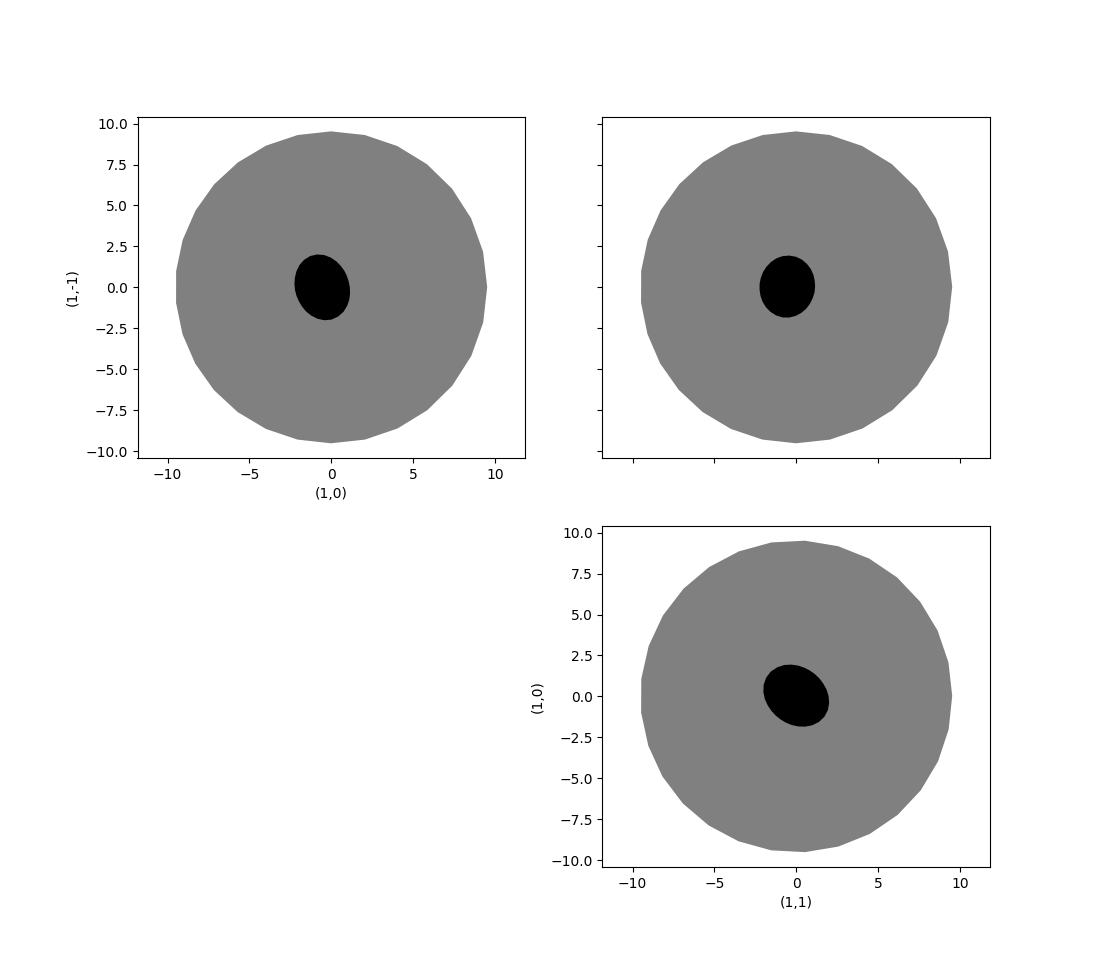}
  \caption{
        As for Fig.\ref{fig:bene2d}, but showing constraints on the degree one
    coefficients once data errors are considered. For these calculations
    the Sobolev space parameters and  prior norm bound were taken equal to
    those used in the error-free case,  while a confidence level of $0.9$
    was selected.
 }  
  \label{fig:bewe2d}
\end{figure}

In Fig.\ref{fig:NLL} we illustrate graphically  the root finding required to solve
the optimality condition in eq.(\ref{eq:lag2}). For each value of $\eta > 0$, the function
$u\mapsto J(u,\eta)$ can be minimised using a slight variant of the gradient-based methods
applied in the error-free case. Notably, in this case the optimisation problem has a unique
minimum, and hence the initial value within the iterative process does not matter.
In particular, once we have solved the problem for a certain value of $\eta$, we could
use this solution as the initial guess for a different but nearby $\eta$, and hence
improve the efficiency of the method. 
In doing all this we need to fix the Sobolev parameters, and in this
example the values  $s = 2$ and $\lambda = 0.25$ were selected.
The figure then shows the variation of the minimum value of the
negative log-likelihood as a function of $\eta$, and it can be seen that the second equality
in eq.(\ref{eq:lag2}) is satisfied for a suitable value of the argument.  In practice, of course,
this root is not found graphically, but obtained using a  bisection
algorithm.

As explained above, minimum norm solutions subject to given constraints on the
property vector can be obtained in a near identical manner.
For a given value of the property vector we can, therefore,
determine the minimum norm value for the model that satisfies this constraint and
fits the data in a statistical sense. If this norm value lies below the prior norm
bound, the property vector is accepted, and if not, it is ruled out. An example of this process can be
seen within Fig.\ref{fig:becmn}. Here the property space comprises just the $(1,-1)$
spherical harmonic coefficient, and we plot the variation of the squared minimum norm
associated with different values of this coefficient. The horizontal dot-dashed line shows
the squared-value of the chosen prior norm bound, and hence the interval in which the $(1,-1)$ coefficient
must lie is delineated. Again, in practice, this interval is more efficiently determined
using a bisection method.  In Fig.\ref{fig:bewe1d} we show the bounds obtained
on each of the degree one spherical harmonic coefficients through this process. To aid
in comparison with the earlier results, the prior norm bound was taken to
have the same value as in the error free-case.

From a computational point
of view it is worth emphasising that the incorporation of data errors comes at a
fairly substantial cost. This is because  the
property vector no longer lies within an hyperellipsoidally bounded region, and hence
a larger number of constrained optimisation problems must be solved to
determine its form. This is especially true if simultaneous bounds
on multiple directions in property space are sought as shown in
Fig.\ref{fig:bewe2d}. There is, however, scope for making
this process more efficient. For example, bounds
along different directions in the property space are independent of one another, and hence
the process can be trivially parallelised.

\subsection{Incorporating data errors into Backus-Gilbert estimators}

To conclude this section we show how data errors can be handled in the context
of Backus-Gilbert estimators. While the resulting theory is less general
and less precise than that just discussed, it is easier to implement
numerically and so might sometimes be preferred. There are also interesting  links
with the probabilistic
methods to be developed in the sequel; see
\cite{valentine2020gaussianI,valentine2020gaussianII}  discussion
of some related ideas. Starting with the general case, we
use eq.(\ref{eq:dmer}) to write the property vector,  $w = Bu$, in the form 
\begin{equation}
 w =  \tilde{w} + H u + C z, 
\end{equation}
for an arbitrary $C \in \Hom(F,G)$, where $\tilde{w} = C v$, and
$H = B-CA \in \Hom(E,G)$.  As previously, we introduce a  constraint set, $U \subseteq E$,
and a confidence set,  $V \subseteq F$, and hence obtain the inclusion
\begin{equation}
 w \in  (\tilde{w} + H U + C V)\cap BU.
\end{equation}
Different values of $C \in \Hom(E,F)$ lead to different bounds on the property vector,
and so having decided what constitutes a good subset, we can seek to optimise the choice
of this linear mapping.

To illustrate how this can be done practically we restrict attention
to  Hilbertable problems, take the constraint set to be $B_{r}(0)$ for some $r > 0$,
and assume  Gaussian errors with zero expectation and covariance $R \in \Hom(F)$.
The confidence set in eq.(\ref{eq:lscon}) can then be used, and comprises
those data vectors for which
\begin{equation}
\frac{1}{2}\cbraket{R^{-1} v}{v}_{F} \le s^{2}, 
\end{equation}
where $s>0$ is fixed by the confidence level. As shown earlier,
the image of the closed ball $ B_{r}(0)$ under the  affine mapping
$u \mapsto \tilde{w} + Hu$ is a closed set in $G$ with hyperellipsoidal boundary
whose elements satisfy
\begin{equation}
  \cbraket{(HH^{*})^{-1}(w-\tilde{w})}{w-\tilde{w}}_{G} \le r^{2}.
\end{equation}
Similarly, the image of the confidence set under $v \mapsto Cv$
is a  closed set in $G$ with hyperellipsoidal boundary
whose elements satisfy
\begin{equation}
\frac{1}{2}\cbraket{(CRC^{*})^{-1} w}{w}_{G} \le s^{2}.
\end{equation}
The property vector  lies in the sum of these two  subsets, but in general
the result will not have a  hyperellipsoidal boundary.
Nonetheless, because we want this subset to be suitably small,
a reasonable quantity to minimise is
\begin{equation}
  \label{eq:bgmis}
  J(C) = r^{2}\,\mathrm{tr}[(B-CA)(B-CA)^{*}] +  2 s^{2} \,\mathrm{tr}[ C R C^{*}], 
\end{equation}
which is proportional to the sum of the squared-lengths of the
principle axes for the two hyperellipsoids.  A simple calculation then shows that the chosen
functional has a unique minimum at
\begin{equation}
  C = B A^{*} \left(
  AA^{*} + 2s^{2}r^{-2} R
  \right)^{-1},
\end{equation}
which  generalises  eq.(\ref{eq:bgoptr}).  Importantly, the inclusion of the data
covariance guarantees that the inverse operator on the right hand side exists even
if the data mapping fails to be surjective. To apply this method practically,
we need only repeatedly act $\left(
AA^{*} + 2s^{2}r^{-2} R   \right)^{-1}$ on data vectors, and this can be
readily done either directly when $\dim F$ is small,  or using standard iterative
methods such as conjugate gradients. In total, this inverse operator needs
to be acted $4 \dim G + 1$ times. The first is to find $\tilde{w} = Cv$,
while $2 \dim G$ further actions are required for the components
of either $HH^{*}$ and $CRC^{*}$. This cost is around twice that
of the error-free case, and is likely to be substantially
less than that required for the Backus estimates discussed above.


\section{Discussion}

\label{sec:dis}

\subsection{Summary of the main ideas}

In spite of the  technicality of this paper, the essential ideas
are rather simple. Within an inference problem we are given
data that have been produced from an unknown model. From this data
we then wish to infer some other numerical properties of the model.
In almost all cases the data alone will not be sufficient to
place constraints on the property vector, and hence some form of prior constraint
on the model must  be assumed. Within this work we have considered only deterministic
constraints, which is to say that the model is assumed to lie within a
certain subset of the model space. With the prior constraint in place, we
can first ask whether there exist models within the constraint set that fit the data
in a statistically plausible manner. If the
answer is no, then there is likely something wrong with either the prior constraint or
the  physical formulation of the problem. Assuming that there are 
models compatible with both the data and the prior constraint, we can
proceed to test different values of the property vector as follows. We select
a  property vector of interest, and restrict attention
to models for which this value is obtained. We then ask whether it  is still possible
to  fit the data subject to the prior constraint. If it is, then
we accept the  value of the property vector, and if not we reject it
as being inconsistent with our information. Repeating this process for
different choices of  property vector, we can delineate the region
of the property space in which the true value must lie. It is
worth emphasising that this general philosophy depends in no way on
the inference problem being linear, and so applies equally well
to  non-linear problems. It is also important to note that  though models are
built within this process using methods similar to those for the solution of  inverse problems,
this is only as an intermediate step, and  these models have no intrinsic interest. 

To implement the theory, the main task is determining whether models exist that
fit the data subject to given constraints; both prior and those imposed in testing
a particular value for property vector.  Doing this requires the solution of constrained
optimisation problems, and it is here that great care is required. In the situation
considered in most detail within this paper, the prior constraint took the form of a
norm bound. To proceed, we  sought the minimum norm value for models
consistent with the constraints imposed by both the data and the property vector being tested.
If the minimum norm value was smaller than the assumed bound, the property vector
was accepted, and if not it was rejected. Thus, if the correct  minimum 
norm value is not obtained within our calculations, we might  reject viable
property vectors or accept those that are not. It was to avoid this possibility that detailed
mathematical analysis was necessary, with, in particular, the problem being posed
on an appropriate function space, and the numerical methods being shown to be
convergent. It is worth emphasising  that in these optimisation problems
it is only the minimum \emph{value} of the functional that is of interest,
and not whether this value is obtained by a \emph{unique model}. Within linear
inference problems it has been shown that this condition is always met, but
this need not be so within non-linear problems.

 The need to work in practice within Hilbert spaces requires
some comment. As noted earlier, most geophysical inference problems
are  naturally posed on Banachable spaces, and it is also in such
spaces that prior constraints are  likely to be expressed. For example,
with   parameters like density or seismic wave speed  we are far more likely to have information
on the magnitude of their point-values and not on their norms  in
some exotic sounding Hilbert space. While the general theory has been developed on
such Banachable spaces, we are quickly led to optimisation problems
that are not tractable numerically. The move
to Hilbert spaces is, therefore, done  pragmatically, but
Section \ref{ssec:norm_bound_res} showed that
this step can be rigorously justified so long as the prior
constraints take a  suitable form.  An important practical question
is  the extent to which different types of  prior information can  be converted into a
norm bound in a suitable Hilbert space. To give an indication of how this might be done,
we return, as ever, to our motivating spectral estimation problem. In this case, suppose
that we believe that  the function is continuously differentiable and  have the
following bound on its point values
\begin{equation}
  \|u\|_{C^{0}(\mathbb{S}^{2})} \le r, 
\end{equation}
for some $r > 0$. The Sobolev embedding theory tells us that $H^{s}(\mathbb{S}^{2})$
for $s > 2$ is comprised of continuously differentiable functions, while using
eq.(\ref{eq:c0bnd}) we can bound the supremum norm of its elements in
terms of their Sobolev norm.  It follows that the space  $H^{s}(\mathbb{S}^{2})$ for $s > 2$
along with the prior norm bound
  \begin{equation}
    \|u\|_{H^{s}(\mathbb{S}^{2})} \le  \left(\sum_{l\in \mathbb{N}}\frac{2l+1}{4\pi}\mult{l}{\lambda}^{-2s}\right)^{-\frac{1}{2}}\, r,
  \end{equation}
provides a  Hilbert space  setting for the problem consistent with our
prior information. The estimate within eq.(\ref{eq:ckbnd}) can be
similarly applied  in cases where pointwise bounds on the functions derivatives
are also given. Clearly the passage to this Hilbert space is not free
from ambiguity, with, for example, the values of $s$ nor $\lambda$ fixed in the above
discussion. Moreover, Sobolev embeddings are strict, and so there will
exist viable models that have been removed from the constraint set.
Until, however, suitable progress is made with computational optimisation in
Banach spaces, a certain amount of pragmatism is required.

\subsection{Extensions and applications}

To conclude it is worth commenting briefly on possible extensions and applications of these ideas.
First, while Theorem \ref{thm:bbackus3}
is  general, the only type of constraint set that has been investigated in
detail is that associated with  a prior norm bound. One area
of  interest would be the use of prior bounds on a semi-norm. This would be relevant, for
example, in situations where information on a function's  derivatives are given.

Next, it is possible to develop an analogous theory for situations in which
the constraints are probabilistic. What is meant by this is that
we are told that the unknown model is a realisation of a specified
``prior'' distribution on the model space. One cannot then, in general,
say anything concrete about the  property vector corresponding
to the realised model. But it is possible to regard this property
vector as a realisation of a random variable whose statistical properties can be determined. What is then done
with this information depends, of course, on both application and whether one is a frequenist or Bayesian.

Putting such issues aside,  mathematically the problem is just one of
conditional distributions. Both the data and property vectors
are viewed as realisations of  underlying random variables whose distributions
are obtained by pushing forward that on the model space.  The conditional distribution for the
property vector given the realised value of the data vector can then be found by standard means. In particular,
if the prior distribution on the model space is Gaussian, then the desired conditional
distribution is also Gaussian and can be readily computed. 
The result is, of course, identical to first applying Bayes theorem to
determine the conditional distribution of the model given the data  \citep[e.g.][]{stuart2010inverse},
and then pushing forward this posterior distribution to the property space.
The suggested method is simpler, however, because it does not require the calculation of a
distribution on an infinite-dimensional space.  Note that while this outline  has not considered
data errors,  these can be built into the theory in an easy and
natural manner. It is also possible to formulate a probablistic version of
Backus-Gilbert estimators, and show that for a Gaussian prior the results
agree with that obtained from Bayes theorem.

To do any of this in practice, of course, one needs to specify the
prior distribution on the model space. In doing this the use of abstract measure
theory is essential because there is no analogue of the Lebesgue measure
on infinite-dimensional spaces \citep[e.g][]{vakhania1987probability}, and hence one cannot
 work solely with probability density functions. Applications
are also likely to be limited to the case of Gaussian measures
\citep[e.g][]{kuo1975gaussian,bogachev1996gaussian}
for which the  calculations are tractable. As with the finite-dimensional case, these distributions
are fully characterised by their mean and covariance, but here the covariance is required
to be trace-class operator on the model space. This is a strong restriction which has no analogue within finite-dimensional
problems.  It should be noted that broadly similar results can be obtained using Gaussian processes -- see
\cite{valentine2020gaussianI,   valentine2020gaussianII} for a discussion in a geophysical context. While such an
approach is a  less general,  it can at least be understood and implemented using only elementary mathematics.

Finally, the extension of the methods to
non-linear inference problems would be of great interest. As noted above,
the basic ideas carry over directly,
while techniques like the adjoint method \citep[e.g.][]{tapeliutromp,lijackson,crawford2018}
provide a practical means for solving the resulting optimisation problems. The difficulty,
however, is that with non-linear problems there is no certainty that global minimum values will be
obtained. Nonetheless,  the present methods can already be applied to non-linear  problems, and so
long as the results are regarded tentatively they might prove useful. Moreover,
the risk of missing   global minimum values can be mitigated through practical steps such as performing
multiple optimisations using different starting values. 

Beyond extensions of the  theory, there are a range of potential practical applications
for the methods presented in this paper. Indeed, the theory has been written so as to be
applicable to a very wide class of linear inference problems. To apply the ideas 
to a given problem the main obstacle is finding a suitable Hilbert space to work in, with
this choice being guided  by the prior constraints that are believed. For very many
geophysical applications the functions comprising the model space are likely to be
(piece-wise) continuous or continuously differentiable up to some finite-order. In such cases
the use of an appropriate Sobolev space would seem sensible. Within this paper
only Sobolev spaces on the unit sphere have been considered, with spherical harmonic
methods rendering their  use almost trivial. In more
complicated domains the same basic ideas apply, but the implementation becomes more a little more
complicated and will be discussed in future work. See \cite{zuberi2017mitigating} for an interesting
discussion of some  related ideas.

\begin{acknowledgments}

  This work has benefited from discussions with many people. I would especially like to thank
  Andy Jackson,  Mark Hoggard,  Ophelia Crawford, Harriet Lau, Matthew Maitra, Frank Syvret,
  and  Zhi Li. I acknowledge the time and effort of  two anonymous reviewers, the editor, Prof.  Simons, along
  with additional input from Dr Valentine  and Prof. Agnew.

\end{acknowledgments}

\section*{Data availability}

No data has been used or generated  as part of this study. The software used for all calculations within in this
article will be shared on reasonable request to the author.

\bibliographystyle{gji}
\bibliography{references}

\appendix

\section{Functional analysis}

\label{sec:funal}

This appendix summarises the  definitions and notations from
functional analysis that are used within the main body of the paper.
It is assumed that the reader is  familiar with the
basic notations of set theory and  the definition of a vector space.

\subsection{Topological spaces}

 A topological space $X$ is a set along
with a collection of its subsets $\mathcal{T}$ subject
to the following axioms:
\begin{enumerate}
\item both $X$ and the empty set $\emptyset$ belong to $\mathcal{T}$;
\item the union of any collection of subsets in
  $\mathcal{T}$ belongs to $\mathcal{T}$;
\item the intersection of a finite collection of subsets in
  $\mathcal{T}$ belongs to $\mathcal{T}$.
\end{enumerate}
Subsets belong to $\mathcal{T}$ are said to be \emph{open}, while
the complement of an open set is \emph{closed}.
Let $(X,\mathcal{T})$ and $(Y,\mathcal{S})$ be two
topological spaces. A mapping $f:X\rightarrow Y$ is said
to be continuous if and only if for each $V \in \mathcal{S}$ we have
$f^{-1}(V) \in \mathcal{T}$. In words, a function is continuous
if the inverse image of each open set is open. Using the identity $f^{-1}(Y\setminus V)
= X\setminus f^{-1}(V)$ which holds for any $V \subseteq Y$, it follows that we can
equivalently say that $f$  is continuous if and only if the inverse image of each
closed set is closed. A topological space
is said to be \emph{Hausdorff} if for any two
points $x_{1}\ne x_{2}$ there exist disjoint open sets
$U_{1}$ and $U_{2}$ such that $x_{1} \in U_{1}$ and
$x_{2} \in U_{2}$. Clearly  within a Hausdorff space each subset  containing
a single element is closed.  For a subset  $U \subseteq X$, 
its interior, $\mathring{U}$, is defined to be the largest open set contained
within $U$, while its closure, $\bar{U}$, is the smallest
closed set that contains it. We say that such  a subset is \emph{dense}
in $X$  if $\bar{U} = X$.

As a simple example, the standard topology on the real line $\mathbb{R}$
can be defined as follows. About a point $x \in \mathbb{R}$
we define the open ball with radius $r > 0$ to be
\begin{equation}
  B_{r}(x) = \{y \in \mathbb{R} \,|\, |x-y| < r\}.
\end{equation}
A subset $U \subset \mathbb{R}$ is then declared open if and only
if for each $x \in U$ we can find an $r > 0$ such that
$B_{r}(x) \subseteq U$. It is readily checked that this definition
 satisfies the above axioms, while the  definition
of continuity for a function $f:\mathbb{R}\rightarrow \mathbb{R}$
reduces to Cauchy's familiar $\epsilon$-$\delta$ definition used within
elementary analysis.

A \emph{neighbourhood} of a point $x$ within a topological
space $(X,\mathcal{T})$ is any subset that contains an open set containing
$x$. It follows that an open set is characterised as being a neighbourhood
of each of its points. It is also possible to take neighbourhoods of
a point as a primitive notion, and \emph{define} open sets
through the property just noted. This process mimics rather closely
what was done for $\mathbb{R}$ above, with details  found,
for example, in  \citet[][Chapter 1]{treves}. In fact, one can make do
with specifying only a so-called
\emph{basis} of neighbourhoods at each point from which the full collection
can be generated. Specifically, a basis of neighbourhoods at $x \in X$ comprises
a collection of subsets that are subject  to certain axioms, and such
that any subset that contains a member of this basis is
 a neighbourhood of $x$. It is this latter approach that
is most useful in the case of topological vector spaces.

Let $X$ be a set, and $\mathcal{T}_{1}$ and $\mathcal{T}_{2}$ two
collections of its subsets that are consistent with the above axioms.
Both $(X,\mathcal{T}_{1})$ and $(X,\mathcal{T}_{2})$ are then
topological spaces. Let us suppose that each subset in $\mathcal{T}_{2}$
also belongs to $\mathcal{T}_{1}$. We then say that the topology on  $(X,\mathcal{T}_{2})$
is \emph{finer} than that on  $(X,\mathcal{T}_{1})$, or conversely that
the latter space has the \emph{coarser} topology. This
defines a partial ordering for topologies on the same underlying
set. The finest topology comprises all subsets of $X$ and is known
as the \emph{discrete topology}, while the coarsest is the \emph{trivial
topology} whose only open sets are $X$ and $\emptyset$; neither
 topology is useful in practice.  Note that  continuity of the  identity
mapping from  $(X,\mathcal{T}_{2})$ onto  $(X,\mathcal{T}_{1})$ is
necessary and sufficient for $(X,\mathcal{T}_{2})$ to have the
finer topology. As an application of these ideas, let
$(X,\mathcal{T})$ and $(Y,\mathcal{S})$ be two topological spaces,
and consider the Cartesian product $X\times Y$. The \emph{product
  topology} on this latter set is defined to be the coarsest topology
such that the two projections $(x,y)\mapsto x$ and $(x,y)\mapsto y$
are continuous. A concrete description of its open sets can
be obtained from this definition if desired, this being most readily
done in terms of a basis of neighbourhoods. 

\subsection{Topological vector spaces}

Let $E$ be a real vector space, this definition being assumed
known \citep[e.g.][Chapter 2]{treves}. To make $E$ a topological
vector space we need to specify a topology such that
addition and scalar multiplication are continuous
(note that $\mathbb{R}$ is always assumed to carry its
standard topology). To do this it is sufficient
to specify a basis of neighbourhoods at the origin, with
continuity of addition implying that the neighbourhoods at
any other point can be obtained by translation. Within this paper there is no need to discuss
the most general  topological vector spaces, but only those that are \emph{normable}. A \emph{norm} $\|\cdot\|_{E}:E\rightarrow \mathbb{R}$ is a mapping such that
\begin{enumerate}
\item $\|u\|_{E} = 0$,  implies $u = 0$;
\item $\|\alpha u\|_{E} = |\alpha|\|u\|_{E}$;
\item $\|u_{1} + u_{2}\|_{E} \le \|u_{1}\|_{E} + \|u_{2}\|_{E}$, 
\end{enumerate}
with the latter property known as the triangle inequality. Given a norm on $E$, we can
define a countable basis of neighbourhoods at the origin using the open balls
\begin{equation}
  B_{1/n}(0) = \{ u \in E \,|\, \|u\|_{E} < 1/n\}, 
\end{equation}
with $n$ ranging over the positive integers.  The resulting topological vector
space is said to be normable, and is readily seen to be
Hausdorff. A second norm $\|\cdot\|'_{E}$ on $E$ is
\emph{equivalent} to the first if there exist positive constants $c < C$ such that
\begin{equation}
  \label{eq:nequiv}
  c\|u\|_{E} \le \|u\|'_{E} \le C \|u\|_{E}, 
\end{equation}
for all $u \in E$.  Each open ball with respect to the first
norm must, therefore,  contain an open ball with respect to the second (and conversely),
which implies that the  norms define  identical topologies on $E$. 
A special case of this construction arises from an inner product,
this being a positive-definite and symmetric bilinear mapping
$(\cdot,\cdot)_{E}:E\times E\rightarrow \mathbb{R}$
such that $\cbraket{u}{u}_{E} = 0$  implies $u = 0$. The inner product
induces a norm $\|u\|_{E} = \sqrt{\cbraket{u}{u}_{E}}$, and then everything proceeds as before.

Let $E$ be a normable vector space. A sequence $\{u_{i}\}_{i\in \mathbb{N}}$ in $E$ is said to be \emph{Cauchy} if for each neighbourhood $U$ of the origin there exists an
$n$ such that $i,j > n$ implies $u_{i}-u_{j} \in U$. Clearly every
convergent sequence is Cauchy, but the converse need not be true.
A normable vector space is said to be \emph{complete} if and only if
each of its Cauchy sequences converges. From a non-complete
normable vector space there is a standard procedure for
obtaining a complete space into which the original is densely embedded,
with the result being unique up to an isomorphism
\citep[e.g.][Chapter 5]{treves}. A complete normable vector space
is said to be \emph{Banachable}, while if its norm is defined
in terms of an inner product it is \emph{Hilbertable}. If
one specific norm $\|\cdot\|_{E}$ for a Banachable space $E$ is chosen, then the
resulting pair $(E,\|\cdot\|_{E})$ is said to be a \emph{Banach space}, while
Hilbert spaces are defined in an analogous manner by picking out
an inner product.

\subsection{Linear subspaces and direct sums}
\label{ssec:linspace}

A \emph{linear subspace} $U$ within a topological  vector space is a subset
$U\subseteq E$ such if $u_{1},u_{2} \in U$ then so is
$\alpha_{1}u_{1}+\alpha_{2}u_{2}$ for any $\alpha_{1},\alpha_{2} \in \mathbb{R}$.
As an example, given elements $u_{1},\dots,u_{n} \in E$ we define their
\emph{span} by
\begin{equation}
  \mathrm{span}\{u_{1},\dots,u_{n}\} = \{\alpha_{1}u_{1}+\cdots + \alpha_{n}u_{n}\,|\,
  \alpha_{1},\dots,\alpha_{n} \in \mathbb{R}\}.
\end{equation}
A \emph{affine subspace}  in $E$ is a subset of the form
\begin{equation}
  u + U = \{u + u'\,|\, u' \in U\},
\end{equation}
with $u$ a fixed vector and $U$  a linear subspace. A linear (or affine) subspace may or may not be
closed, but this is always true if it is finite-dimensional or if it has a finite-dimensional
complementary subspace (in which case we say is has finite codimension). 

Given two linear subspaces $U_{1}$ and $U_{2}$ of $E$ with $U_{1}\cap U_{2} = \{0\}$, we can define their \emph{direct sum} by
\begin{equation}
  U_{1}\oplus U_{2} = \{u_{1} + u_{2}\,|\, u_{1}\in U_{1},\, u_{2}\in U_{2}\}, 
\end{equation}
which is again  a linear subspace.  If $E$ can be written in the form
\begin{equation}
  E = U_{1} \oplus U_{2}, 
\end{equation}
we say that $U_{2}$ is a \emph{complementary subspace} to $U_{1}$ and visa versa. Given a
linear subspace $U_{1} \subseteq E$, it is not necessarily true that
there exists a complementary subspace $U_{2}$ such that $E = U_{1}\oplus U_{2}$.
Such a  subspace does, however, always exist if $U_{1}$ has  finite dimension
or codimension \citep[e.g.][Proposition 9.3]{treves}.

The direct sum, $E\oplus F$, of two Banachable spaces $E$ and $F$
can be defined as follows. It is comprised of ordered pairs
$(u,v)$ with $u \in E$ and $v \in F$, with addition
and scalar multiplication defined by
\begin{equation}
  \lambda\,(u,v) = (\lambda u,\lambda v), \quad
  (u_{1},v_{1}) + (u_{2},v_{2}) = (u_{1}+u_{2},v_{1}+v_{2}).
\end{equation}
By convention,  the element $(u,v)$ of this space is  denoted
by  $u \oplus v$. The topology on the direct sum is set using the norm
\begin{equation}
  \label{eq:prodnorm}
  \|u \oplus v\|_{E\oplus F} = \|u\|_{E} + \|v\|_{F}, 
\end{equation}
where any compatible norms for $E$ and $F$ have been selected. It is readily
shown that $E\oplus F$ is complete, and hence Banachable. 
In the case of two Hilbert spaces this 
construction immediately  applies, but here it is useful to define the inner product
\begin{equation}
  \label{eq:dsumprod}
  \cbraket{u_{1}\oplus v_{1}}{u_{2}\oplus v_{2}}_{E\oplus F}
  = \cbraket{u_{1}}{u_{2}}_{E} + \cbraket{v_{1}}{v_{2}}_{F}.
\end{equation}
While the induced norm does not coincide with that in
eq.(\ref{eq:prodnorm}), they are readily shown to be equivalent. 

The two forms of the direct sum just described are related but distinct.
In the first case both subspaces lie within the same larger vector space,
and the condition that they have a trivial intersection is required
for the  decomposition of a vector to  be unique. In the second case, however,
the direct sum is formed from two potentially unrelated vector spaces (but defined
over the same field). This is done by  forming their Cartesian product, and then defining
a suitable vector space structure. The relation between the two is
as follows. Having formed $E\oplus F$ from two distinct spaces, we can identify within
it two linear subspaces $E\oplus \{0\}$ and $\{0\}\oplus F$. These subspaces
clearly have trivial intersection, while their direct sum $(E\oplus \{0\}) \oplus
(\{0\}\oplus F)$, made in the  first sense, is equal to $E\oplus F$.
That the same terminology is used in both cases is standard, and should cause
no confusion in context. No one would think to question
the identity $\mathbb{R}\oplus \mathbb{R} = \mathbb{R}^{2}$, for example.

\subsection{Linear operators and dual spaces}

\label{ssec:linop}

Let $E$ and $F$ be Banachable vector spaces, with $\|\cdot\|_{E}$ and $\|\cdot\|_{F}$
 compatible norms. A linear mapping
$A:E\rightarrow F$ is continuous if and only if there is a constant
$c > 0$ such that
\begin{equation}
  \|Au\|_{F} \le c \,\|u\|_{E}, 
\end{equation}
 for all $u \in E$.
The collection of continuous linear mappings from $E$ into $F$ is
denoted by $\Hom(E,F)$, and is itself  Banachable.
To describe its topology we  define the  \emph{operator norm}
\begin{equation}
  \label{eq:op_norm}
  \|A\|_{\Hom(E,F)} = \sup_{u \in E\setminus \{0\}}\frac{\|Au\|_{F}}{\|u\|_{E}}.
\end{equation}
It is readily checked that equivalent choices of norms on $E$ and $F$
lead to equivalent operator norms, and hence a unique Banachable structure.
As special cases we first note  that the space $\Hom(E,E)$ is
usually abbreviated to $\Hom(E)$. Next, $\mathbb{R}$  is itself  Banachable,
and so we can define   $E' = \Hom(E,\mathbb{R})$ which is known
as the \emph{dual space} of $E$. The action of a dual vector
$u'\in E'$ on a vector $u \in E$  will be written
$\braket{u'}{u}$. For each $u \in E$ we can
consider the linear mapping
\begin{equation}
E'\ni  u' \mapsto \braket{u'}{u}, 
\end{equation}
which is clearly continuous.  Each element of
$E$ can, therefore, be identified with a unique point in the
\emph{bidual} $E'' = (E')'$. If the inclusion of $E$
into $E''$ is an isomorphism we say that $E$ is \emph{reflexive}.
For two Banachable spaces $E_{1}$ and $E_{2}$, it can 
be shown that the dual $(E_{1}\oplus E_{2})'$ of their direct sum is
canonically isomorphic to $E_{1}'\oplus E_{2}'$. To each
element of $ (E_{1}\oplus E_{2})'$ there, therefore, exists unique
$u_{1}' \in E_{1}'$ and $u_{2}'\in E_{2}'$ such that its action
 on $u_{1}\oplus u_{2}$ takes the form
\begin{equation}
  \braket{u_{1}'\oplus u_{2}'}{u_{1}\oplus u_{2}}= \braket{u_{1}'}{u_{1}} + \braket{u_{2}'}{u_{2}}.
\end{equation}

Let $A \in \Hom(E,F)$. We define its \emph{kernel} to be
the linear subspace
\begin{equation}
  \ker A = \{u \in E \,|\, Au = 0\}, 
\end{equation}
which is closed because it is the inverse image of a closed set under
a continuous mapping.   A mapping for which the
kernel is trivial is said to be \emph{injective}. A related
subspace is the \emph{image} of $A$ defined through
\begin{equation}
  \image A = \{Au\in F\,|\, u \in E\}.
\end{equation}
The image of a continuous linear mapping need not be closed. If
$\image A = F$ we say the mapping is \emph{surjective}.

\subsection{Calculus on Banachable spaces}

Let $E$ and $F$ be  Banachable spaces, and $U\subseteq E$ an open subset.
A mapping $f:U\rightarrow F$ is \emph{Fr\'{e}chet differentiable} at the point
$u \in U$ if there exists a linear mapping $A \in \Hom(E,F)$ such that
\begin{equation}
  f(u+u') = f(u) + A  u' + O(\|u'\|_{E}^{2}), 
\end{equation}
for all sufficiently small $u' \in E$. When this holds we write
$Df(u)$ for the linear mapping and call it the \emph{Fr\'{e}chet derivative}
of $f$ at $u$, or usually just the derivative. Assuming that
$f$ is  \emph{Fr\'{e}chet differentiable} at each point in its domain,
we can consider the mapping $u \mapsto Df(u) \in \Hom(E,F)$. If this
mapping is continuous relative to the topology defined
using eq.(\ref{eq:op_norm}), we say that $f$ is continuously
Fr\'{e}chet differentiable. Higher order derivatives are defined
inductively so long as they exist. For example, we write
$D^{2}f(u)$ for the second derivative of $f$ at $u$, this
object taking values in $\Hom(E,\Hom(E,F))$. Partial
derivatives are defined in the obvious manner for
mappings  on product spaces, with subscripts used
to indicate the variable with respect to which the derivative
has been taken.

\section{Sobolev  spaces on $\mathbb{S}^{2}$}

\label{sec:fspace}

In this appendix we recall the definition and relevant properties of
the Sobolev spaces $H^{s}(\mathbb{S}^{2})$. Several equivalent definitions
for these spaces exist, while the ideas and results can be  extended to more general domains
\citep[e.g.][]{lions2012non,shubin1987pseudodifferential,taylor1996partial}.
The definition  given here is in terms of spherical harmonic expansions, this
being the simplest method available when working on $\mathbb{S}^{2}$. In detail
this is an instance of an approach based on fractional powers of self-adjoint
elliptic operators \citep[e.g.][]{shubin1987pseudodifferential}. The equivalence of this approach
with other methods can be found in \cite{lions2012non}.
A somewhat similar discussion based on spherical harmonic expansions  can be found in \cite{freeden1995survey}  along
with many other works by Freeden.  There is no claim to
originality in any of the material in this appendix.  Proofs are only given
were a result  was required in the course of this work, but for  which a
convenient reference was not known to the author.

\subsection{Square-integrable functions and spherical harmonic expansions}

\label{ssec:sqint}

A measurable  function $u:\mathbb{S}^{2}\rightarrow\mathbb{R}$ is said to be square-integrable if
\begin{equation}
  \int_{\mathbb{S}^{2}} u^{2} \dd S < \infty. 
\end{equation}
The space of such functions forms the Hilbert space $L^{2}(\mathbb{S}^{2})$, with inner product
given by
\begin{equation}
  \cbraket{u}{v}_{L^{2}(\mathbb{S}^{2})} = \int_{\mathbb{S}^{2}} u\,v \dd S, 
\end{equation}
and associated norm  $\|\cdot\|_{L^{2}(\mathbb{S}^{2})}$. As noted in the main text, this
space also can be obtained by completing $C^{0}(\mathbb{S}^{2})$ relative
to this choice of inner product. A technical but important remark is that
elements of $L^{2}(\mathbb{S}^{2})$ are actually defined as equivalence classes of
square-integrable functions on $\mathbb{S}^{2}$, with two such functions
being equivalent if they are equal everywhere except a set of null measure. It follows,
in particular, that the point-values of an element of $L^{2}(\mathbb{S}^{2})$ cannot be defined.

Let $C^{\infty}(\mathbb{S}^{2})$ denote the space of smooth functions on $\mathbb{S}^{2}$.
The Laplace-Beltrami operator, written here as $\Delta$, is defined by
\begin{equation}
    \Delta u = -\frac{1}{\sin \theta} \frac{\partial}{\partial \theta}
  \left(\sin \theta \frac{\partial u}{\partial \theta}
  \right)
  - \frac{1}{\sin^{2}\theta} \frac{\partial^{2} u}{\partial \varphi^{2}}, 
\end{equation}
for smooth $u$, with $(\theta,\varphi)$ spherical coordinates. It can be  verified that $\Delta$ is
formally self-adjoint, this meaning that
\begin{equation}
  \cbraket{\Delta u}{v}_{L^{2}(\mathbb{S}^{2})} =   \cbraket{ u}{\Delta v}_{L^{2}(\mathbb{S}^{2})}, 
\end{equation}
for all $u,v \in C^{\infty}(\mathbb{S}^{2})$, while,  due to our sign convention, it is
non-negative in the sense that
\begin{equation}
  \cbraket{\Delta u}{u}_{L^{2}(\mathbb{S}^{2})} \ge 0,
\end{equation}
for all $u \in C^{\infty}(\mathbb{S}^{2})$.  Extending $\Delta$ to a
densely defined unbounded operator on $L^{2}(\mathbb{S}^{2})$,  we can consider the
associated eigenvalue problem. This  leads  to the introduction of the spherical harmonics
$\{Y_{lm} \,|\, l \in \mathbb{N}, \,-l\le m\le l\}$ which satisfy
\begin{equation}  \Delta Y_{lm} = l(l+1)Y_{lm}.
\end{equation}
Each spherical harmonic can be shown to be smooth, while together they form an
a  orthonormal basis for $L^{2}(\mathbb{S}^{2})$. This latter idea
can be most conveniently expressed in terms of the finite-dimensional subspaces
\begin{equation}
  \mathcal{H}_{l} = \mathrm{span}\{Y_{lm} \,|\, -l\le m\le l\}, 
\end{equation}
for each degree $l \in \mathbb{N}$. 
We can then write $L^{2}(\mathbb{S}^{2})$ as the orthogonal direct sum
\begin{equation}
  \label{eq:l2decomp}  
  L^{2}(\mathbb{S}^{2}) = \bigoplus_{l\in \mathbb{N}} \mathcal{H}_{l}.
\end{equation}
The projection operator 
mapping $L^{2}(\mathbb{S}^{2})$ onto $\mathcal{H}_{l}$ is given by
\begin{equation}
  \label{eq:projl}
  \mathbb{P}_{l}u = \sum_{m=-l}^{l} \cbraket{Y_{lm}}{u}_{L^{2}(\mathbb{S}^{2})} Y_{lm}, 
\end{equation}
and  eq.(\ref{eq:l2decomp}) is equivalent to the condition
\begin{equation}
  \label{eq:resid}
  \sum_{l\in \mathbb{N}} \mathbb{P}_{l} = 1.
\end{equation}
It is readily seen that these projection operators satisfy the following conditions
\begin{equation}
  \label{eq:projid}
  \mathbb{P}_{l}\mathbb{P}_{l'} = \delta_{ll'}\mathbb{P}_{l}, \quad
  \mathbb{P}_{l}^{*} = \mathbb{P}_{l}.
\end{equation}
Using eq.(\ref{eq:l2decomp}) and (\ref{eq:resid}), any function  $u \in L^{2}(\mathbb{S}^{2})$ can be expressed in the form
\begin{equation}
  \label{eq:sphexp}
  u = \sum_{l\in \mathbb{N}} \mathbb{P}_{l}u, 
\end{equation}
and taking the squared-norm of this orthogonal decomposition we obtain
\begin{equation}
  \|u\|^{2}_{L^{2}(\mathbb{S}^{2})} = \sum_{l\in \mathbb{N}} \|\mathbb{P}_{l}u\|_{L^{2}(\mathbb{S}^{2})}^{2}, 
\end{equation}
which implies that the sum on the right hand side is finite. 
Consider a sequence $\{u_{l}\}_{l\in \mathbb{N}}$ of functions such that $u_{l} \in \mathcal{H}_{l}$
for each $l \in \mathbb{N}$. The Riesz-Fischer theorem shows conversely that if
\begin{equation}
  \sum_{l\in \mathbb{N}} \|u_{l}\|^{2}_{L^{2}(\mathbb{S}^{2})} < \infty,
\end{equation}
then $u = \sum_{l\in \mathbb{N}} u_{l}$ is a well-defined element of
$L^{2}(\mathbb{S}^{2})$. This fact will play a central role in what follows. 

It will be useful to recall briefly how spherical harmonics transform under the
action of the rotation group, $\mathrm{SO}(3)$. For a given rotation matrix
$R \in \mathrm{SO}(3)$ we define a linear mapping
\begin{equation}
  \label{eq:rotrep}
  (T_{R}u)(x) = u(R^{*}x), 
\end{equation}
which acts on measurable functions $u:\mathbb{S}^{2}\rightarrow \mathbb{R}$. This mapping
can be extended to elements of $L^{2}(\mathbb{S}^{2})$ and   forms
a unitary representation of $\mathrm{SO}(3)$. 
This implies, in particular, that for all $u,v \in L^{2}(\mathbb{S}^{2})$ we have
\begin{equation}
  \label{eq:invar}
  \cbraket{T_{R}u}{T_{R}v}_{L^{2}(\mathbb{S}^{2})} = \cbraket{u}{v}_{L^{2}(\mathbb{S}^{2})}, 
\end{equation}
and hence  the $L^{2}(\mathbb{S}^{2})$-norm
is invariant under this action of the rotation group. Moreover, it can be shown
that the subspaces $\mathcal{H}_{l}$ for each $l \in \mathbb{N}$ are invariant
and irreducible under this representation. In fact, eq.(\ref{eq:l2decomp})
is precisely what the Peter-Weyl theorem yields in this situation \citep[e.g.][]{bump2004lie}. For our
purposes,  the key point is that the following commutation relation holds
\begin{equation}
  \label{eq:procom}
  [T_{R},\mathbb{P}_{l}] = 0,
\end{equation}
for all $R \in \mathrm{SO}(3)$. Suppose that $A$ is a linear
operator mapping $L^{2}(\mathbb{S}^{2})$ into itself which satisfies 
\begin{equation}
  [T_{R},A] = 0,
\end{equation}
for all $R \in \mathrm{SO}(3)$. Schur's lemma \citep[e.g.][]{bump2004lie} then  implies that for some
function $\mathbb{N} \ni l \mapsto a_{l} \in \mathbb{R}$ we  have
\begin{equation}
  A \mathbb{P}_{l} =  \mathbb{P}_{l} A = a_{l} \mathbb{P}_{l}, 
\end{equation}
and hence using eq.(\ref{eq:resid}) we obtain the general form
\begin{equation}
  \label{eq:rotop}
  A = \sum_{l \in \mathbb{N}} a_{l} \mathbb{P}_{l},
\end{equation}
of a rotationally invariant linear operator on $L^{2}(\mathbb{S}^{2})$.

\subsection{Definition of the Sobolev space $H^{s}(\mathbb{S}^{2})$ with non-negative exponent}

For a fixed positive real number $\lambda$, we introduce the function
\begin{equation}
  \mathbb{N} \ni l \mapsto \mult{l}{\lambda} = \sqrt{1+ \lambda^{2}\,l(l+1)}.
  \end{equation}
The Sobolev space $H^{s}(\mathbb{S}^{2})$ with exponent $s\ge 0$ is  defined by
\begin{equation}
  \label{eq:sobdef}
  H^{s}(\mathbb{S}^{2}) = \left\{ u \in L^{2}(\mathbb{S}^{2}) \left|\right.
  \sum_{l\in \mathbb{N}} \mult{l}{\lambda}^{2 s} \|\mathbb{P}_{l}u\|_{L^{2}(\mathbb{S}^{2})}^{2} < \infty
  \right\}. 
\end{equation}
Clearly   $H^{0}(\mathbb{S}^{2}) = L^{2}(\mathbb{S}^{2})$. The following results summarise some
further properties  that will be required. 

\begin{prop} 
  For each $s \ge 0$ and $\lambda > 0$,  $H^{s}(\mathbb{S}^{2})$ is a Hilbert space  with
  respect to the   inner product
  \begin{equation}
    \cbraket{u}{v}_{H^{s}(\mathbb{S}^{2})} = \sum_{l\in \mathbb{N}}\mult{l}{\lambda}^{2 s}
    \cbraket{\mathbb{P}_{l}u}{\mathbb{P}_{l}v}_{L^{2}(\mathbb{S}^{2})}.
  \end{equation}
  Moreover, this inner product is invariant under the action of the rotation group.
\end{prop}

\begin{prop}
  \label{prop:smooth}
  The set   $C^{\infty}(\mathbb{S}^{2})$ of smooth real-valued functions on $\mathbb{S}^{2}$ is contained
  densely    in $H^{s}(\mathbb{S}^{2})$ for all $s \ge 0$.
\end{prop}

\begin{prop}
  \label{prop:sobexp}
  When $s > t\ge 0$ there is a continuous, dense, and proper,
 embedding $H^{s}(\mathbb{S}^{2}) \hookrightarrow H^{t}(\mathbb{S}^{2})$.
\end{prop}

\begin{prop}
  \label{prop:stopo}
  The topological structure of $H^{s}(\mathbb{S}^{2})$ is independent of $\lambda > 0$.
\end{prop}

\emph{Proof:} We will temporarily use the notation $H^{s}_{\lambda}(\mathbb{S}^{2})$ to
emphasise the choice of parameter $\lambda > 0$ within the definition of the
Sobolev space. Let $u \in H^{s}_{\lambda}(\mathbb{S}^{2})$ for some $\lambda > 0$, and $\mu > 0$
be given. From the definition of the respective norms we have
\begin{equation}
  \label{eq:eqtmp1}
\|u\|_{H^{s}_{\lambda}(\mathbb{S}^{2})}^{2} =     \sum_{lm}\mult{l}{\lambda}^{2s}\mult{l}{\mu}^{-2s}
  \mult{l}{\mu}^{2s} \|\mathbb{P}_{l}u\|^{2}_{L^{2}(\mathbb{S}^{2})}
  \le \max(1,\lambda^{2s}/\mu^{2s})\|u\|_{H^{s}_{\mu}(\mathbb{S}^{2})}^{2}, 
\end{equation}
where we have use the inequality
\begin{equation}
  \mult{l}{\lambda}^{2s}\mult{l}{\mu}^{-2s} =   \left[\frac{1+\lambda^{2}l(l+1)}{1+\mu^{2}l(l+1)}\right]^{s}
  \le \max(1,\lambda^{2s}/\mu^{2s}),
  \quad l \in \mathbb{N}.
\end{equation}
It follows that there is  a continuous embedding $H^{s}_{\mu}(\mathbb{S}^{2}) \hookrightarrow
H^{s}_{\lambda}(\mathbb{S}^{2})$, while an identical argument shows
$H^{s}_{\lambda}(\mathbb{S}^{2}) \hookrightarrow H^{s}_{\mu}(\mathbb{S}^{2})$.
The two Hilbert spaces are, therefore, isomorphic, and hence
identical from a topological perspective. \eproof

The results  of this section show that the Sobolev spaces are Hilbertable, with their
topological properties  set by the exponent $s\ge 0$. Varying the parameter $\lambda > 0$
merely changes the form of the  inner product. Indeed, there are other equivalent ways of defining these spaces which lead to quite different looking inner products. It is for this reason that the notation
$H^{s}(\mathbb{S}^{2})$ emphasises  the Sobolev exponent, but leaves the specific choice of
inner product implicit.

\subsection{Dual Sobolev spaces}

The Riesz representation theorem  states that
each Hilbert space is isometrically isomorphic to its dual. In the case of
$L^{2}(\mathbb{S}^{2})$, for example, this means to each $u'
\in L^{2}(\mathbb{S}^{2})'$ there is a unique $v \in L^{2}(\mathbb{S}^{2})$
such $\|u'\|_{L^{2}(\mathbb{S}^{2})'} = \|v\|_{L^{2}(\mathbb{S}^{2})}$, while also
\begin{equation}
  \braket{u'}{u} = \cbraket{v}{u}_{L^{2}(\mathbb{S}^{2})}, 
\end{equation}
for all $u \in L^{2}(\mathbb{S}^{2})$. Here  $\|\cdot\|_{L^{2}(\mathbb{S}^{2})'}$
denotes the dual norm that is defined by
\begin{equation}
  \|u'\|_{L^{2}(\mathbb{S}^{2})'} = \sup_{u \in L^{2}(\mathbb{S}^{2})} \frac{\braket{u'}{u}}{\|u\|_{L^{2}(\mathbb{S}^{2})}}.
\end{equation}
This theorem  applies similarly to $H^{s}(\mathbb{S}^{2})$,
but in this case it will be useful to determine the concrete relationship between
dual vectors and their representations
in $H^{s}(\mathbb{S}^{2})$. We begin with the following  characterisation of  dual
Sobolev spaces:

\begin{prop}
  \label{prop:sobd}
  Let $\{u_{l}'\}_{l\in\mathbb{N}}$ be a sequence of functions with $u_{l}' \in \mathcal{H}_{l}$
  for each $l \in \mathbb{N}$, and suppose that
  \begin{equation}
    \label{eq:sobdcon}
    \sum_{l\in \mathbb{N}} \mult{l}{\lambda}^{-2s} \|u'_{l}\|_{L^{2}(\mathbb{S}^{2})}^{2} < \infty,
  \end{equation}
  for $s > 0$.  The linear functional $u'$ defined by
  \begin{equation}
    \label{eq:fundef}
   \braket{u'}{u} = \sum_{l\in\mathbb{N}}\cbraket{u'_{l}}
    {\mathbb{P}_{l}u}_{L^{2}(\mathbb{S}^{2})},
  \end{equation}
  is continuous on $H^{s}(\mathbb{S}^{2})$, with dual norm  given by
  \begin{equation}
    \label{eq:sobdn}
    \|u'\|_{H^{s}(\mathbb{S}^{2})'} = \left(\sum_{l\in \mathbb{N}}\mult{l}{\lambda}^{-2s} \|u'_{l}\|_{L^{2}(\mathbb{S}^{2})}^{2}\right)^{\frac{1}{2}}.
  \end{equation}
  Conversely, for each $u'\in H^{s}(\mathbb{S}^{2})'$ there is
  a sequence $\{u_{l}'\}_{l\in\mathbb{N}}$ of functions with the above  properties  such that
  eq.(\ref{eq:fundef}) holds.
  \end{prop}

Motivated by  Proposition \ref{prop:sobd}, the domain of the orthogonal
projection operators $\mathbb{P}_{l}$ for $l \in \mathbb{N}$ can
be usefully extended to include $H^{s}(\mathbb{S}^{2})'$ through the definition
\begin{equation}
  \mathbb{P}_{l}u' = \sum_{m=-l}^{l} \braket{u'}{Y_{lm}} Y_{lm} \in \mathcal{H}_{l}.
\end{equation}
Proposition \ref{prop:sobd}  then shows that an
element $u' \in H^{s}(\mathbb{S}^{2})'$ can be identified with its formal
spherical harmonic decomposition
\begin{equation}
  u' = \sum_{l\in \mathbb{N}}\mathbb{P}_{l}u', 
\end{equation}
subject to the convergence condition in eq.(\ref{eq:sobdcon}) which we rewrite as
\begin{equation}
  \label{eq:sobdcon2}
  \sum_{l\in \mathbb{N}} \mult{l}{\lambda}^{-2s} \|\mathbb{P}_{l}u'\|_{L^{2}(\mathbb{S}^{2})}^{2} < \infty.
\end{equation}
 The collection of all such formal spherical harmonic decompositions
 \emph{defines}  the Sobolev space  $H^{-s}(\mathbb{S}^{2})$ with negative exponent.
In this manner, the definition of $H^{s}(\mathbb{S}^{2})$ is extended to all real exponents, 
and we have established the isomorphism
\begin{equation}
  H^{s}(\mathbb{S}^{2})' \cong H^{-s}(\mathbb{S}^{2}), 
\end{equation}
which is consistent with the dual norm in eq.(\ref{eq:sobdn}). Moreover,
Proposition \ref{prop:smooth} can also be shown to be valid for
negative $s$, and hence $C^{\infty}(\mathbb{S}^{2})$ is dense
in $H^{s}(\mathbb{S}^{2})$ for all $s \in \mathbb{R}$.
We can now state the main result of this subsection:

\begin{thm}
  \label{prop:sobrep}
  For each $u' \in H^{s}(\mathbb{S}^{2})$, the unique element $v \in H^{s}(\mathbb{S}^{2})$
  such that  $\|u'\|_{H^{s}(\mathbb{S}^{2})'} = \|v\|_{H^{s}(\mathbb{S}^{2})}$ and
  \begin{equation}
    \label{eq:sobrep1}
    \braket{u'}{u} = \cbraket{v}{u}_{H^{s}(\mathbb{S}^{2})},
  \end{equation}
  for all $u \in H^{s}(\mathbb{S}^{2})$  is given by
  \begin{equation}
    v = \sum_{l \in \mathbb{N}} \mult{l}{\lambda}^{-2s} \mathbb{P}_{l}u'.
  \end{equation}
\end{thm}

\emph{Proof:} Using the spherical harmonic decomposition  of the dual vector, we can write
\begin{equation}
  \braket{u'}{u} = \sum_{l\in \mathbb{N}}\cbraket{\mathbb{P}_{l}u'}{\mathbb{P}_{l}u}_{L^{2}(\mathbb{S}^{2})}
   = \sum_{l\in \mathbb{N}}\mult{l}{\lambda}^{2s}\cbraket{\mult{l}{\lambda}^{-2s}\mathbb{P}_{l}u'}{\mathbb{P}_{l}u}_{L^{2}(\mathbb{S}^{2})}.
\end{equation}
Defining $v_{l} = \mult{l}{\lambda}^{-2s}\mathbb{P}_{l}u'$, we see  from eq.(\ref{eq:sobdcon2}) that
\begin{equation}
  \sum_{l\in \mathbb{N}} \mult{l}{\lambda}^{2s}\|v_{l}\|^{2}_{L^{2}(\mathbb{S}^{2})}
  = \sum_{l\in \mathbb{N}} \mult{l}{\lambda}^{-2s}\|\mathbb{P}_{l}u'\|^{2}_{L^{2}(\mathbb{S}^{2})} < \infty.
\end{equation}
It follows that $v = \sum_{l\in \mathbb{N}} v_{l} \in H^{s}(\mathbb{S}^{2})$ and eq.(\ref{eq:sobrep1}) holds,
while trivially  $\|u'\|_{H^{s}(\mathbb{S}^{2})'} = \|v\|_{H^{s}(\mathbb{S}^{2})}$.
\eproof

It is worth emphasising that while the topological structure of $H^{s}(\mathbb{S}^{2})$
is independent of the specific inner product chosen (i.e. corresponding to
different values of $\lambda > 0$), the representation of a dual vector given by Theorem \ref{prop:sobrep}
is affected. Numerical examples  at the end of this section are used to illustrate this point.

\begin{figure}
  \centering
  \includegraphics[width=0.48\textwidth]{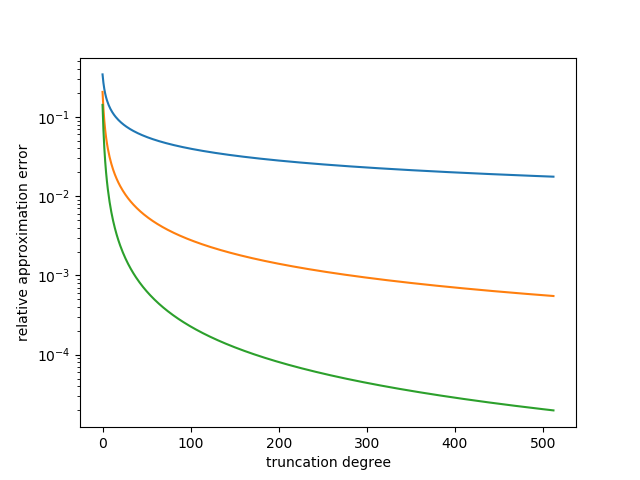}
  \includegraphics[width=0.48\textwidth]{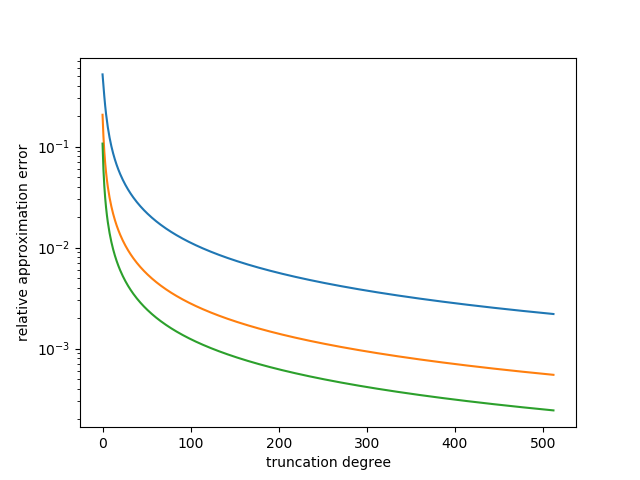}
  \caption{Variation with truncation degree of the relative approximation error $\left\|\hat{\delta}_{x}-\tilde{\hat{\delta}}_{x} \right\|_{H^{s}(\mathbb{S}^{2})}$  for point evaluation of a function.
    The left hand panel show the relative error for fixed $\lambda = 1.0$ for  three  values of
    the Sobolev exponent: 1.5 in blue, 2.0 in orange, and 2.5 in green.
    The right hand panel then fixes $s = 2.0$,
    and plots the relative error for $\lambda$ equal to: 0.5 in blue, 1.0 in orange, and 1.5 in green. 
 }  
  \label{fig:sob_trunc}
\end{figure}

\subsection{The Sobolev embedding theorem}

Elements of $H^{s}(\mathbb{S}^{2})$ for $s > 0$ form a subset of $L^{2}(\mathbb{S}^{2})$,
and so their point-values need not be defined. However, the famous Sobolev embedding theorem
shows that if the Sobolev exponent is sufficiently large, then elements of
$H^{s}(\mathbb{S}^{2})$ do have well-defined point values, and may also be
continuously differentiable up to a finite-order. A proof of this result will not be given, but one
can be found in Chapter 4 of \cite{taylor1996partial} which is valid for any compact manifold. Here it
should be noted again that there are several different means by which Sobolev spaces can be defined, but
this does not change either the validity nor form of the embedding theorem.

\begin{thm}
  \label{thm:sobemb}
  The Sobolev space  $H^{s}(\mathbb{S}^{2})$ for $s > k+1$
  is continuously, densely, and properly embedded into  $C^{k}(\mathbb{S}^{2})$.
\end{thm}

Assuming that $s > 1$,  we can now conclude that the Dirac measure $\delta_{x}$ belongs
to $H^{s}(\mathbb{S}^{2})'$, and so Theorem \ref{prop:sobrep} implies
\begin{equation}
  \label{eq:delrep1}
  u(x) = \cbraket{\hat{\delta}_{x}}{u}_{H^{s}(\mathbb{S}^{2})}, 
\end{equation}
for all $u \in H^{s}(\mathbb{S}^{2})$, where the $H^{s}(\mathbb{S}^{2})$-representation
of $\delta_{x}$ is given by
\begin{equation}
  \label{eq:delrep2}
  \hat{\delta}_{x} = \sum_{l\in \mathbb{N}} \mult{l}{\lambda}^{-2s} \sum_{m=-l}^{l}Y_{lm}(x) Y_{lm}.
\end{equation}
The Cauchy-Schwarz inequality leads to the  sharp bound
\begin{equation}
  \label{eq:cauchy_tmp}
  |u(x)| \le \|\hat{\delta}_{x}\|_{H^{s}(\mathbb{S}^{2})} \|u\|_{H^{s}(\mathbb{S}^{2})},
\end{equation}
where  $\|\hat{\delta}_{x}\|_{H^{s}(\mathbb{S}^{2})}$ can be determined using
the spherical harmonic addition theorem \citep[e.g.][Section B.6]{DT}
\begin{equation}
  \|\hat{\delta}_{x}\|^{2}_{H^{s}(\mathbb{S}^{2})} = \sum_{l \in \mathbb{N}}\mult{l}{\lambda}^{-2s}
  \sum_{m=-l}^{l}Y_{lm}(x)Y_{lm}(x) 
  = \sum_{l\in \mathbb{N}}\frac{2l+1}{4\pi}\mult{l}{\lambda}^{-2s}.
\end{equation}
Taking the supremum over $x \in \mathbb{S}^{2}$  in  eq.(\ref{eq:cauchy_tmp})
we   arrive at the the useful estimate
\begin{equation}
  \label{eq:c0bnd}
    \|u\|_{C^{0}(\mathbb{S}^{2})} \le \left(\sum_{l\in \mathbb{N}}\frac{2l+1}{4\pi}\mult{l}{\lambda}^{-2s}\right)^{\frac{1}{2}}
    \|u\|_{H^{s}(\mathbb{S}^{2})}, 
  \end{equation} 
for $u \in H^{s}(\mathbb{S}^{2})$ for $s > 1$. In a similar manner is can be shown that
when $s > k+1$ we have
\begin{equation}
  \label{eq:ckbnd}
  \|\Delta^{\frac{k}{2}} u\|_{C^{0}(\mathbb{S}^{2})}
  \le \left( \sum_{l\in \mathbb{N}}\frac{2l+1}{4\pi}\mult{l}{\lambda}^{-2s}
    [l(l+1)]^{k} \right)^{\frac{1}{2}}\|u\|_{H^{s}(\mathbb{S}^{2})}.
\end{equation}
Here $\Delta^{k/2}$ denotes a fractional power of the Laplace-Beltrami operator, this
being defined through its action
\begin{equation}
    \Delta^{\frac{k}{2}} u = \sum_{l \in \mathbb{N}} [l(l+1)]^{\frac{k}{2}} \mathbb{P}_{l}u, 
\end{equation}
for a suitably regular function. The utility of eq.(\ref{eq:ckbnd}) is that the  left hand
side provides a rotationally invariant, and computationally simple, proxy for the
magnitude of a functions $k$th derivatives.

\begin{figure}
  \centering
  \includegraphics[width=0.85\textwidth]{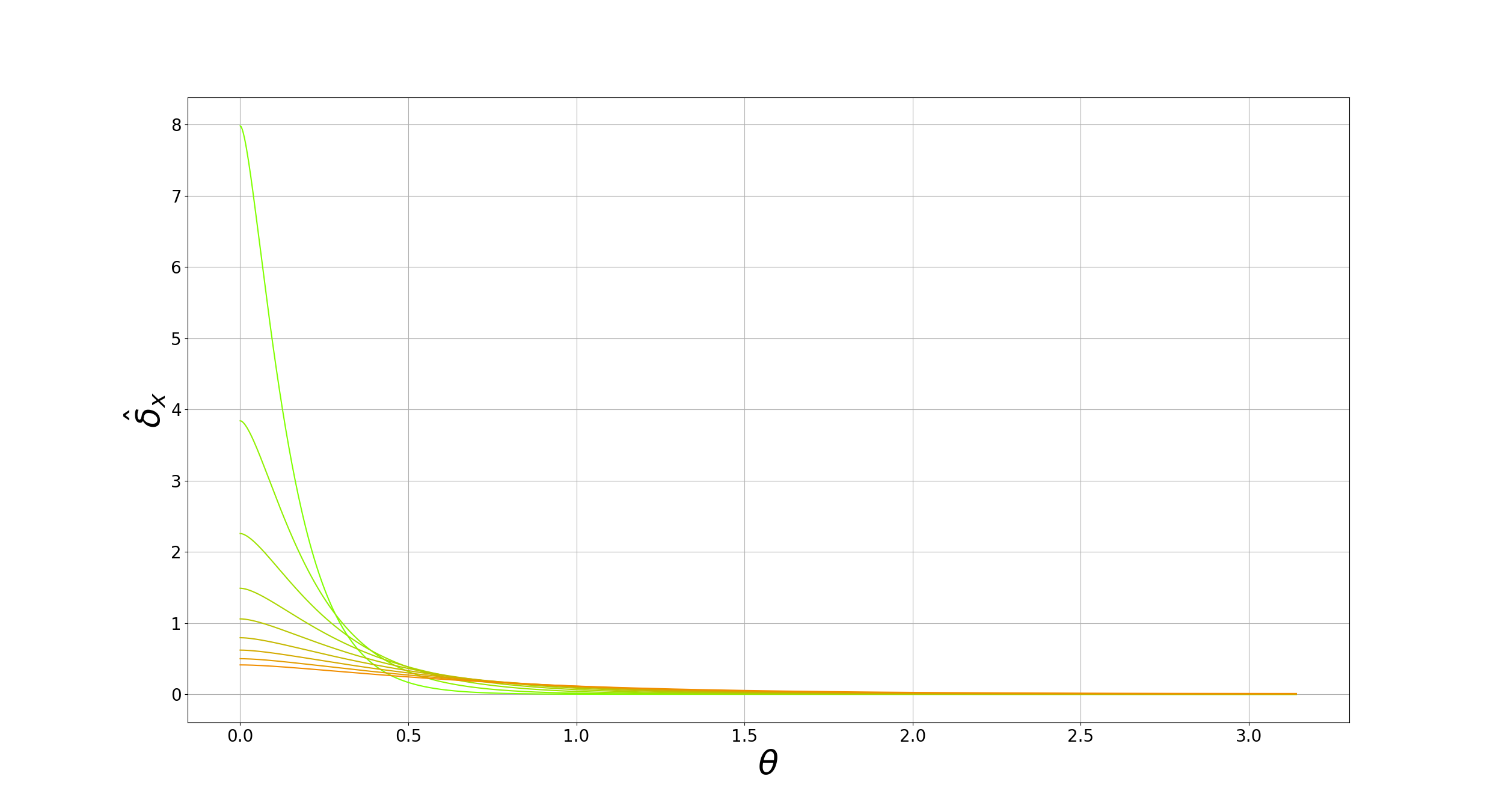}
  \includegraphics[width=0.85\textwidth]{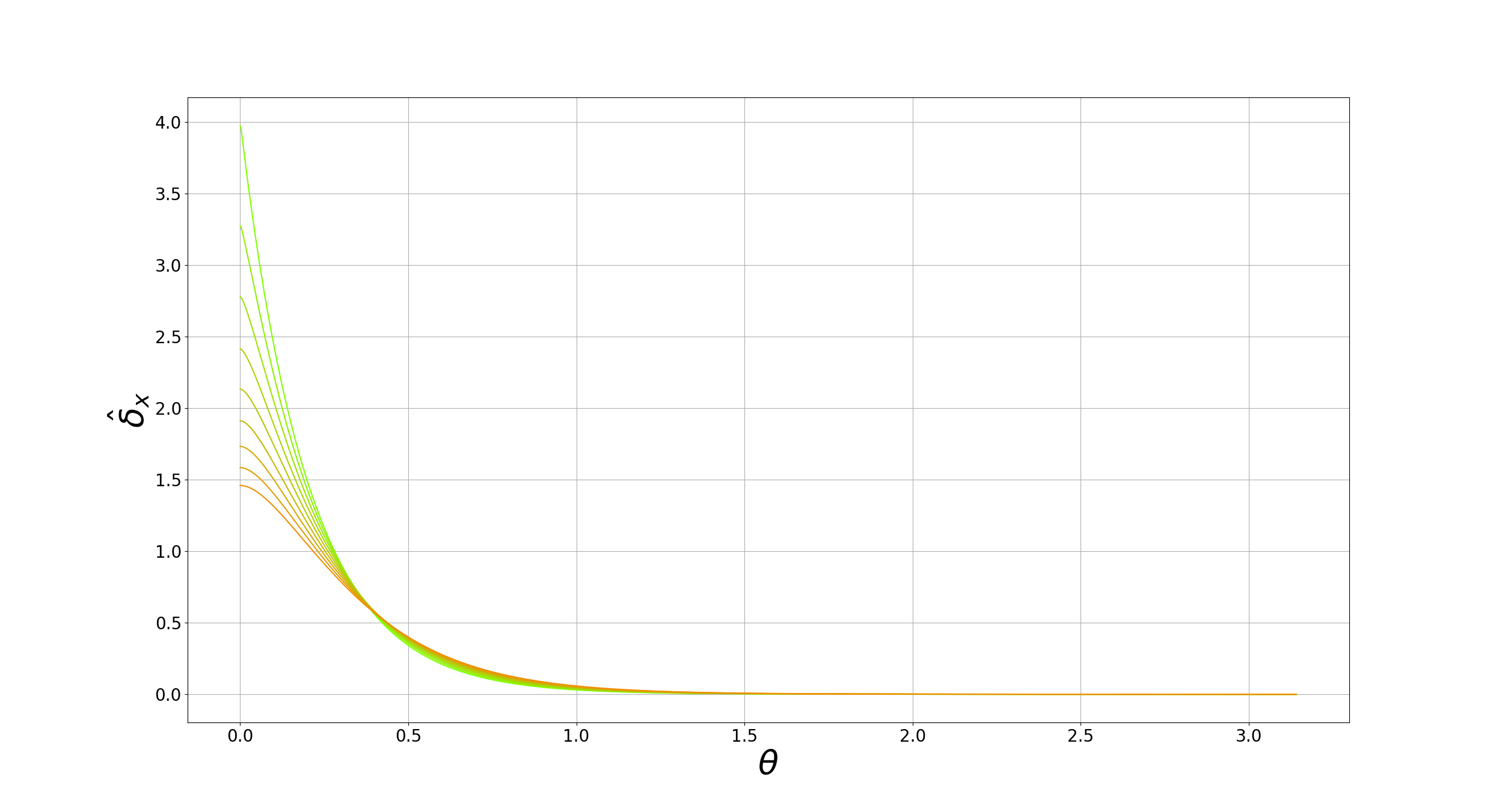}
  \caption{The $H^{s}(\mathbb{S}^{2})$-representation of a Dirac measure $\hat{\delta}_{x}$
    as a function of the angle $\theta$ defined within eq.(\ref{eq:delta_simp}).
    Within the upper plot the value of the Sobolev exponent $s$ is fixed at $2$,
    while $\lambda$ is spaced equally between $0.1$ (green) and
    $0.5$ (orange), with the colours grading smoothly between these limits.
    The lower plot is made in the same manner, but with fixed $\lambda = 0.2$ and
    $s$ ranging from $1.5$ (green) to $2.5$ (orange).
 }  
  \label{fig:delta_plot}
\end{figure}

\subsection{Numerical implementation}

\label{ssec:sobnum1}

The Sobolev spaces $H^{s}(\mathbb{S}^{2})$ are infinite-dimensional, and so their elements
must be suitably approximated within any numerical work. An obvious way to do this is with
truncated spherical harmonic expansions up to some maximum degree $L$. Thus, given
a function $u \in H^{s}(\mathbb{S}^{2})$, its truncated expansion to degree $L$ is defined by
\begin{equation}
  u_{L} = \sum_{l = 0}^{L}\mathbb{P}_{l}u.
\end{equation}
Consider a linear functional $u' \in H^{s}(\mathbb{S}^{2})'$, and let $v$
denote its $H^{s}(\mathbb{S}^{2})$-representation such that
\begin{equation}
  \braket{u'}{u} = \cbraket{v}{u}_{H^{s}(\mathbb{S}^{2})}, 
\end{equation}
for all $u \in H^{s}(\mathbb{S}^{2})$. Acting such a functional on $u_{L}$ we obtain
\begin{equation}
  \braket{u'}{u_{L}} =\sum_{l=0}^{L} \cbraket{v}{\mathbb{P}_{l}u}_{H^{s}(\mathbb{S}^{2})}
  =\sum_{l=0}^{L} \cbraket{\mathbb{P}_{l}v}{u}_{H^{s}(\mathbb{S}^{2})}
  = \cbraket{v_{L}}{u}_{H^{s}(\mathbb{S}^{2})}, 
\end{equation}
where $v_{L} $ denotes the truncated expansion of $v$.  The Cauchy-Schwarz inequality
leads to the sharp error-estimate
\begin{equation}
  |\braket{u'}{u} - \braket{u'}{u_{L}}| \le \|v-v_{L}\|_{H^{s}(\mathbb{S}^{2})} \|u\|_{H^{s}(\mathbb{S}^{2})}.
\end{equation}
It follows that if a prior bound on $\|u\|_{H^{s}(\mathbb{S}^{2})}$ is known, we can
choose a truncation degree $L$ such that the absolute  error in $\braket{u'}{u}$ is as small as desired.  This
method  trivially extends to  any finite collection of linear functionals, and in this manner
we can ensure that the numerical discretisation of a linear inference problem is sufficiently accurate.
To illustrate this idea, consider point-evaluation of a function. Here the
relevant functional is $\delta_{x}$ for some $x \in \mathbb{S}^{2}$, and
we  write $\hat{\delta}_{x}$ for its $H^{s}(\mathbb{S}^{2})$-representation
whose concrete form is given by  eq.(\ref{eq:delrep2}). Applying the spherical
harmonic addition theorem we obtain 
\begin{equation}
  \label{eq:delanorm}
  \left\|\hat{\delta}_{x}-\hat{\delta}_{x ,L} \right\|
     _{H^{s}(\mathbb{S}^{2})} = \left(\sum_{l=L+1}^{\infty}\frac{2l+1}{4\pi} \mult{l}{\lambda}^{-2s}\right)^{\frac{1}{2}}, 
\end{equation}
where $\hat{\delta}_{x ,L}$ denotes the truncated expansion of $\hat{\delta}_{x}$. Using this
expression, we plot in  Fig.\ref{fig:sob_trunc}  the variation of the relative approximation error with
truncation degree $L$  for a range of values of $s$ and $\lambda$. In practice, however, these
 worse-case error estimates are overly conservative, with convergence usually being
obtained at substantially lower truncation degrees.

Building on this example, we can examine visually   the dependence of the  $H^{s}(\mathbb{S}^{2})$-representation
of the Dirac measure $\delta_{x}$ on the parameters $s>1$ and $\lambda > 0$. Again using the
spherical harmonic addition theorem, eq.(\ref{eq:delrep2}) can be simplified to
\begin{equation}
  \label{eq:delta_simp}
  \hat{\delta}_{x}(y) = \sum_{l \in \mathbb{N}} \mult{l}{\lambda}^{-2s}\frac{2l+1}{4\pi} P_{l}(\cos \theta), 
\end{equation}
where $P_{l}$ is a Legendre polynomial of degree $l$, and $\theta$ is the angle between
 $x,y\in \mathbb{S}^{2}$. Plots of this function are shown in Fig.\ref{fig:delta_plot}
for a range of values of $s$ and $\lambda$. Within these calculations the truncation degree $L$
was chosen so  that the relative approximation error always lies below $10^{-5}$ which is
sufficient for visual inspection.
As a general trend, it is seen that as either $s>1$ or $\lambda>0$ decreases, the function
becomes more sharply peaked at $\theta = 0$. Nonetheless, in each case the
Dirac measure is, in accordance with the Sobolev embedding theorem, represented by a continuous function
that can be accurately captured by truncated spherical harmonic expansions.

\end{document}